\documentclass[aps,prapplied,longbibliography,twocolumn,10pt,showpacs,superscriptaddress]{revtex4-2}
\usepackage{amsmath,amssymb,amsfonts,mathtools}
\usepackage{graphicx}
\usepackage{textcomp}
\usepackage{braket}
\usepackage{physics}
\usepackage{tabularx}
\usepackage{xcolor}
\usepackage{multirow}
\usepackage{hyperref}
\usepackage{rotating}
\usepackage{booktabs}
\usepackage{dirtytalk}
\usepackage[table]{xcolor}
\usepackage[noend]{algpseudocode}
\usepackage{comment}
\usepackage[T1]{fontenc}
\usepackage{makecell}
\usepackage{svg}
\usepackage[utf8]{inputenc}
\usepackage[T1]{fontenc}
\usepackage{bm}
\usepackage{xr}
\usepackage{tikz}
\usetikzlibrary{arrows.meta, positioning}

\tikzset{
  box/.style={rectangle, rounded corners=4pt, minimum width=7.5cm, minimum height=1.2cm,
              text centered, draw=black, fill=blue!8, font=\small, align=center},
  decision/.style={diamond, aspect=2.5, minimum width=3cm, minimum height=1cm,
                   text centered, draw=black, fill=green!10, font=\small},
  output/.style={rectangle, rounded corners=4pt, minimum width=7.5cm, minimum height=1.4cm,
                 text centered, draw=black, fill=violet!10, font=\small, align=center},
  arrow/.style={-{Stealth}, thick},
}

\newcolumntype{Y}{>{\centering\arraybackslash}X}

\begin{document}
\definecolor{darkteal}{RGB}{0, 96, 100}
\definecolor{lightteal}{RGB}{178, 223, 219}
\definecolor{white}{RGB}{255,255,255}
\definecolor{black}{RGB}{0,0,0}

\title{Fault-tolerant modular quantum computing with surface codes using single-shot emission-based hardware}

\author{Siddhant Singh}
\email{siddhant.singh@tudelft.nl}
\affiliation{QuTech, Delft University of Technology, Lorentzweg 1, 2628 CJ Delft, The Netherlands}

\author{Rikiya Kashiwagi}
\affiliation{Graduate School of Informatics, Kyoto University, 36-1 Yoshida-Honmachi, Sakyo-ku, Kyoto 606-8501, Japan}
\affiliation{Department of Electronics and Electrical Engineering, Keio University, 3-14-1 Hiyoshi, Kohoku-ku, Yokohama 223-8522, Japan}

\author{Kazufumi Tanji}
\affiliation{Department of Electronics and Electrical Engineering, Keio University, 3-14-1 Hiyoshi, Kohoku-ku, Yokohama 223-8522, Japan}

\author{Wojciech Roga}
\affiliation{Department of Electronics and Electrical Engineering, Keio University, 3-14-1 Hiyoshi, Kohoku-ku, Yokohama 223-8522, Japan}

\author{Daniel Bhatti}
\affiliation{Networked Quantum Devices Unit, Okinawa Institute of Science and Technology Graduate University, Okinawa, Japan}

\author{Masahiro Takeoka}
\email{takeoka@elec.keio.ac.jp}
\affiliation{Department of Electronics and Electrical Engineering, Keio University, 3-14-1 Hiyoshi, Kohoku-ku, Yokohama 223-8522, Japan}
\affiliation{National Institute of Information and Communications Technology (NICT), Koganei, Tokyo 184-8795, Japan}

\author{David Elkouss}
\email{david.elkouss@oist.jp}
\affiliation{QuTech, Delft University of Technology, Lorentzweg 1, 2628 CJ Delft, The Netherlands}
\affiliation{Networked Quantum Devices Unit, Okinawa Institute of Science and Technology Graduate University, Okinawa, Japan}

\date{\today}

\begin{abstract}
Fault-tolerant modular quantum computing requires stabilizer measurements across the modules in a quantum network. For this, entangled states of high quality and rate must be distributed. Currently, two main types of entanglement distribution protocols exist, namely emission-based and scattering-based, each with its own advantages and drawbacks.
On the one hand, scattering-based protocols with cavities or waveguides are fast but demand stringent hardware such as high-efficiency integrated circulators or strong waveguide coupling. On the other hand, emission-based platforms are experimentally feasible but so far rely on Bell-pair fusion with extensive use of slow two-qubit memory gates, limiting thresholds to $\approx 0.16\%$.
Here, we consider a fully distributed surface code using emission-based entanglement schemes that generate GHZ states in a single shot, i.e., without the need for Bell-pair fusions. 
We show that our optical setup produces Bell pairs, W states, and GHZ states, enabling both memory-based and optical protocols for distilling high-fidelity GHZ states with significantly improved success rates.
Furthermore, we introduce protocols that completely eliminate the need for memory-based two-qubit gates, achieving thresholds of $\approx 0.19\%$ with modest hardware enhancements, increasing to above $\approx 0.24\%$ with photon-number-resolving detectors. These results show the feasibility of emission-based architectures for scalable fault-tolerant operation.
\end{abstract}

\maketitle

\section{\label{sec:introduction}Introduction}
Scaling quantum computation to execute useful algorithms requires millions of physical qubits~\cite{Gidney2021howtofactorbit, gidney2025factor2048bitrsa}. The modular quantum computing paradigm, comprising independent yet interconnected quantum modules, addresses this scalability challenge~\cite{mohseni2024build, Megrant2025,PhysRevX.14.041030,singh2024modulararchitecturesentanglementschemes,CALEFFI2024110672, Main2025, yoder2025tourgrossmodularquantum, AghaeeRad2025}. To enable this modularity in practice, several methods have been proposed allowing qubits in separate modules to interact with high success rates and fidelity~\cite{monroeLargescaleModularQuantumcomputer2014,10.1145/3233188.3233224,8910635,7562346,https://doi.org/10.1049/iet-qtc.2020.0002,doi:10.1126/science.aam9288,sunamiScalableNetworkingNeutralAtom2025}. These architectures must incorporate quantum error correction (QEC) to suppress errors~\cite{Roffe_2019,Lidar_Brun_2013,fujii2015quantumcomputationtopologicalcodes,Campbell2017,Acharya2025,Lacroix2025,eickbusch2025demonstratingdynamicsurfacecodes,Bravyi2024,Bluvstein2024,Xu2024,doi:10.1126/science.adr7075,PhysRevLett.133.240602}. %While different error correction codes exist~\cite{Bravyi2024,landahl2011faulttolerantquantumcomputingcolor,dinur2022goodquantumldpccodes,PhysRevX.13.041052,PhysRevX.11.041058,Brown2016,Alam2025}, the surface code has emerged as a leading candidate~\cite{Kitaev2003,Fowler2012,Dennis2002, Acharya2023,Litinski_2019,McEwen_2023}. 
Due to its high error correction threshold close to half a percent, the surface code has emerged as a leading candidate~\cite{Kitaev2003,Fowler2012,Dennis2002, Acharya2023,Litinski_2019,McEwen_2023}, guiding both algorithm development and hardware co-design~\cite{PhysRevA.106.062428,Watkins_2024,Erhard2021,delfosse2020hierarchicaldecodingreducehardware,das2020scalabledecodermicroarchitecturefaulttolerant}.
Recent literature has extended these QEC techniques to modular systems, proposing a diverse range of implementations~\cite{Nickerson2013,PhysRevX.4.041041,10.1116/5.0200190,PRXQuantum.4.020321,de_Gliniasty_2024,bombin2021interleavingmodulararchitecturesfaulttolerant,bombín2023modulardecodingparallelizablerealtime,PhysRevResearch.7.013313,sutcliffe2025distributedquantumerrorcorrection,stack2025assessingteleportationlogicalqubits,doi:10.1126/science.adr7075}.

In this context, various entanglement generation schemes have been introduced and partially demonstrated~\cite{singh2024modulararchitecturesentanglementschemes,PhysRevLett.123.063601,Pompili_2021,Hermans2022,Ruskuc2025,Evans_2018,Knaut2024, Awschalom2018,Hofmann2012,Hucul2015,Stephenson2020,vanleentEntanglingSingleAtoms2022,sahaHighfidelityRemoteEntanglement2025a,liHighRateHighFidelityModular2024,kikuraPassiveQuantumInterconnects2025} in the context of color‑centers~\cite{inc2024distributedquantumcomputingsilicon,Bradley2022}, neutral atoms~\cite{Henriet_2020,Ramette2022,PRXQuantum.5.020363}, and trapped‑ions~\cite{ryananderson2024highfidelityfaulttolerantteleportationlogical,PhysRevX.11.041058}. 
One such scheme, already demonstrated experimentally, is the emission‑based approach~\cite{PhysRevA.59.1025,Barrett_2005}. It generates entanglement between different modules by measuring photons emitted from each module's emitter/communication qubit and transmitted through an optical fiber. Then one builds a modular QEC architecture by using the module's memory qubits as code data qubits. Typically, these memory qubits are also controlled via the emitter qubit.

\begin{figure*}
    \centering
    \includegraphics[width=\textwidth]{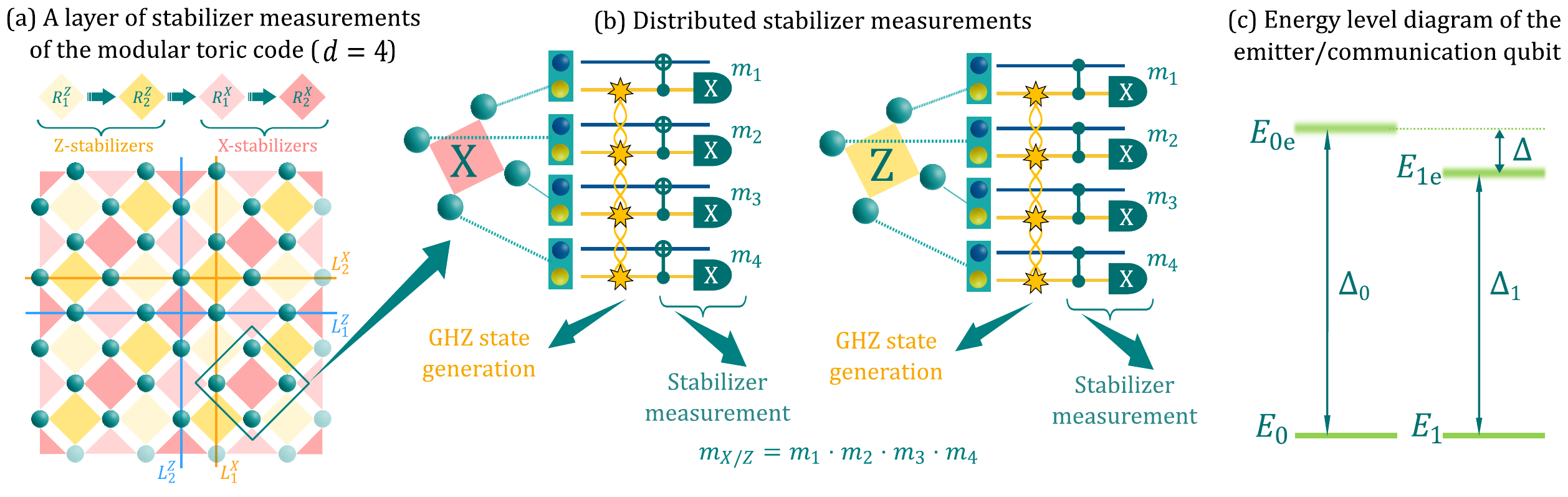}
    \caption{Modular toric surface code. (a) A direct translation of the monolithic surface code of distance $d=4$ into a modular architecture, where each module (teal circles) hosts a data qubit and a communication qubit (dark blue and yellow circles in (b),  respectively). Each stabilizer spans four modules, which share GHZ states generated using the proposed single-shot emission-based protocols. As each module has only one communication qubit (yellow), simultaneous measurement of neighboring stabilizers is not possible. Thus, the stabilizers are measured in four sequential rounds $R^{X,Z}_{1,2}$, as shown. Logical operators are also shown for both encoded logical qubits. (b) Distributed stabilizer measurement circuits using a GHZ state in the form $\ket{\Phi^+_4}=\tfrac{1}{\sqrt{2}}(\ket{0000}+\ket{1111})$. The stars (in yellow) connected via links denote entanglement generation on those communication qubits, either a raw GHZ state or a distilled GHZ state using distillation protocols. The joint parity is extracted by local operations within each module, followed by classical post-processing. The measurement outcomes for the stabilizer syndrome are calculated as products of individual outcomes. (c) Energy level diagram of the communication (emitter) qubit, showing the ground states $E_0$ and $E_1$ (corresponding to dark and bright states) and the excited states $E_{0e}$ and $E_{1e}$, along with their associated linewidths. For the noise model (see App.~\ref{app:Noise_model}), this is the emitter model we have followed.
}
    \label{fig:modular_surface_code}
\end{figure*}

A fully distributed surface‑code experiment requires degree-four nearest neighbor connectivity to measure stabilizers~\cite{Nickerson2013,10.1116/5.0200190,singh2024modulararchitecturesentanglementschemes}. One approach is to distill and fuse Bell‑pairs to create high‑fidelity four-partite GHZ states as a resource~\cite{Nickerson2013,10.1116/5.0200190,singh2024modulararchitecturesentanglementschemes}. 
However, the two‑qubit gates required for fusion and distillation are a bottleneck in modular architectures because emitter-qubit gates can be orders of magnitude faster than memory-qubit gates, as is the case in color‑centers~\cite{inc2024distributedquantumcomputingsilicon,Bradley2022}. 
%This is the case even when multiplexing entanglement generation~\cite{liHighRateHighFidelityModular2024,PhysRevResearch.3.043154}.
Although memory-qubit gate speed can also be improved through optimization~\cite{schaferFastQuantumLogic2018a} or alternative gate protocols~\cite{chewUltrafastEnergyExchange2022a}, enhancing gate speed while keeping high fidelity is more difficult.
%However, the two‑qubit gates required for fusion and distillation are a bottleneck in modular architectures because memory‑qubit gates are orders of magnitude slower than emitter‑qubit gates, as is the case in color‑centers~\cite{inc2024distributedquantumcomputingsilicon,Bradley2022}, neutral atoms~\cite{Henriet_2020,Ramette2022,PRXQuantum.5.020363}, and trapped‑ions~\cite{ryananderson2024highfidelityfaulttolerantteleportationlogical,PhysRevX.11.041058}. 
Consequently, the performance of these fusion-based schemes was estimated to saturate around a gate error threshold of $0.16\%$ even with hardware improvements~\cite{singh2024modulararchitecturesentanglementschemes}. As an alternative, several scattering-based schemes have been proposed to avoid this limitation, but they demand highly complex hardware that is unlikely to be available in the near term~\cite{singh2024modulararchitecturesentanglementschemes}.

To overcome these bottlenecks, here we consider a recent generalization of the bipartite entanglement emission-based scheme to multiple parties~\cite{Bartolucci2021,rogaEfficientDickestateDistribution2023,Shimizu2025}. Such extensions avoid Bell‑pair fusion and enable the direct distribution of multipartite entanglement in a single shot. However, the performance of these direct schemes on realistic hardware remains unknown, as does the potential for more efficient protocols based on this single-shot setup that exploit the existing setup to produce high‑fidelity GHZ states. We pose the following questions to investigate whether the single-shot scheme can outperform the fusion-based schemes. Which near-term upgrade yields the largest threshold gain? Do thresholds scale continuously with hardware, unlike the fusion-based schemes? And what concrete requirements enable early fault-tolerant demonstrations?

In this work, we put forward these single-shot protocols by considering a four‑module setup designed to generate entangled states such as raw GHZ states, Bell pairs, and W states between modules in the single-shot run of the setup. We propose protocols that use these states to produce high‑fidelity GHZ states either through limited distillation using quantum memory gates or through purely optical distillation without the use of memory gates. We benchmark the best protocols for modular error correction by identifying fault‑tolerant regions for realistic hardware parameters of color centers in diamond. These parameters include losses in optical components, photon emission and detection quality, and circuit‑level noise in qubit operations.

\begin{figure*}
    \centering
    \includegraphics[width=\textwidth]{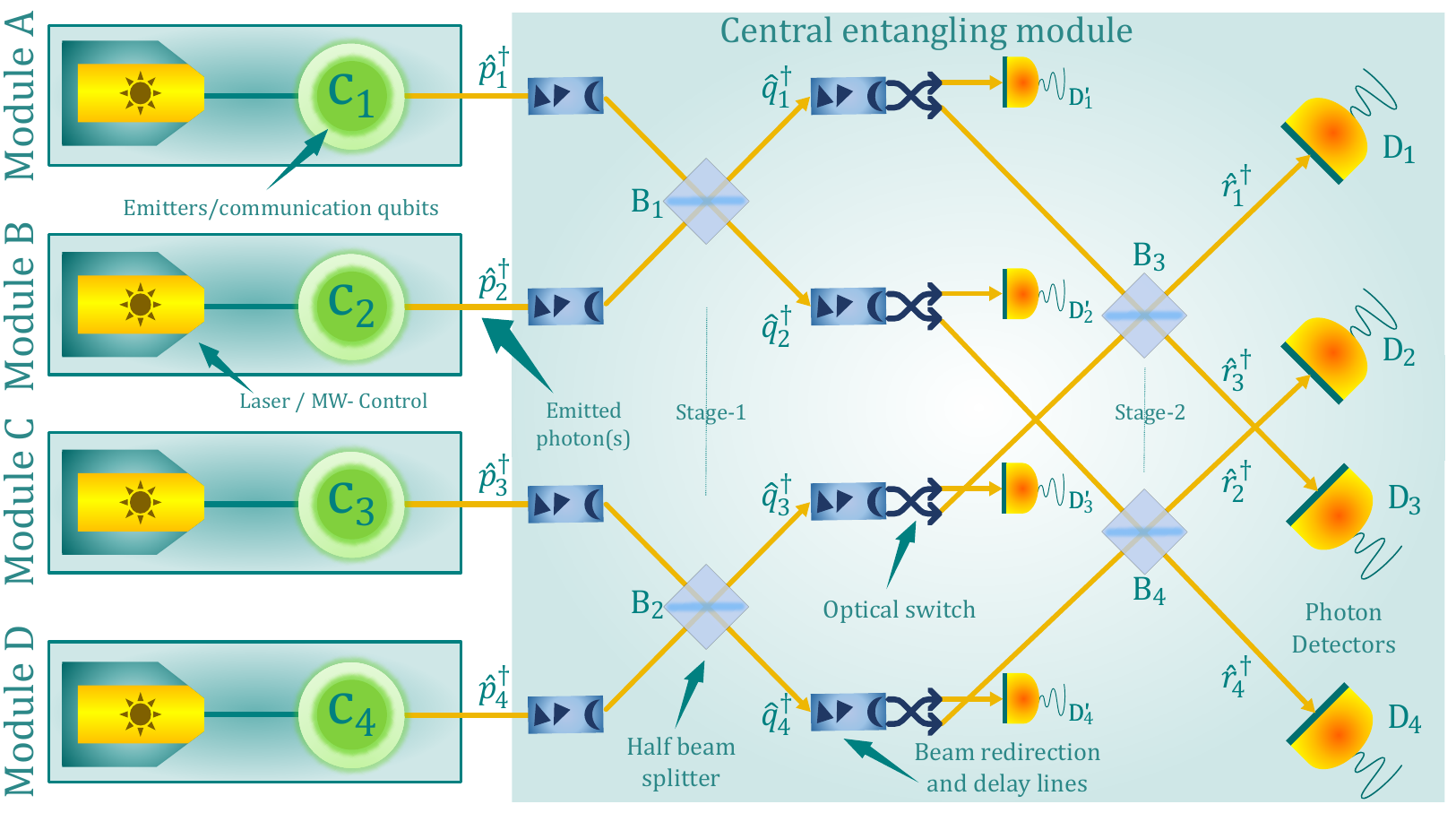}
    \caption{Hardware setup for the emission-based scheme that generates entanglement links—Bell pairs, W states, and GHZ states of weight-4. Emitter (communication) qubits $c_1,c_2,c_3,c_4$ are excited with a laser pulse to emit photons. These photons propagate through a network of beam splitters (stage-1 and stage-2), which mix the modes of all emitted photons. We observe detection patterns as photon counts at detectors $\text{D}_1,\text{D}_2,\text{D}_3,\text{D}_4$ or at detectors $\text{D}_1',\text{D}_2',\text{D}_3',\text{D}_4'$ depending on which states are created. Post-selecting for specific patterns generates the target states. Creation operators for photonic number states $\hat{p}_i^{\dagger},\hat{q}_i^{\dagger}$ and $\hat{r}_i^{\dagger}$ ($i=1,2,3,4$) are indicated at various input and output ports of the beam splitters.
}
    \label{fig:direct_emission_setup}
\end{figure*}

Our results show that the existing emission‑based hardware can support modular error correction and clarify the hardware requirements to achieve the surface code error thresholds competitive with monolithic architectures. Notably, our optical distillation protocols are fast and enable direct GHZ generation without fusion, yielding code thresholds of $0.19\%$ for non‑photon‑resolving detectors and $0.24\%$ for number‑resolving detectors with only modest improvements in hardware parameters. These thresholds scale with improving hardware parameters, unlike the Bell-pair fusion-based schemes. This establishes a clear pathway toward scalable modular architectures with emission‑based hardware. This article is organized as follows. We present the physical setup for modular QEC in Sec.~\ref{sec:Physical_setup}. Following that, Sec.~\ref{sec:GHZ_state_generation} describes the novel protocols in the noiseless regime that are based on this single-shot emission setup for high-fidelity GHZ protocols. Finally, we present the protocols' hardware performance with full noise, fault-tolerance hardware requirements, and threshold results in Sec .~\ref {sec:results}.

\section{\label{sec:Physical_setup}Physical setup for modular QEC with single-shot emission-based hardware}

\subsection{\label{subsec:arch_overview_comparison}Architecture comparison and overview}

Recent progress in modular fault-tolerant quantum computing has produced a diverse landscape of architectural proposals, which differ substantially in their resource assumptions and operational regimes. Lattice surgery and transversal gate methods~\cite{marqversen2025faulttolerantinterfacesmodularquantum} and constant-rate entanglement distillation schemes~\cite{pattison2024fastquantuminterconnectsconstantrate} adopt a semi-distributed architecture. Each module hosts a significant portion of a logical qubit or an entire logical block, and nonlocal operations are performed at the logical level between modules. This substantially relaxes the demands on intermodule link fidelity and rate, at the cost of requiring many physical qubits, high-fidelity two-qubit gates, and large local memory resources per module. Similarly, the fault-tolerant optical interconnect scheme of Ref.~\cite{PhysRevResearch.7.013313} operates at the logical level on neutral-atom arrays with Rydberg gates, targeting the high-local-resource regime.

The present work operates in a fundamentally different regime. We consider a fully distributed modular architecture in which each module hosts exactly one data qubit and a small number of distillation-ancilla memory qubits, with the surface code distributed at the single-qubit level and all stabilizer measurements requiring intermodule entanglement. Crucially, our formalism reduces modular fault tolerance to the reliable generation of high-fidelity multipartite entangled states for stabilizer measurements, which is the fundamental primitive underlying all higher-level operations, including lattice surgery and transversal gates. Any modular code architecture that requires nonlocal stabilizer measurements can therefore directly leverage the protocols introduced here. This minimal-resource-per-module regime directly targets color-center platforms: the fault-tolerant operation of a logical qubit within a single NV-center device has already been demonstrated~\cite{Abobeih2022}, and the present work is the natural extension of that result to the modular setting. This addresses the intermodule entanglement generation bottleneck that must be overcome to scale beyond a single module. Our single-shot emission-based protocols are specifically designed for this regime, eliminating memory-gate requirements entirely in the optical distillation protocols, with hardware requirements fully compatible with demonstrated NV-center devices.

\subsection{\label{subsec:our_arch}Single-shot modular architecture}
We consider a physical architecture in which multiple modules are interconnected through optical links that enable entanglement generation on demand. For clarity, we describe the setup in two parts: (i) the quantum error correction machinery implementing the surface code in the memory, and (ii) the optical interface that establishes entanglement between modules to support these operations. We consider a distributed quantum architecture in which multiple modules are interconnected through optical links that generate entanglement on demand. These links are assumed to be short-distance links or rack-size assembled quantum computer scale or integrated photonic waveguides. Each module contains a communication qubit, responsible for emitting photons, mediating inter-module entanglement, and acting as a control for local gates, together with several memory qubits. The memory qubits serve either as data qubits storing the logical information or as auxiliary qubits assisting in GHZ-state generation, depending on the protocol~\cite{10.1116/5.0200190,singh2024modulararchitecturesentanglementschemes,Nickerson2013,PhysRevX.4.041041}. Only the communication qubit can be directly measured, while memory qubits are read out indirectly via \textsc{swap} operations. The physical platform we consider is inspired by color centers in diamond~\cite{PhysRevX.9.031045,Bradley2022}. By arranging such modules into a two-dimensional code layout, one realizes a modular surface code capable of performing stabilizer measurements through distributed multi-qubit entanglement (see Fig.~\ref{fig:modular_surface_code}). Our construction follows the weight-4 architecture of Ref.~\cite{singh2024modulararchitecturesentanglementschemes}, exemplified by a modular toric code with periodic boundaries; planar variants without periodicity can be defined analogously. Stabilizer measurements are carried out using GHZ states distributed across modules via the optical links. Measuring the GHZ qubits yields correlated outcomes corresponding to parity checks, which are then processed by the Union-Find decoder to identify and apply the appropriate corrections.

For the optical interfaces, we consider a setup that allows GHZ generation for stabilizer measurements described above. We refer to this setup as a ``single-shot'' emission-based scheme. As we will show, this setup allows GHZ state generation between the communication qubits in one single run of the experiment. The focus is on four modules (see Fig.~\ref{fig:direct_emission_setup}) unit cell that is crucial for all stabilizer measurements. 
We first describe the hardware parameters and the relevant error sources in the emission-based setup. Subsequently, we present the different types of entanglement links that can be generated in the experiment. Each module is equipped with a local laser or microwave control, which can manipulate the communication qubit by applying local operations or by exciting it for emission. The initialization parameter for the communication-qubit state is the bright-state parameter $\alpha$, which controls the population of the ground state $\ket{0}$ (with energy $E_0$) and the excited (bright) state $\ket{1}$ (with energy $E_1$) (energy level diagram of the emitter is shown in Fig.~\ref{fig:modular_surface_code}(c)). The local communication-qubit state can be initialized with a fidelity $F_\text{prep}$ and amplitude controlled by $\alpha$. A subsequent local control pulse can then be used to emit a photon from the excited state, which becomes entangled with the communication qubit. Ideally, this process generates the emitter–photon entangled state $\sqrt{1-\alpha}\,\ket{0}\ket{0_{\text{ph}}} + \sqrt{\alpha}\,\ket{1}\ket{1_{\text{ph}}}$, with $\ket{0_{\text{ph}}}$ ($\ket{1_{\text{ph}}}$) being the zero-photon (single-photon) Fock state. The parameter $\alpha$ is tunable in the experiment.

During an excitation attempt, more than one photon may be emitted, resulting in a double-excitation error. If this unwanted photon is lost to the environment, it can induce dephasing in the state of the emitter. This effect was modeled in Ref.~\cite{singh2024modulararchitecturesentanglementschemes} by a dephasing probability $p_\text{EE}$. In this work, we neglect this parameter under the following assumptions: The emitter is placed in an optical resonator such as a cavity~\cite{PRXQuantum.5.010202}, the excitation pulse is shorter than the cavity-enhanced decay time, and the excitation pulse is optimized so that other transitions undergo a full cycle and return to the ground state~\cite{PhysRevA.104.052604,mirambell2019fidelity}.

As every photon propagates through an optical fiber, it may be lost due to transmission inefficiencies or may not be registered by an imperfect detector. We capture both effects using an effective photon detection efficiency, $\eta_\text{ph}$. Because the physical distance from the emitter modules to the central entangling module is much larger than the distance from the beam splitters to the detectors, we assume photon loss occurs before the beam splitters. Finally, photons emitted from different modules are not perfectly identical, so their interference at the beam splitter is not ideal. We model this effect with a two-photon visibility factor $\mu$. Full details of these noisy hardware parameters, their modeling, and their effects on the quantum states are provided in App.~\ref{app:Noise_model}. We note that detectors' capability for number resolving is also taken as a hardware parameter, and we model both photon-number-resolving (PNR) detectors and non-PNR detectors.

\section{\label{sec:GHZ_state_generation}GHZ protocols for single-shot emission scheme}

Based on the emission-based setup described in the previous section, we propose several entanglement-generation protocols for high-fidelity GHZ states. In this section, we present the noiseless working principles of these protocols, and in App.~\ref{app:Noise_model} we provide a recipe to construct full noisy analytical simulations by explaining the noise sources and the error channels we used for their modeling. First, we show how the elementary entangled states can be generated using this setup, and then we describe the high-fidelity GHZ protocols in detail and calculate their success probabilities and the fidelities of the resulting GHZ states. We classify the protocols into two categories: i) Memory-based distillation (MD) protocols that apply two-qubit gates on the memory qubits of the modules to perform distillation. ii) Optical distillation (OD) protocols that avoid any two-qubit gates on the memory qubits and instead rely on multiple rounds of emission operations and local qubit rotations on the communication qubits.

Throughout this paper, we adopt a unified naming convention for all protocols. The Bell, W, or GHZ elementary links are referred to by their names. Memory distillation protocols are labeled MD-XY, where XY encodes the two input resource states. The first state is the distilling state, and the second is the distilled state. Optical distillation protocols are labeled OD-XY, where XY denotes the states used in the first and second rounds, respectively.  Since all OD and MD protocols produce a GHZ output, we do not explicitly state this in the protocol naming. The full protocol index is given in Table~\ref{tab:protocol_naming}.

\begin{table}[h]
\centering
\caption{Protocol naming convention.}
\label{tab:protocol_naming}
\scriptsize

\begin{tabular}{lll}
\hline
\textbf{Short Label} & \textbf{Full Name} & \textbf{Type} \\
\hline
SC-Bell   & Single-click Bell link   & Elementary \\
DC-Bell   & Double-click Bell link   & Elementary \\
Raw-W     & Raw W link               & Elementary \\
Raw-GHZ   & Raw GHZ link             & Elementary \\
MD-BG(SC) & MD: SC-Bell + GHZ        & Memory distillation \\
MD-BG(DC) & MD: DC-Bell + GHZ        & Memory distillation \\
MD-WG     & MD: W + GHZ              & Memory distillation \\
MD-WW     & MD: W + W                & Memory distillation \\
MD-GG     & MD: GHZ + GHZ            & Memory distillation \\
OD-GHZ    & OD: GHZ + GHZ            & Optical distillation \\
OD-W      & OD: W + W                & Optical distillation \\
\hline
\end{tabular}
\end{table}

\subsection{\label{subsec:raw_protocol}Creating elementary states}
\subsubsection{\label{subsubsec:create_bell}Creating a Bell state}
As shown in Fig.~\ref{fig:direct_emission_setup}, one can use detectors $\text{D}_1',\text{D}_2'$ ($\text{D}_3',\text{D}_4'$) to generate entangled Bell pair links between the subsystems A and B (C and D). The optical switches re-route the incoming photons to the selected set of detectors. We consider both the single-click and double-click protocols for generating Bell-pair links between these modules. Both protocols were studied in Ref.~\cite{singh2024modulararchitecturesentanglementschemes}, and here we restate only their noiseless working principles.

First, each module A and B (C and D) generates a local emitter–photon state. The total state is
\begin{equation}
  |\psi_\mathrm{init}\rangle=\left(  \sqrt{1-\alpha}\,\ket{00_{\text{ph}}} + \sqrt{\alpha}\,\ket{11_{\text{ph}}}\right)^{\otimes 2},
  \label{eqn:emitter_photon_bell}
\end{equation}
where the subscript `ph' labels the number state of the photon(s).
Going through a balanced two-mode beam splitter, the transformation for the photonic creation operators can be modeled as
\begin{equation}
\begin{pmatrix}
\hat p_1^\dagger \\
\hat p_2^\dagger
\end{pmatrix}
=
\frac{1}{\sqrt{2}}
\begin{pmatrix}
1 & 1 \\
1 & -1
\end{pmatrix}
\begin{pmatrix}
\hat q_1^\dagger \\
\hat q_2^\dagger
\end{pmatrix},
\end{equation}
where $\hat{p}_i^\dagger$ and $\hat{q}_j^\dagger$ are photonic creation operators for input and output ports for stage-1 beam splitters (see Fig.~\ref{fig:direct_emission_setup}).
This transformation maps the photon number states from input ports to output ports as follows:
\begin{align}
\ket{0_\text{ph}0_\text{ph}}_{\mathrm{in}} &\;\rightarrow\; \ket{0_\text{ph}0_\text{ph}}_{\mathrm{out}}, \\
\ket{0_\text{ph}1_\text{ph}}_{\mathrm{in}} &\;\rightarrow\; \tfrac{1}{\sqrt{2}}\!\left( \ket{1_\text{ph}0_\text{ph}}_{\mathrm{out}}-\ket{0_\text{ph}1_\text{ph}}_{\mathrm{out}}\right), \\
\ket{1_\text{ph}0_\text{ph}}_{\mathrm{in}} &\;\rightarrow\; \tfrac{1}{\sqrt{2}}\!\left(\ket{0_\text{ph}1_\text{ph}}_{\mathrm{out}} + \ket{1_\text{ph}0_\text{ph}}_{\mathrm{out}}\right), \\
\ket{1_\text{ph}1_\text{ph}}_{\mathrm{in}} &\;\rightarrow\;\tfrac{1}{\sqrt{2}} \!\left(\ket{2_\text{ph}0_\text{ph}}_{\mathrm{out}}-\ket{0_\text{ph}2_\text{ph}}_{\mathrm{out}} \!\right).
\end{align}
Therefore, the joint emitter-photon state of the two modules becomes
\begin{equation}
\begin{aligned}
\ket{\psi_\mathrm{final}} & = \,(1-\alpha)\ket{00}\ket{0_\mathrm{ph}0_\mathrm{ph}} \\
&\phantom{{}={}} + \sqrt{\tfrac{\alpha(1-\alpha)}{2}}\!\left(\ket{01}+\ket{10}\right)\ket{0_\mathrm{ph}1_\mathrm{ph}} \\
&\phantom{{}={}} - \sqrt{\tfrac{\alpha(1-\alpha)}{2}}\!\left(\ket{01}-\ket{10}\right)\ket{1_\mathrm{ph}0_\mathrm{ph}} \\
&\phantom{{}={}}+ \tfrac{\alpha}{\sqrt{2}}\ket{11}\!\left(\ket{2_\mathrm{ph}0_\mathrm{ph}}-\ket{0_\mathrm{ph}2_\mathrm{ph}}\right).
\end{aligned}
\end{equation}

Observe that if we now obtain a single-photon click in detector $\mathrm{D}_1'$ ($\mathrm{D}_2'$), we obtain the perfect Bell state of the emitters as $\ket{\psi_\mathrm{emitters,Bell}^\mathrm{PNR}}=\ket{\Psi^\mp} = \tfrac{1}{\sqrt{2}}\!\left(\ket{01} \mp \ket{10}\right)$. For any other detection pattern, we abort the protocol and repeat until success (RUS). As pointed out above, the detectors may not be photon-number resolving (PNR), and two photons can be registered as a false single click, which adds the term $\ket{11}$ to the emitters' density matrix. In the case of non-PNR detectors, for instance, when $\mathrm{D}_2'$ clicks, we obtain a mixed state (reduced density matrix state after tracing out the photons)
\begin{equation}
\rho_\mathrm{emitters,Bell}^\mathrm{non-PNR}=(1-\alpha)\ket{\Psi^+}\bra{\Psi^+}+\alpha \ket{11}\bra{11}.
\label{eqn:bell_single_click_state}
\end{equation}
Note that here we have neglected dark counts, which correspond to detectors registering a false click. Since we assume sufficiently fast remote entanglement generation, the probability of dark counts is very small and can be safely neglected~\cite{shibataUltimateLowSystem2015}. Looking at Eq.~(\ref{eqn:bell_single_click_state}), it is clear that in the non-PNR case, one needs a vanishingly small $\alpha$ to achieve a high-fidelity Bell pair.

To further improve the Bell-pair fidelity with non-PNR detectors, we employ the double-click (DC) protocol (also known as Barrett–Kok protocol~\cite{Barrett_2005}). Local bit-flip operations are applied to the emitters, flipping the undesired two-photon-emitting term ($\ket{11}\bra{11}$) into the non-emitting term ($\ket{00}\bra{00}$), leaving the term $\ket{\Psi^+}\bra{\Psi^+}$ unchanged. Another round of emission is then executed, yielding a perfect Bell pair upon another single click in one of the detectors, even with non-PNR detectors. Note that this requires two successive rounds of detection patterns, implemented in an RUS manner. And the overall success rate will depend quadratically on the probability of such single clicks. Full noisy analytical results for Bell-pair creation were previously reported in Ref.~\cite{singh2024modulararchitecturesentanglementschemes}, and we follow this model identically with the same noise model.

\subsubsection{\label{subsubsec:create_w}Creating a W state}
Now we focus on how to craft weight-4 W states using the same experimental setup. In this case, the optical switches re-route the incoming photons to the stage-2 beam splitters, after which they are detected at the final detectors $\mathrm{D_1},\mathrm{D_2},\mathrm{D_3},\mathrm{D_4}$ (see Fig.~\ref{fig:direct_emission_setup}). Local emitter–photon states are created within each module A, B, C, and D, analogous to Eq.~(\ref{eqn:emitter_photon_bell}). The full unitary operation of all the beam splitters, mapping the input creation operators $\hat{p}_i^\dagger$ to the output operators $\hat{r}_j^\dagger$, can be modeled as (see App.~\ref{app:povm})~\cite{rogaEfficientDickestateDistribution2023}:
\begin{equation}
\begin{pmatrix}
\hat p_1^\dagger \\
\hat p_2^\dagger \\
\hat p_3^\dagger \\
\hat p_4^\dagger 
\end{pmatrix}
=
\frac{1}{2}
\begin{pmatrix}
1 & 1 & 1 &1 \\
1 & -1 & 1 & -1 \\
1 & 1 & -1 & -1 \\
1 & -1 & -1 & 1
\end{pmatrix}
\begin{pmatrix}
\hat r_1^\dagger \\
\hat r_2^\dagger \\
\hat r_3^\dagger \\
\hat r_4^\dagger 
\end{pmatrix} .
\label{eqn:4_beamsplitter_transformation}
\end{equation}
The emitter-photon state for all modules is
\begin{equation}
    |\psi_\mathrm{init}\rangle=\left(  \sqrt{1-\alpha}\,\ket{00_{\text{ph}}} + \sqrt{\alpha}\,\ket{11_{\text{ph}}}\right)^{\otimes 4}.
\label{eqn:3_module_emitter_photon_state}
\end{equation}
The photonic terms from this input state can be written in the shorthand notation as
\begin{equation}
(\hat{p}_1^\dagger)^{p_1}(\hat{p}_2^\dagger)^{p_2}(\hat{p}_3^\dagger)^{p_3}(\hat{p}_4^\dagger)^{p_4}\ket{0_\mathrm{ph}0_\mathrm{ph}0_\mathrm{ph}0_\mathrm{ph}}_{\mathrm{in}},
\label{eqn:4_module_input}
\end{equation}
where $p_1,p_2,p_3,p_4\in\{0,1\}$ are integers that define the number of photons in each input mode. We assume that at most one photon is emitted per module and contributes to the output detection pattern. Any additional photon from a double-excitation is either lost due to dephasing or occurs with vanishing probability, as justified by the physical model. Therefore, double-excitation errors can be safely ignored.

Inducing the transformation of the beam splitter, we find the photonic output state
\begin{equation}
    \begin{aligned}
      & (\hat{p}_1^\dagger)^{p_1}(\hat{p}_2^\dagger)^{p_2}(\hat{p}_3^\dagger)^{p_3}(\hat{p}_4^\dagger)^{p_4}\ket{0_\mathrm{ph}0_\mathrm{ph}0_\mathrm{ph}0_\mathrm{ph}}_\mathrm{in} \\
      &\rightarrow\; \frac{1}{2^{p_1+p_2+p_3+p_4}}(\hat{r}_1^\dagger+\hat{r}_2^\dagger+\hat{r}_3^\dagger+\hat{r}_4^\dagger)^{p_1}\\
      &\phantom{\rightarrow\;{}}\times(\hat{r}_1^\dagger-\hat{r}_2^\dagger+\hat{r}_3^\dagger-\hat{r}_4^\dagger)^{p_2}(\hat{r}_1^\dagger+\hat{r}_2^\dagger-\hat{r}_3^\dagger-\hat{r}_4^\dagger)^{p_3}\\
      &\phantom{\rightarrow\;{}}\times(\hat{r}_1^\dagger-\hat{r}_2^\dagger-\hat{r}_3^\dagger+\hat{r}_4^\dagger)^{p_4}\ket{0_\mathrm{ph}0_\mathrm{ph}0_\mathrm{ph}0_\mathrm{ph}}_\mathrm{out} ,
    \end{aligned}
    \label{eqn:4_module_input_output}
\end{equation}
where we have ignored the prefactors depending on $\alpha$ and only show the photonic degrees of freedom (see Eq.~\eqref{eqn:4_module_input}).
In general, many detection patterns are possible, corresponding to various input combinations.
% Here, for simplicity, we ignore the prefactors that depend on $\alpha$. For each mode with a photon present, the output state acquires a prefactor of $\sqrt{\alpha}$, and for each mode with no photon, the output state acquires a factor of $\sqrt{1-\alpha}$.
However, to create a W state, we only post-select on the detection pattern where we get a single click in one of the detectors. Note that these detection patterns and success rates are actually modelled via the POVM-based Kraus operators described in App.~\ref{app:povm}, which takes into account photon indistinguishability factors as described in App.~\ref{appsubsubsec:photon_indistinguishability}. 

Say, we get a click from a single photon in detector $\mathrm{D}_1$, meaning we get the output state $\hat{r}_1^\dagger\ket{0_\mathrm{ph}0_\mathrm{ph}0_\mathrm{ph}0_\mathrm{ph}}_\mathrm{out}$. This pattern is possible via a superposition of different single excitations from various emitter states (using Eq.~\eqref{eqn:4_module_input_output}), creating the entangled state of the emitters
\begin{equation}
\begin{aligned}
&\ket{\psi_\mathrm{emitters,W}^\mathrm{PNR}}\\
&=\frac{1}{2}\left(\ket{0001}+\ket{0010}+\ket{0100}+\ket{1000} \right) = \ket{W_4}.
\label{eqn:pnr_w_state_output}
\end{aligned}
\end{equation}
When the detectors are perfectly PNR, a single click in $\mathrm{D_1}$ (or any single click in one of the detectors with appropriate post-detection local rotations) yields the desired state. With non-PNR detectors, unwanted higher-photon terms also contribute to the final state. For example, the detector $\mathrm{D_1}$ cannot distinguish between the photon-number terms $r_1^\dagger$, $(r_1^\dagger)^2$, $(r_1^\dagger)^3$, and $(r_1^\dagger)^4$. This introduces noise, and we obtain the following state:
\begin{equation}
    \begin{aligned}
&\rho_\mathrm{emitters,W}^\mathrm{non-PNR}\\
&=\frac{1}{\mathcal{N}_W}\bigg(\alpha(1-\alpha)^3\ket{W_4}\bra{W_4} +\frac{3\alpha^2(1-\alpha)^2}{4}\ket{E_2}\bra{E_2}    \\
& \phantom{{}={}}+\frac{3\alpha^3(1-\alpha)}{8}\ket{W'_4}\bra{W'_4}+\frac{3\alpha^4}{32}\ket{1111}\bra{1111}\bigg),
        \label{eqn:non_pnr_w_output}
    \end{aligned}
\end{equation}
where the first term is the desired W state, resulting from a true single-photon detection in $\mathrm{D_1}$ and $\mathcal{N}_W$ denotes a normalization factor. The second term comes from a single click in $\mathrm{D_1}$ triggered by two unresolved photons, with $\ket{E_2}=\tfrac{1}{\sqrt{6}}(\ket{0011}+\ket{1100}+\ket{0101}+\ket{1010}+\ket{0110}+\ket{1001})$, the third term comes from three unresolved photons in $\mathrm{D_1}$, with $\ket{W_4'}=X_1X_2X_3X_4\ket{W_4}$, and the fourth term comes from four unresolved photons in $\mathrm{D_1}$. In this case, the fidelity of the desired W state is relatively lower and dependent upon the bright-state parameter. We get the highest fidelity with the lowest value of $\alpha$.

\subsubsection{\label{subsubsec:create_GHZ}Creating a GHZ state}
Creating a GHZ state works analogously to the creation of the W state. The difference is in the detection pattern that we post-select on. To create a GHZ state, we demand exactly two single-photon detections in different detectors, e.g, $\mathrm{D}_1$ and $\mathrm{D}_2$. 
% Based on Eq.~\eqref{eqn:4_module_input_output} we find output terms that exactly have one photon each in output modes via the product $\hat{r}_1^\dagger\hat{r}_2^\dagger$, based on the relative signs that occur in the products of sums on the right-hand side. This is possible with the combination of values $(p_1,p_2,p_3,p_4)=(1,0,1,0),(0,1,0,1)$.
Based on Eq.~(\ref{eqn:4_module_input_output}), we find two two-photon input terms that can lead to this detection pattern, i.e., $(p_1,p_2,p_3,p_4)=(1,0,1,0),(0,1,0,1)$.
Explicitly, this results in the terms
\begin{equation}
    \begin{aligned}
&\hat{p}_1^\dagger\hat{p}_3^\dagger\ket{0_\mathrm{ph}0_\mathrm{ph}0_\mathrm{ph}0_\mathrm{ph}}_\mathrm{in}\;\rightarrow\;\frac{1}{4}(\hat{r}_1^{\dagger 2}+\hat{r}_2^{\dagger 2}-\hat{r}_3^{\dagger 2}-\hat{r}_4^{\dagger 2}\\ &+2\hat{r}_1^\dagger \hat{r}_2^\dagger-2\hat{r}_3^\dagger \hat{r}_4^\dagger )\ket{0_\mathrm{ph}0_\mathrm{ph}0_\mathrm{ph}0_\mathrm{ph}}_\mathrm{out}, \\
& \hat{p}_2^\dagger\hat{p}_4^\dagger\ket{0_\mathrm{ph}0_\mathrm{ph}0_\mathrm{ph}0_\mathrm{ph}}_\mathrm{in}\;\rightarrow\;\frac{1}{4}(\hat{r}_1^{\dagger 2}+\hat{r}_2^{\dagger 2}-\hat{r}_3^{\dagger 2}-\hat{r}_4^{\dagger 2}\\ & -2\hat{r}_1^\dagger \hat{r}_2^\dagger+2\hat{r}_3^\dagger \hat{r}_4^\dagger )\ket{0_\mathrm{ph}0_\mathrm{ph}0_\mathrm{ph}0_\mathrm{ph}}_\mathrm{out}.
    \end{aligned}
\end{equation}
Post-selecting based on this detection pattern prepares the emitters in the perfect GHZ state
\begin{equation}
\ket{\psi_\mathrm{emitters,GHZ}^\mathrm{PNR}} = \frac{1}{\sqrt{2}}\!\left(\ket{0101}-\ket{1010}\right)=\ket{\Psi^-_4},
\end{equation}
which can be converted to the $\ket{\Phi^+_4}$ GHZ state with local operations. We call this the ``Raw GHZ'' state. Note that this perfect GHZ generation is only possible in the noiseless case with PNR detectors.

When we have non-PNR detectors, we must also include contributions from the output terms corresponding to $\hat{r}_1^{\dagger 2}\hat{r}_2^\dagger$, $\hat{r}_1^{\dagger 3}\hat{r}_2^\dagger$, $\hat{r}_1^\dagger\hat{r}_2^{\dagger 2},\hat{r}_1^\dagger\hat{r}_2^{\dagger 3}$, and $\hat{r}_1^{\dagger 2}\hat{r}_2^{\dagger 2}$ which correspond to detecting up to a total of four photons in both detectors combined. We collect the contribution due to all these terms on the emitters' state, which gives us the mixed state
\begin{equation}
    \begin{aligned}
&\rho_\mathrm{emitters,GHZ}^\mathrm{non-PNR}\\
&= \frac{1}{\mathcal{N}_{GHZ}}\bigg( \frac{\alpha^2 (1-\alpha)^2}{2}\ket{\Psi^-_4}\bra{\Psi^-_4} +\frac{\alpha^3(1-\alpha)}{4}\rho_{1,2} \\
&\phantom{={}}  +\frac{\alpha^4}{16}  \ket{1111}\bra{1111}\bigg).
\label{eqn:non_pnr_GHZ_state}
    \end{aligned}
\end{equation}
where the fidelity is dependent on $\alpha$ and we get higher fidelity with lower $\alpha$, up to some normalization factor $\mathcal{N}_{GHZ}$. The second term corresponds to the noise term resulting from detecting a single-photon click and an unresolved two-photon click, with $\rho_{1,2}$ containing only terms with exactly three emitters in the excited state, and the third term results from measuring four photons unresolved at the two detectors. Three variations of the four-party GHZ state can be generated based on $^4C_2=6$ two-detector-click patterns. All of them can be transformed into a raw $\ket{\Phi^+_4}$ GHZ state, which is needed (in our convention) for the modular stabilizer measurement.

\subsection{\label{subsec:distil_protocols}Memory distillation protocols}
Having established reliable methods for creating elementary entangled states, we now introduce a set of distillation protocols that build on these states. Ref.~\cite{singh2024modulararchitecturesentanglementschemes} demonstrated that fusion-based architectures fail to achieve high thresholds because they rely on an excessive number of memory-based two-qubit gates. In contrast, the extent to which distillation can be effectively incorporated in an emission-based architecture has remained unclear. To address this gap, we investigate minimal distillation strategies for generating GHZ states from the elementary links described above. The proposed memory distillation protocols are specifically adapted to the architectural quantum circuits of diamond color‑center hardware. In each protocol, we first generate a base state in the form of an elementary link and swap it onto the auxiliary memory qubits within each module. We then generate an additional elementary link and employ a dedicated distillation circuit that, with a certain success probability, purifies the stored base state to yield a high‑fidelity GHZ state.

\subsubsection{\label{subsubsec:bell_to_ghz}MD: Bell + GHZ protocol}
\begin{figure*}
    \centering
    \includegraphics[width=0.85\textwidth]{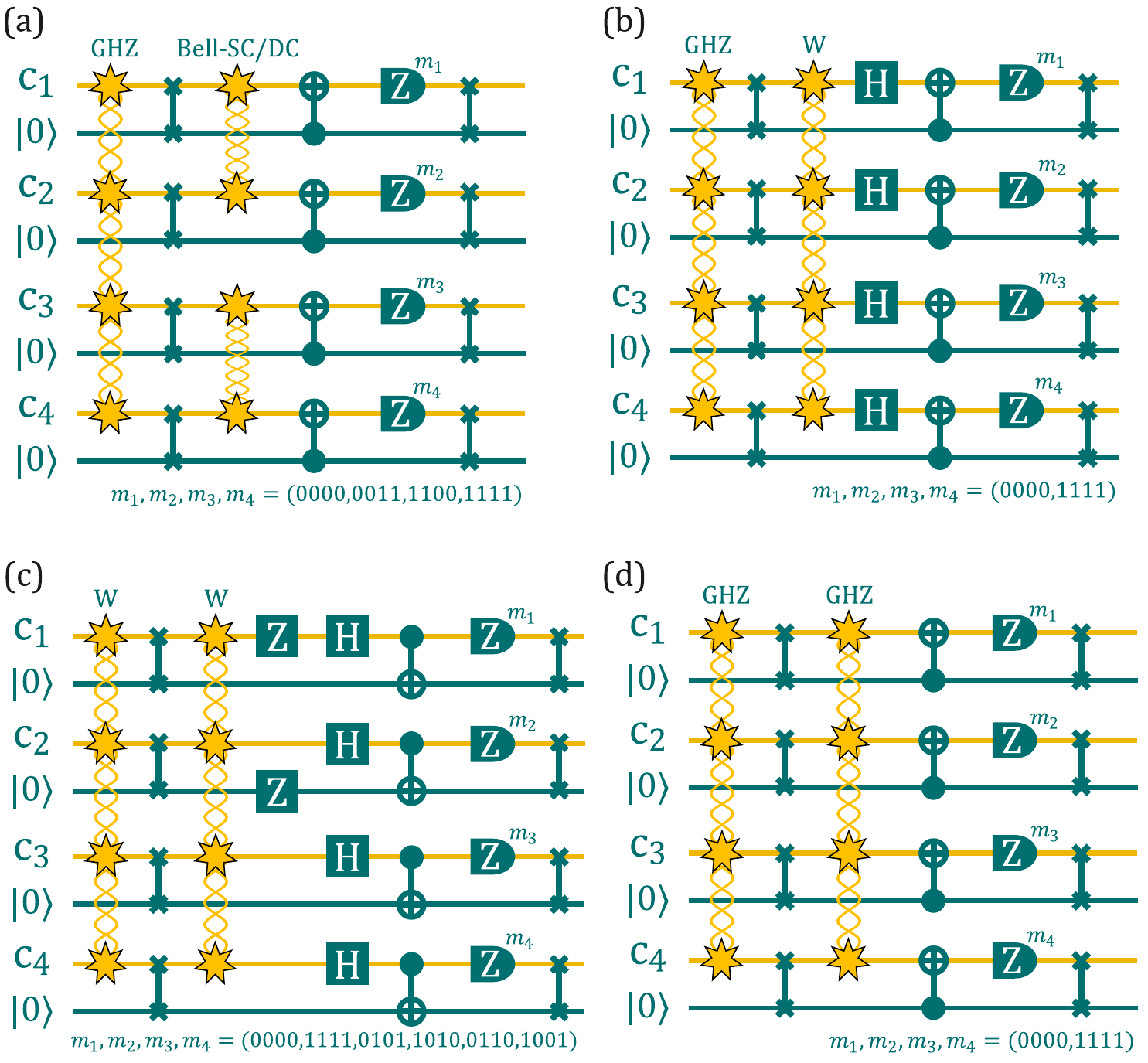}
\caption{Quantum circuits for memory-assisted distillation protocols. 
Communication qubits (yellow wires) act as photon-emitting qubits, which create entangled states, while auxiliary memory qubits (teal wires) assist with the storage of entangled states for the memory-assisted distillation.  
Elementary entanglement links are indicated by starred connections, labeled with the raw state produced (GHZ, Bell, or W).  
Post-selection is performed on measurement outcomes $m_1,m_2,m_3,m_4$ with the indicated valid patterns, and the protocols are executed in the RUS manner as shown in the sub-figures.
(a) \textbf{MD: SC/DC-Bell + GHZ:} a base GHZ link is distilled with auxiliary Bell pairs.
(b) \textbf{MD: W + GHZ:} a base GHZ link is distilled with auxiliary $W$ state.
(c) \textbf{MD: W + W:} two raw $W$ states are locally rotated and distillation projects the base state into a GHZ state.
(d) \textbf{MD: GHZ + GHZ :} two raw GHZ states are created, and the new one is used to distill the prior one.  }
\label{fig:distillation_circuits}
\end{figure*}

Having created a raw GHZ state, minimal‑resource distillation can be achieved by employing Bell states. Bell states have higher success probabilities than directly creating a raw GHZ state. We generate two Bell states simultaneously in a RUS procedure between modules A–B and C–D to distill the raw GHZ state. Either the single‑click or double‑click protocol can be used for creating these Bell states, and we name the protocols accordingly. The raw GHZ state is first created and swapped to the auxiliary qubits, freeing the communication qubits. Simultaneous Bell‑pair generation is then performed between modules A–B and C–D. 
Once both Bell pairs are successfully created, after potentially different waiting times, the distillation protocol is executed (see Fig.~\ref{fig:distillation_circuits}(a) for the quantum circuit). First, \textsc{cnot} gates are applied from the communication qubits to the auxiliary qubits, implementing a joint $XX$ stabilizer measurement that projects the raw GHZ state. The communication qubits are then measured in the $Z$ basis, with post‑selection on the required measurement outcomes and appropriate post‑measurement corrections. These desired measurement patterns are shown at the bottom of Fig.~\ref{fig:distillation_circuits}(a). Finally, the distilled GHZ state is swapped back to the communication qubits to proceed with the stabilizer measurement of the surface code.

We can understand that this protocol works from looking at the noiseless raw GHZ and Bell state (see Eqs.~(\ref{eqn:bell_single_click_state}) and (\ref{eqn:non_pnr_GHZ_state})). Postselecting for 1111 after applying the \textsc{cnot} gates can only result from having generated a perfect GHZ state and two perfect Bell states, distilling a perfect GHZ state. On top of that, post-selecting for 0000, 1100, and 0011 as they are valid measurement patterns also increases the success rate of the GHZ by yielding different GHZ states that can be converted to $\ket{\Phi^+_4}$ with appropriate local operations. Note that if the value of $\alpha$ is large, with non-PNR detectors as the only noise source, then post-selecting on the 1111 measurement pattern gives the highest fidelity, as it eliminates the higher excitation terms via distillation. However, in our simulations (see Sec.\ref{sec:results}), since we have multiple sources of noise with a very low value of $\alpha$ (for the highest fidelity of the elementary states), we choose other measurement patterns to increase the protocol success rate.

\subsubsection{\label{subsubsec:w_to_ghz}MD: W + GHZ protocol}
In this protocol, we use a W state to distil a raw GHZ state.
The raw state is first created using the RUS procedure and is then swapped to the auxiliary memory qubits, freeing the communication qubits for another round of elementary link generation. A W state is subsequently created on the communication qubits, again using RUS.
We perform a basis transformation on the W‑state qubits using Hadamard gates
\begin{equation}
\begin{aligned}
\label{eqn:rotated_w_state_hadamards}
\ket{\tilde{W}_4} & = H_1 H_2 H_3 H_4 \ket{W_4} \\
& = \frac{1}{2}\ket{W_4} - \frac{1}{2} \ket{W_4'} + \frac{1}{\sqrt{2}} \ket{\Phi_4^{-}}.
\end{aligned}
\end{equation}
One can directly see that the rotated W state contains the GHZ state $\ket{\Phi_4^{-}}$ and can therefore be used for GHZ state distillation.
Again, the distillation protocol consists of applying \textsc{cnot} gates from the auxiliary qubits to the communication qubits, followed by final measurements on the communication qubits. This sequence completes the protocol when the desired measurement pattern is obtained.
Since both the rotated W state and the raw GHZ state (see Eq.~(\ref{eqn:non_pnr_GHZ_state})) contain terms with odd numbers of excitations, it becomes clear that this protocol cannot yield perfect GHZ states. Its advantage lies in the higher success rates for generating W states instead of GHZ states, which allows for lower values of $\alpha$ and, thereby, an increase in fidelity.
The schematic of the protocol is depicted in Fig.~\ref{fig:distillation_circuits}(b).

\subsubsection{\label{subsubsec:w_to_w}MD: W + W protocol}
This memory-based distillation protocol distills a GHZ state from solely using two W states. Similar to the above protocols, we first generate two raw W states (RUS). Then the states are locally rotated with $Z$ and Hadamard rotations (as shown in Fig.~\ref{fig:distillation_circuits}(c)). Assuming perfect W states, the rotated states take the forms
\begin{align}
H_1 H_2 H_3 H_4 Z_1 \ket{W_4} = X_1 \ket{\tilde{W}_4}, \\
H_1 H_2 H_3 H_4 Z_2 \ket{W_4} = X_2 \ket{\tilde{W}_4}.
\end{align}
From Eq.~(\ref{eqn:rotated_w_state_hadamards}), one can understand that both rotated W states contain all possible terms with even numbers of excitations coming from $X_i (\ket{W_4} - \ket{W_4'})$. At the same time, they contain nonequivalent terms with odd numbers of excitations coming from $X_i \ket{\Phi^{-}_4}$. Applying \textsc{cnot} gates and postselecting for six different patterns with even numbers of 0s and 1s, yields all terms with even numbers of excitations, which are identical to a rotated GHZ state.
Finally, rotating the distilled GHZ state transforms it into $\ket{\Phi^{+}_4}$.
The schematic of the full protocol is depicted in Fig.~\ref{fig:distillation_circuits}(c).

\subsubsection{\label{subsubsec:ghz_to_ghz}MD: GHZ + GHZ protocol}
The most intuitive distillation protocol consists of generating two raw GHZ states successively and distilling one from the other~\cite{Shimizu2025}.
This can be understood from looking at the form of the noiseless raw GHZ state without PNR (see Eq.~\eqref{eqn:non_pnr_GHZ_state}). Applying \textsc{cnot} gates between two copies of this state and postselecting for 0000 or 1111 outcomes directly increases the weight of the GHZ state terms. Post-selecting only for 1111 would even generate a perfect GHZ state in the noiseless scenario. We show the quantum circuit in Fig.~\ref{fig:distillation_circuits}(d) along with the measurement outcome patterns. Similar to the case of the Bell state to GHZ distillation protocol, the 1111 measurement pattern gives the highest fidelity when $\alpha$ is large, but due to the regime of our simulation parameters, we also include the 0000 measurement pattern.

\subsection{\label{subsec:optical_distil_protocols}Optical distillation protocols}
We have proposed protocols that require minimal memory distillation in the previous section, meaning that each protocol involves only a single round of distillation. When gates are very noisy or slow, it becomes necessary to explore alternative approaches that can produce high-fidelity GHZ states without relying on memory-based two-qubit gates. Optical distillation protocols, as the DC protocol, enable this by performing two successive rounds of photon emission and detection to filter out the desired state. The downside is that the overall success rate becomes the square of the individual success rates, which is disadvantageous in regimes with high loss or low effective photon detection efficiency. Nevertheless, we show that these protocols yield high-fidelity GHZ states and offer promising performance for modest values of photon detection efficiency.

\subsubsection{\label{subsubsec:dc_ghz}OD: GHZ + GHZ protocol}
Analogous to the DC version of the Bell-pair generation~\cite{Barrett_2005}, we can generalize the single-shot raw-GHZ state generation protocol to a DC setup. Here, DC would refer to the `double' rounds of optical detection clicks with respective desired patterns. For this, we start by generating a raw GHZ state as discussed in Sec.~\ref{subsubsec:create_GHZ}. From Eq.~(\ref{eqn:non_pnr_GHZ_state}), we observe that the raw GHZ state only contains noise terms with three or four excited communication qubits. Applying local bit-flip operations to all communication qubits, i.e., $X_1 X_2 X_3 X_4$, changes the raw GHZ state to
\begin{equation}
    \begin{aligned}
& \rho_\mathrm{emitters,GHZ}^\mathrm{non-PNR} \ \xrightarrow{X_1 X_2 X_3 X_4}\\
& \frac{1}{\mathcal{N}_{GHZ}}\bigg( \frac{\alpha^2 (1-\alpha)^2}{2}\ket{\Psi^-_4}\bra{\Psi^-_4} +\frac{\alpha^3(1-\alpha)}{4}\rho'_{1,2} \\
&\phantom{={}}  +\frac{\alpha^4}{16}  \ket{0000}\bra{0000}\bigg).
\label{eqn:non_pnr_GHZ_state_flip}
    \end{aligned}
\end{equation}
From this state, one can see that the GHZ terms remain unchanged. However, the undesired noise terms, i.e., $\rho_{1,2}$ and $\ket{1111}\bra{1111}$, are flipped to $\rho'_{1,2}$ and $\ket{0000}\bra{0000}$, which contain only one and zero excited qubits, respectively, and cannot emit more than a single photon. Executing a second round of emission and post-selecting for the same two-detector patterns as in the raw state generation round generates a perfect GHZ state, even without using PNR detectors. Using PNR detectors increases the theoretical success probability by a factor of 2 since one can also post-select for both photons arriving at the same detector. The protocol is summarized in Fig.~\ref{fig:dc_ghz_flow}.

\begin{figure}[h]
\centering
\begin{tikzpicture}[node distance=0.35cm,
  box/.style={rectangle, rounded corners=2pt, minimum width=4.2cm,
              minimum height=0.75cm, text centered, draw=black,
              fill=blue!7, font=\scriptsize, align=center},
  dec/.style={rectangle, rounded corners=2pt, minimum width=4.2cm,
              minimum height=0.75cm, text centered, draw=black,
              fill=green!10, font=\scriptsize},
  outbox/.style={rectangle, rounded corners=2pt, minimum width=4.2cm,
              minimum height=0.75cm, text centered, draw=black,
              fill=violet!10, font=\scriptsize, align=center},
  arrow/.style={-{Stealth}, thick}]

\node (s1) [box] {(1)~Init: $(\sqrt{1{-}\alpha^2}|0\rangle{+}\alpha|1\rangle)^{\otimes 4}$};
\node (s2) [box, below=of s1] {(2)~Emit + post-select (Round 1)};
\node (d1) [dec, below=of s2] {Coincidence? \textbf{No} $\to$ restart};
\node (s3) [box, below=of d1, fill=orange!10] {(3)~Apply $X_1X_2X_3X_4$};
\node (s4) [box, below=of s3] {(4)~Emit + post-select (Round 2)};
\node (d2) [dec, below=of s4] {Coincidence? \textbf{No} $\to$ restart};
\node (s5) [outbox, below=of d2]
  {(5)~$|\Phi^+_4\rangle$,\quad $p_\text{succ}{=}3\alpha^2(1-\alpha)^2$};

\draw[arrow](s1)--(s2);
\draw[arrow](s2)--(d1);
\draw[arrow](d1)--(s3);
\draw[arrow](s3)--(s4);
\draw[arrow](s4)--(d2);
\draw[arrow](d2)--(s5);

\end{tikzpicture}
\caption{OD-GHZ protocol operational summary (RUS). (1) All four communication qubits are initialized and excited to emit photons into the beam-splitter network. (2) A two-detector coincidence post-selection heralds the raw GHZ state (Eq.~\ref{eqn:non_pnr_GHZ_state}). (3) Local $X_1X_2X_3X_4$ gates map the noise terms into the low-excitation subspace (Eq.~\ref{eqn:non_pnr_GHZ_state_flip}). (4) A second emission round is executed on the same device without reconfiguration. (5) Post-selection on the same coincidence pattern heralds the purified GHZ state $|\Phi^+_4\rangle$ with $p_\text{succ} = 3\alpha^2(1-\alpha)^2$. A failed coincidence at either round triggers a full restart.}
\label{fig:dc_ghz_flow}
\end{figure}

\subsubsection{\label{subsubsec:dc_w}OD: W + W protocol}
Remarkably, one can also find a DC protocol generating GHZ states when starting from a raw W state. As in the memory distillation protocols using W states, the reason for this lies in the rotated W state. Here, we are interested in a rotated W state with slightly different phases, i.e.,
\begin{equation}
\begin{aligned}
&H_1 H_2 H_3 H_4 Z_2 Z_4\ket{W_4}=X_2 X_4 \ket{\tilde{W}_4} \\
&= \frac{1}{2} X_2 X_4 (\ket{W_4} - \ket{W'_4}) + \frac{1}{\sqrt{2}} \ket{\Psi_4^{-}}.
\label{eqn:pnr_w_state_output_tilde_rotated}
\end{aligned}
\end{equation}
From this state, we can see that executing another round of emission and post-selecting for two-detector (two-photon) coincidences using non-PNR (PNR) detectors generates a GHZ state. However, looking at Eq.~(\ref{eqn:non_pnr_w_output}), it becomes obvious that this protocol cannot generate a perfect GHZ state under realistic conditions. In contrast to the OD-GHZ protocol, the OD-W protocol cannot suppress the noise terms present in the raw W state completely. Since the Hadamard rotations of the noise terms also contain two excitation GHZ terms, they cannot be filtered out. Only in the case of a perfect raw W state ($\alpha\rightarrow 0$) can this protocol generate a close to perfect GHZ state.

\begin{figure}[h]
\centering
\begin{tikzpicture}[node distance=0.35cm,
  box/.style={rectangle, rounded corners=2pt, minimum width=4.2cm,
              minimum height=0.75cm, text centered, draw=black,
              fill=blue!7, font=\scriptsize, align=center},
  dec/.style={rectangle, rounded corners=2pt, minimum width=4.2cm,
              minimum height=0.75cm, text centered, draw=black,
              fill=green!10, font=\scriptsize},
  outbox/.style={rectangle, rounded corners=2pt, minimum width=4.2cm,
              minimum height=0.75cm, text centered, draw=black,
              fill=violet!10, font=\scriptsize, align=center},
  arrow/.style={-{Stealth}, thick}]

\node (s1) [box] {(1)~Init: $(\sqrt{1{-}\alpha^2}|0\rangle{+}\alpha|1\rangle)^{\otimes 4}$};
\node (s2) [box, below=of s1] {(2)~Emit + post-select (Round 1)};
\node (d1) [dec, below=of s2] {Single click? \textbf{No} $\to$ restart};
\node (s3) [box, below=of d1, fill=orange!10]
  {(3)~Apply $H_1H_2H_3H_4 Z_2 Z_4$};
\node (s4) [box, below=of s3] {(4)~Emit + post-select (Round 2)};
\node (d2) [dec, below=of s4] {Coincidence? \textbf{No} $\to$ restart};
\node (s5) [outbox, below=of d2]
  {(5)~$|\Phi^+_4\rangle$};

\draw[arrow](s1)--(s2);
\draw[arrow](s2)--(d1);
\draw[arrow](d1)--(s3);
\draw[arrow](s3)--(s4);
\draw[arrow](s4)--(d2);
\draw[arrow](d2)--(s5);

\end{tikzpicture}
\caption{OD-W protocol operational summary (RUS). (1) Initialize and emit photons into the beam-splitter network. (2) Single-detector click heralds the raw W state. (3) Local $H_1H_2H_3H_4Z_2Z_4$ rotations expose the GHZ component. (4) Second emission round with two-detector coincidence post-selection. (5) Output approximates $|\Phi^+_4\rangle$; unlike OD-GHZ, a perfect GHZ state requires $\alpha \to 0$ as W-state noise terms cannot be fully suppressed. A failed post-selection at either round triggers a full restart.}
\label{fig:dc_w_flow}
\end{figure}

\subsection{\label{ssubsec:protocol_summary}Summary of all protocols under noiseless hardware }
\begin{table*}[t]
\centering
\small
\setlength{\tabcolsep}{6pt}
\renewcommand{\arraystretch}{2.0}  
\begin{tabular}{lcc}
\hline\hline
\textbf{Protocol (detector type)} & \textbf{Success Rate} & \textbf{Fidelity} \\
\hline
SC-Bell (PNR) \& SC-Bell (non-PNR) & $2\alpha(1-\alpha)$ \& $\alpha(2-\alpha)$ & $1$ \& $\sqrt{\frac{2(1-\alpha)}{2-\alpha}}$ \\
\hline
DC-Bell (both) & $2\alpha(1-\alpha)$ & $1$ \\
\hline
W (PNR) \& W (non-PNR) & $4\alpha(1-\alpha)^3$ \& $\frac{\alpha(32-72\alpha+60\alpha^2-17\alpha^3)}{8}$ & $1$ \& $4\sqrt{\frac{2(1-\alpha)^3}{32-72\alpha+60\alpha^2-17\alpha^3}}$ \\
\hline
GHZ (PNR) \& GHZ (non-PNR) & $3\alpha^2(1-\alpha)^2$ \& $\frac{3\alpha^2(5\alpha^2-12\alpha+8)}{8}$ & $1$ \& $\frac{2\sqrt{2}(1-\alpha)}{\sqrt{5\alpha^2-12\alpha+8}}$ \\
\hline
MD: SC/DC-Bell + GHZ (both) & $\bigg(\frac{n_d}{n_s}\bigg)\bigg/\sum_{i=1}^{n_s}\bigg[n_a^{G}+\max(n_a^{B1},n_a^{B2})\bigg]_i$ & $\frac{1}{n_s}\sum_{i=1}^{n_s}F_{i,G}$ \\
\hline
MD: W + GHZ (both) & $\bigg(\frac{n_d}{n_s}\bigg)\bigg/\sum_{i=1}^{n_s}\bigg[n_a^G+n_a^W\bigg]_i$ & $\frac{1}{n_s}\sum_{i=1}^{n_s}F_{i,G}$ \\
\hline
MD: W + W (both) & $\bigg(\frac{n_d}{n_s}\bigg)\bigg/\sum_{i=1}^{n_s}\bigg[n_a^{W1}+n_a^{W2}\bigg]_i$ & $\frac{1}{n_s}\sum_{i=1}^{n_s}F_{i,G}$ \\
\hline
MD: GHZ + GHZ (both) & $\bigg(\frac{n_d}{n_s}\bigg)\bigg/\sum_{i=1}^{n_s}\bigg[n_a^{G1}+n_a^{G2}\bigg]_i$ & $\frac{1}{n_s}\sum_{i=1}^{n_s}F_{i,G}$ \\
\hline
OD: GHZ + GHZ (both) & $3\alpha^2(1-\alpha)^2$ & $1$  \\
\hline
OD: W + W (PNR) \& (non-PNR) & $2\alpha(1-\alpha)^3$ \& $\frac{\alpha(1152-2288\alpha+1552\alpha^2-377\alpha^3)}{1024}$ & $1$ \& $8\sqrt{\frac{16-48\alpha+54\alpha^2-22\alpha^3}{1152-2288\alpha+1552\alpha^2-377\alpha^3}}$ \\
\hline\hline
\end{tabular}
\caption{Output state success rates and fidelity for various entanglement generation protocols with noiseless hardware. \textit{Abbreviations}: SC, single-click; DC, double-click; PNR, photon-number resolving; $n_s=n_{\mathrm{shots}}$--the total shots of the distillation protocol, $n_d=n_{\mathrm{distil}}$--shots where the respective distillation pattern succeeds, $n_a=n_{\mathrm{attempts}}$--to generate a respective elementary state via RUS (superscripts R, B, W, G denote Raw, Bell, W, GHZ respectively); $F_{i,G}=F_{i,\mathrm{distilled-GHZ}}$.}
\label{tab:protocol_summary}
\end{table*}

We present the output results of all protocols in Tab.~\ref{tab:protocol_summary}. The success rates and fidelities of the output GHZ states are reported as functions of the tunable bright-state parameter $\alpha$, for both PNR and non-PNR detector types. For elementary link generation and optical distillation protocols, we provide analytical expressions. For the expressions of success rates, we have taken into account all detection patterns of each round that contribute to the creation of the desired state, since all these output states can be locally rotated to yield the $\ket{\Phi^+_4}$ GHZ state. For memory distillation protocols, the overall success rate depends on multiple independent RUS events for each link generation. We define the total protocol success rate as the ratio of successful protocol executions ($n_{\rm distil}$) to the total number of simulation shots, $n_\mathrm{shots}$. For each required elementary `state', the number of attempts until success is denoted by $n_\mathrm{attempts}^\mathrm{state}$. For each successful attempt, we compute the fidelity $F_{i,\mathrm{GHZ}}$ of the output state, and report the final protocol fidelity as the numerical average over all successful attempts. In our simulations, we use $n_\mathrm{shots} = 10^4$, yielding statistical errors below $0.1\%$ per data point.

\section{\label{sec:results}Results}

In this section, 
%we evaluate the performance of all proposed single-shot emission-based protocols under full hardware noise and report the modular surface code thresholds achieved with each. 

we assess whether single-shot entanglement generation protocols can outperform fusion-based schemes. 
To this end, we analyze all proposed single-shot emission-based protocols under full hardware noise and report the modular surface-code thresholds achieved with each. In Table~\ref{tab:hardware_emission_sets}, we summarize the hardware parameter sets considered.

To address this comparison, we first investigate which near-term hardware upgrade yields the largest threshold improvement and whether the thresholds saturate, as in fusion-based schemes. Finally, we determine the hardware requirements for near-term fault-tolerant modular demonstrations.

%Four concrete questions structure our analysis. Can the memory-gate bottleneck inherent to fusion-based entanglement generation be eliminated without sacrificing fault-tolerance thresholds? Among all near-term hardware upgrades, which yields the largest threshold improvement? And does the threshold scale continuously with improving hardware parameters, or does it saturate as in fusion-based schemes? And what are the concrete hardware requirements for an early fault-tolerant modular demonstration? 
%\textcolor{blue}{We first present the performance analysis of the GHZ-state generation protocols, comparing each against the relevant noise parameters and identifying the fault-tolerance regions where circuit-level noise thresholds for the fully distributed toric surface code are expected. The parameter sets of Table~\ref{tab:hardware_emission_sets} then provide a transparent map of concrete hardware requirements, serving as supporting evidence for the conclusions drawn from this analysis.}

To ensure fair comparison with Ref.~\cite{singh2024modulararchitecturesentanglementschemes}, we simulate the infidelity and success rates of all protocols, including the physical error probability $p$ for circuit-level noise. Explicitly, this means that all communication qubit measurements and all quantum gates on communication and memory qubits have a failure rate of $p$. In Ref.~\cite{singh2024modulararchitecturesentanglementschemes}, code thresholds were reported that ranged from $0.01 \%$ to $0.4 \%$ including fusion-based emission and scattering schemes. Here, we use $p=0.1\%$ to have error-contribution in the range where we expect the threshold to exist, to compare the output GHZ states from several protocols. For the results on the GHZ state quality, we use $10^4$ shots for each data point for the average output, which results in tiny error zones as presented in the plots. 

\subsection{\label{subsec:results_decoherence_hardware_sets}Decoherence and hardware parameter sets}
Decoherence is a dominant error noise source, which becomes more crucial since the GHZ state generation is probabilistic and memory gates are slow, causing significant decoherence. We model the performance of protocols with active time-tracking decoherence channels for both communication and memory qubits.  We use a representative range of coherence and gate timing parameters which correspond to ``Set-3'' in Ref.~\cite{singh2024modulararchitecturesentanglementschemes}, as summarized in Tab.~\ref{tab:coherence_operation_times}. The quantities $T_\text{link}^\text{dec}$ and $T_\text{idle}^\text{dec}$ denote coherence times of the communication and memory qubits during entanglement attempts and idling, respectively. Operational times include $t_\text{link}$ for one entanglement attempt and $t_\text{meas}$ for communication-qubit measurement. Gate durations are given by $t_{P}^\text{c/m}$ (communication/memory) for single-qubit Pauli ($P$) gates ($P \in \{X, Y, Z, H\}$), and $t_{CZ}, t_{CX}, t_{CY}$ for two-qubit gates on the memory, along with $t_{\textsc{SWAP}}$ for the \textsc{SWAP} operation. While the simulation framework supports separate $T_1$ and $T_2$ coherence times for both communication and memory qubits, we assume $T_\text{1, link}^\text{dec} = T_\text{2, link}^\text{dec}$ and $T_\text{1, idle}^\text{dec} = T_\text{2, idle}^\text{dec}$ throughout, effectively modeling qubit noise as depolarizing. This also makes our surface code unbiased for the hardware errors, since there is no bias against X or Z errors from the hardware via the decoherence mechanism~\cite{singh2024modulararchitecturesentanglementschemes}.

\begin{table}[t]
\centering
\caption{Coherence and operation times for each module. The values are reported with respect to $t_\mathrm{link}$, which is assumed to take one unit of time. For color centers in diamond, $t_\mathrm{link}=6\times 10^{-6}$ s~\cite{singh2024modulararchitecturesentanglementschemes}.}
\label{tab:coherence_operation_times}
\includegraphics[width=0.65\linewidth]{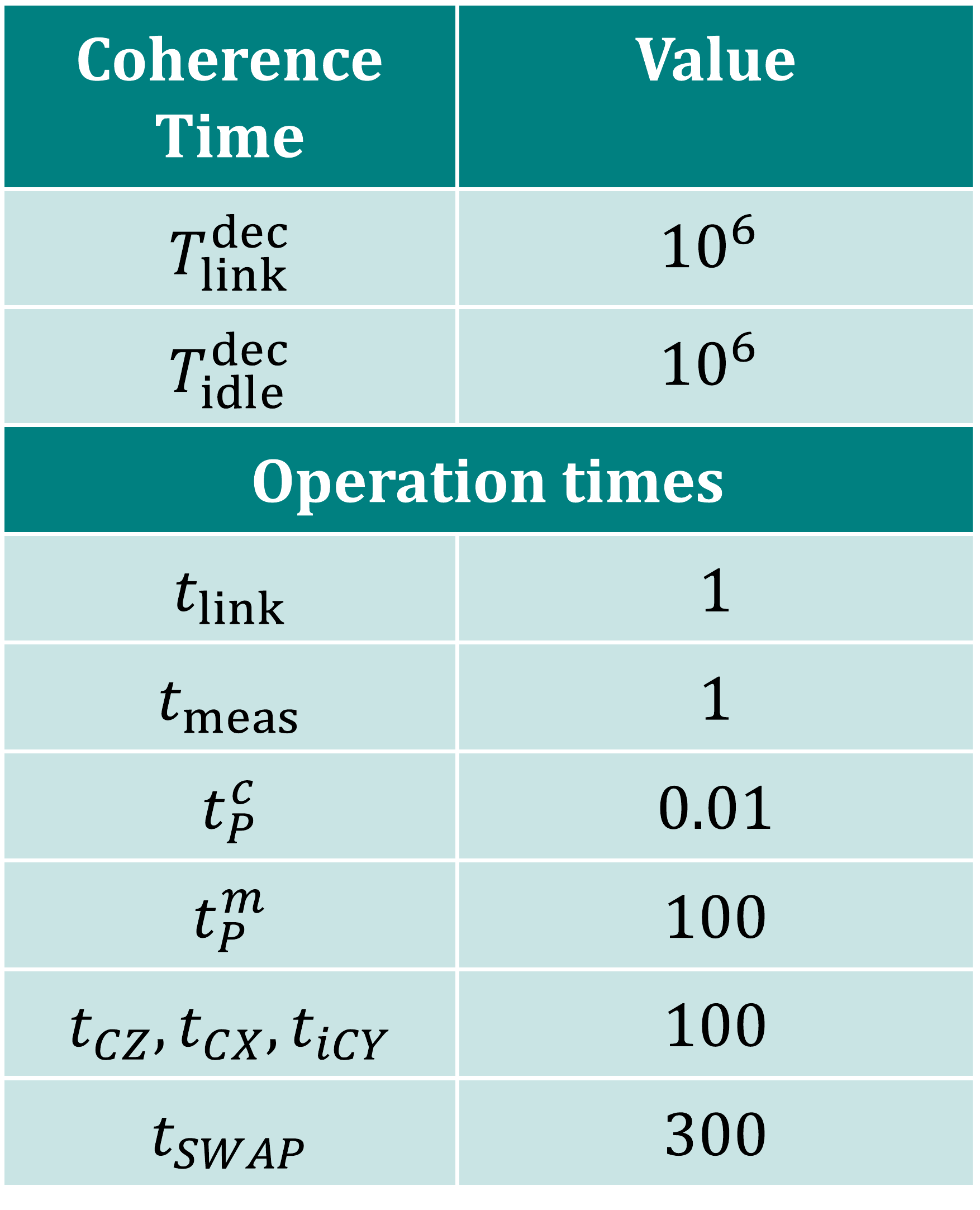}
\end{table}

Next, we describe the emission-scheme hardware parameter sets considered for this work, as presented in Tab.~\ref{tab:hardware_emission_sets}. There are 18 emission-scheme (ES) parameter sets (ES-1 to ES-18) that show gradual improvement towards a noiseless scenario. The first set ES-1 is the same as the future-parameter (FP) set for the emission-based scheme considered in Ref.~\cite{singh2024modulararchitecturesentanglementschemes}. From Set ES-2, we apply our approximation model for the elimination of the double-excitation error under the state assumptions in Sec .~\ref {sec:Physical_setup}.

\begin{table}[t]
\centering
\caption{Hardware improvement parameter sets for the single-shot emission-scheme. The parameters are presented in a total of 18 sets, which show gradual improvement towards an ideal scenario. Set ES-1 corresponds to ``Future parameters (FP)'' considered in Ref.~\cite{singh2024modulararchitecturesentanglementschemes}. The sets trace a path from the near-term hardware baseline toward the near-ideal regime, serving as waypoints to quantify the hardware requirements for each architectural milestone.}
\label{tab:hardware_emission_sets}
\includegraphics[width=0.96\linewidth]{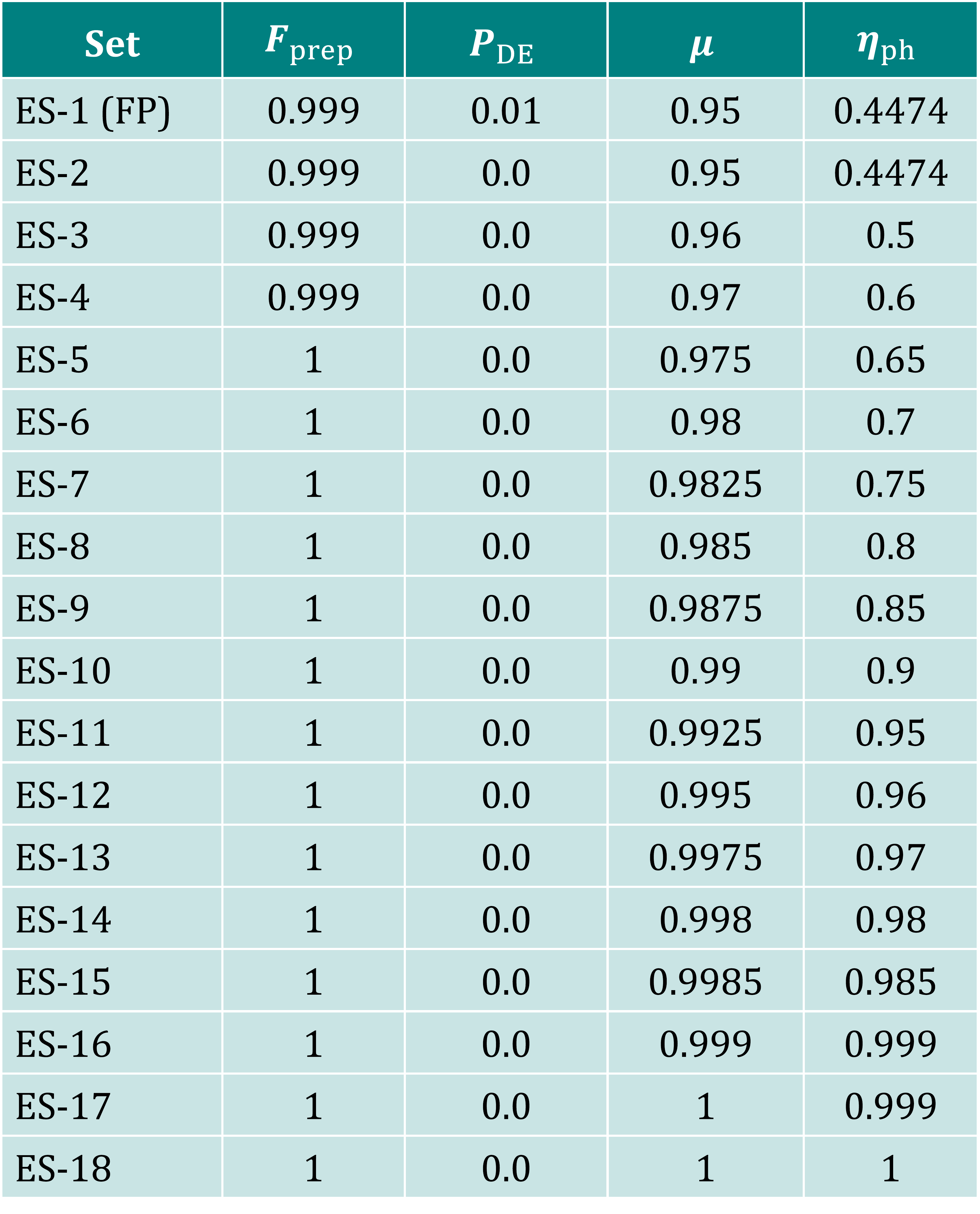}
\end{table}

\subsection{\label{subsec:results_bright_state_parameter}Bright state parameter influence}
Given the full set of hardware, coherence, and gate timing parameters, the bright-state parameter $\alpha$ can be tuned to control both the success probability and the infidelity of the output GHZ state. As shown earlier, in the absence of noise, both quantities are functions of $\alpha$: smaller (larger) values of $\alpha$ yield lower (higher) infidelity and success rates. Under realistic noise, these quantities must be computed explicitly for each protocol to determine the optimal $\alpha$ to determine exact values. As a starting point, we set $\alpha_\mathrm{base} = \alpha_\mathrm{distil} = \alpha$, for the base raw GHZ state and the distilling state, respectively. Following Ref.~\cite{singh2024modulararchitecturesentanglementschemes}, we impose the thresholds $F_\mathrm{GHZ} > 0.98$ and $P_\mathrm{succ} > 10^{-3}$ to identify operating points relevant for toric surface code implementations. For the more general case $\alpha_\mathrm{base} \neq \alpha_\mathrm{distil}$, we present numerical performance of the distillation protocols in App.~\ref{app:trade_off_alpha} when sweeping both $\alpha_\mathrm{base}$, and $\alpha_\mathrm{distil}$.

\begin{figure}[hbtp]
    \centering
    \includegraphics[width=0.48\textwidth]{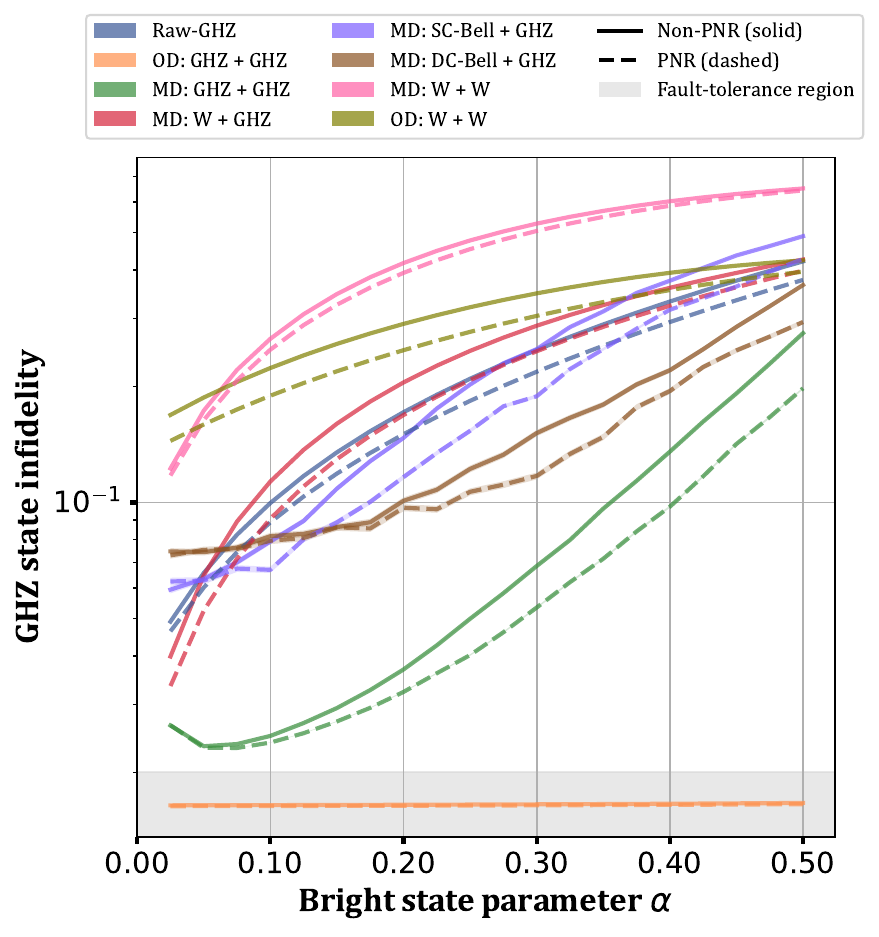}
    \caption{GHZ state infidelity ($1-F_\mathrm{GHZ}$) for varying bright-state parameter ($\alpha$) for ES-2 hardware parameter set, coherence times $T=10^6$, and physical error-rate of $p=10^{-3}$.}
    \label{fig:alpha_infidelity}
\end{figure}

\begin{figure}[hbtp]
    \centering
    \includegraphics[width=0.48\textwidth]{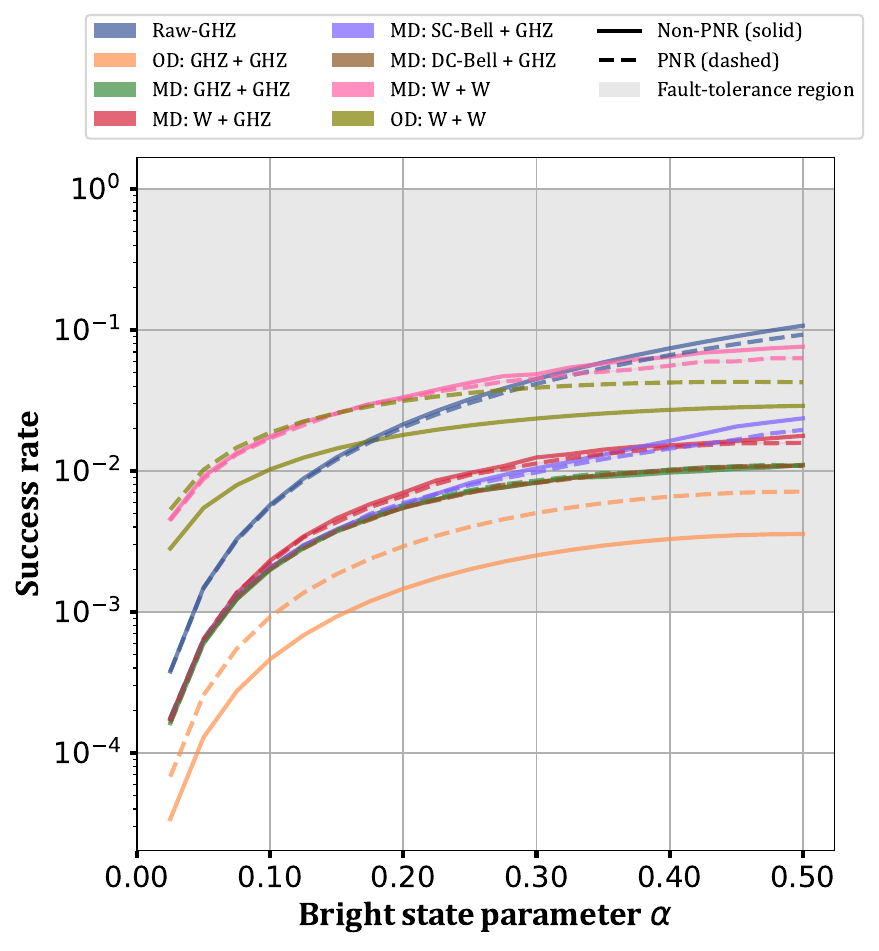}
    \caption{Success rates for GHZ state ($P_\mathrm{succ}$) for varying bright-state parameter ($\alpha$) for ES-2 hardware parameter set, coherence times $T=10^6$, and physical error-rate of $p=10^{-3}$.}
    \label{fig:alpha_success_rates}
\end{figure}

The dependence of infidelity and success probability on $\alpha$ is shown in Fig.~\ref{fig:alpha_infidelity} and Fig.~\ref{fig:alpha_success_rates}, respectively. These plots use the ES-2 hardware parameters. We observe that $P_\mathrm{succ} > 10^{-3}$ is typically achieved for $\alpha > 0.1$ across most protocols. However, only the OD-GHZ protocol meets the fidelity threshold $F_\mathrm{GHZ} > 0.98$. Since the fidelity of the OD-GHZ protocol is independent of $\alpha$, we can choose $\alpha = 0.5$ to maximize its success rate. We now proceed to vary other hardware parameters to explore the full fault-tolerance regions of all protocols. For all other protocols, we choose the lowest value of $\alpha$ to be 0.025 to strongly bias for the lowest infidelity, since that is the bottleneck in the performance based on Fig.~\ref{fig:alpha_infidelity}. For these protocols, we need to lower the fidelity more, with hardware improvements. Unless stated otherwise, we use $\alpha=\alpha_\mathrm{base}=\alpha_\mathrm{distil}=0.025$ for further analysis.

\subsection{\label{subsec:results_coherence_times}Varying the coherence times}
In addition to the fixed coherence times listed in Tab.~\ref{tab:coherence_operation_times}, it is informative to evaluate protocol performance over a wider range of coherence values. We find numerically that the success probabilities are largely insensitive to changes in coherence times. However, the GHZ state fidelity shows a strong dependence. We analyze infidelity as a function of coherence time $T$, sweeping from $10^4$ to $10^7$, which spans the parameter range defined by Set-1 to Set-3 in Ref.~\cite{singh2024modulararchitecturesentanglementschemes} and beyond.

Among the protocols, those involving memory-based distillation are most sensitive to short coherence times due to their reliance on two-qubit memory gates. For $T = 10^4$, these protocols fail to reach meaningful fidelities for quantum error correction. In contrast, setting $T = 10^6$ yields optimistic fidelity values, allowing us to identify fault-tolerant regions for distillation-based protocols.

\begin{figure}[hbtp]
    \centering
    \includegraphics[width=0.48\textwidth]{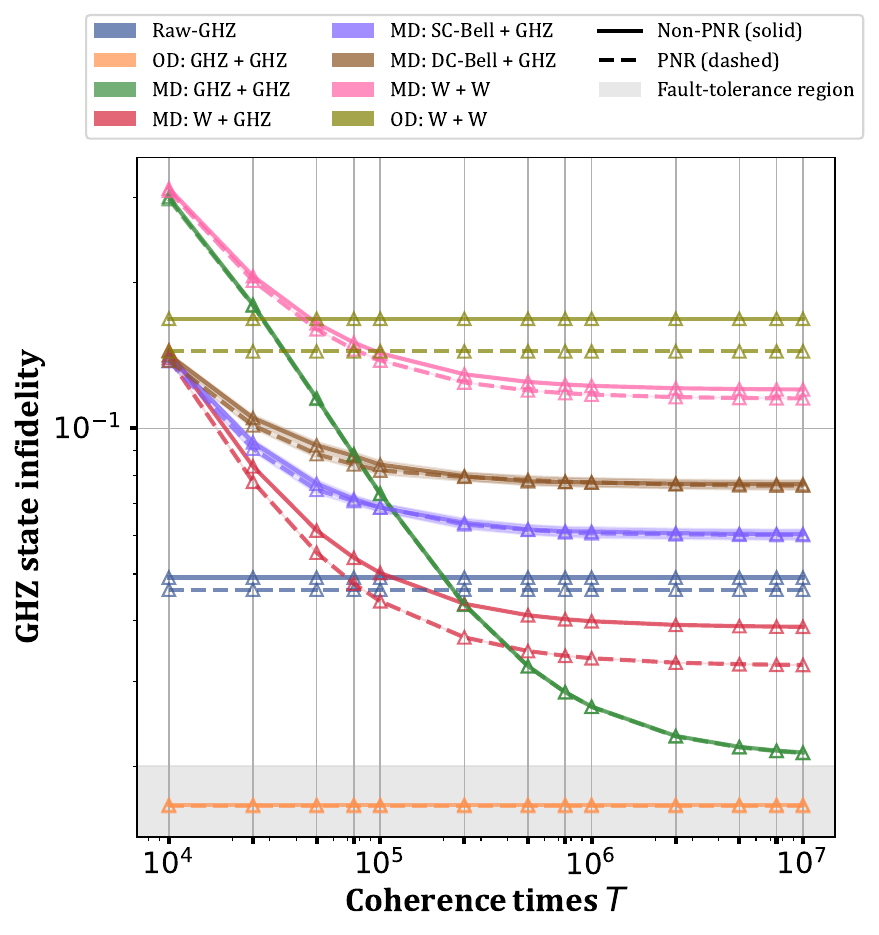}
    \caption{GHZ state infidelity ($1-F_\mathrm{GHZ}$) for varying coherence times ($T=T^\mathrm{dec}_\mathrm{link}=T^\mathrm{dec}_\mathrm{idle}$) for ES-2 hardware parameter set, physical error-rate of $p=10^{-3}$, tuned at $\alpha=0.025$.}
    \label{fig:coherence_infidelity}
\end{figure}

\subsection{\label{subsec:results_hardware_improvements}Hardware parameters roadmap}
\begin{figure}[hbtp]
    \centering
    \includegraphics[width=0.48\textwidth]{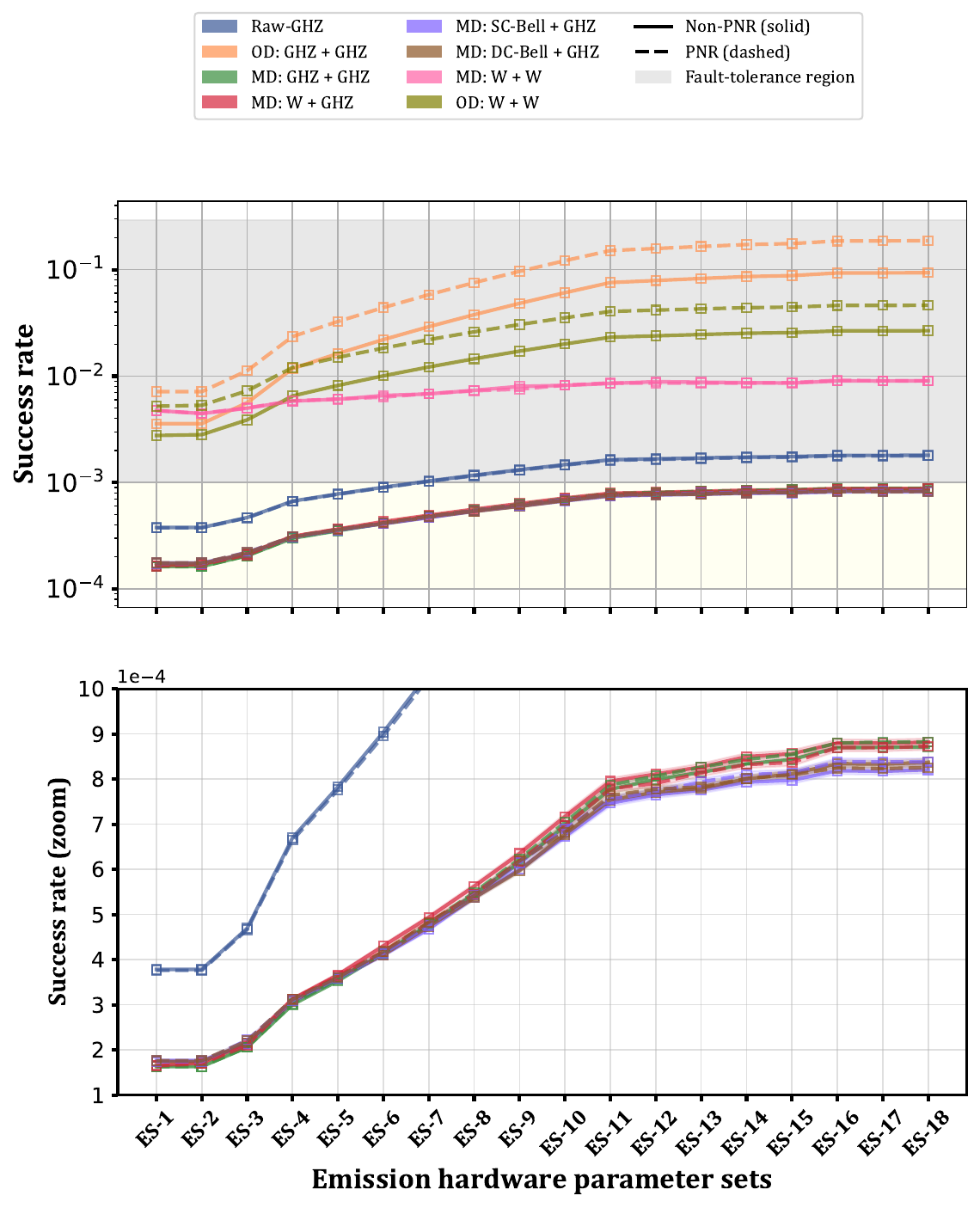}
    \caption{GHZ success rates ($P_\mathrm{succ}$) for varying hardware parameters from ES-1 to ES-18 for coherence times of $T=10^6$, physical-error rates of $p=10^{-3}$ tuned at $\alpha=0.025$ (except OD-GHZ that is biased for $\alpha=0.5$).}
    \label{fig:hardware_success_rates}
\end{figure}

\begin{figure}[hbtp]
    \centering
    \includegraphics[width=0.48\textwidth]{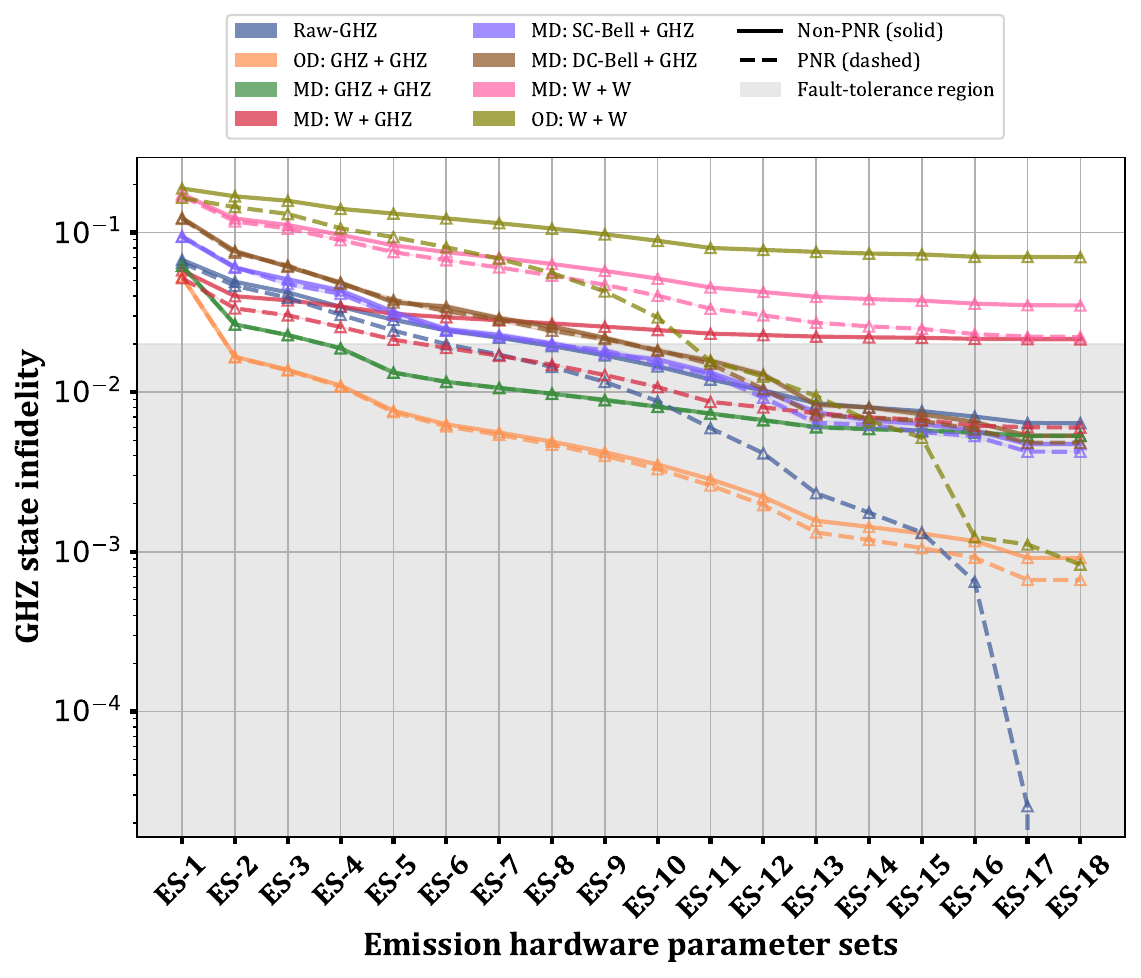}
    \caption{GHZ state infidelity ($1-F_\mathrm{GHZ}$) for varying hardware parameters from ES-1 to ES-18 for coherence times of $T=10^6$, physical-error rates of $p=10^{-3}$ tuned at $\alpha=0.025$ (except OD-GHZ that is biased for $\alpha=0.5$).}
    \label{fig:hardware_infidelity}
\end{figure}

For fixed choices of coherence times, physical error rate, and bright-state parameter, we now examine which hardware improvements are required to meet the fault-tolerance criteria for surface codes. This analysis is shown in Fig.~\ref{fig:hardware_success_rates} and Fig.~\ref{fig:hardware_infidelity}, which display success probabilities and infidelities, respectively based on the parameters presented in Tab.~\ref{tab:hardware_emission_sets}. Rather than serving as primary results, the parameter sets are presented as supporting evidence for three architectural conclusions established in previous subsections: the viability of gate-free optical distillation, the outsized impact of PNR detection, and the continuous scaling of thresholds. Among all protocols, the OD-GHZ protocol yields the lowest infidelity, and it meets the fault-tolerance requirements under hardware set ES-2, for both PNR and non-PNR detectors.

However, success probabilities show a substantial difference between the two detector types. In the PNR case, the OD-GHZ protocol achieves higher success rates compared to the non-PNR case. Acceptable success rates for the OD-GHZ protocol are first observed at hardware set ES-5. Memory distillation protocols begin to reach near usable fidelities only at higher hardware quality, but still below the desired threshold due to the gate noise.

The OD-W protocol shows better success rates than memory-based protocols, but only becomes relevant under PNR detection and with significant hardware improvements. Among all considered approaches, the OD-GHZ protocol consistently performs best across hardware sets, with PNR detectors providing a clear advantage as parameters approach ideal values.

Interestingly, beyond the hardware set ES-14, the fidelity of the raw GHZ state becomes high enough that additional distillation is no longer needed. Nonetheless, the success rate for the raw GHZ protocol remains relatively low due to its dependence on small $\alpha$, and does not match the performance of the OD-GHZ protocol for equivalent fidelity. From the noiseless analysis of raw GHZ generation, we estimate that a value of $\alpha = 0.0762$ is needed to meet the surface code threshold, that is, a fidelity of at least $0.98$. Now, in the presence of noise, the fidelity would drop due to noise, and a much lower value of $\alpha$ would be required to bring the fidelity to $0.98$, in line with what we report. In Fig.~\ref{fig:hardware_infidelity}, the infidelity for protocols other than the raw-GHZ state does not go to zero in the limit of noiseless hardware, because there is a fixed physical error per gate and measurement of $10^{-3}$, which is around the expected surface code threshold. In this analysis, we vary one parameter at a time to find fault-tolerance thresholds. To provide more insight useful to experimentalists for estimating the full hardware performance, we present two-dimensional plots of fault tolerance, sweeping both coherence times and hardware parameter sets in App.~\ref{app:heatmap_coherence_hardware}.

\subsection{\label{subsec:results_thresholds}Thresholds for fault-tolerance}
Based on the hardware parameter plots (Fig.~\ref{fig:hardware_success_rates} and Fig.~\ref{fig:hardware_infidelity}), we find the sets that yield code thresholds. These thresholds are calculated using the Union Find decoder due to its nearly linear time complexity~\cite{Delfosse_2021}. The best fidelity protocol is OD-GHZ, and we search for its threshold performance starting with ES-1, for both PNR and non-PNR detector types. These GHZ states are input to the noisy stabilizer circuits shown in Fig.~\ref{fig:modular_surface_code}(b), and then QEC cycles are repeated to estimate the logical success (or error) rates. Detailed procedure of this calculation is presented in App.~\ref{app:superoperator}. GHZ state generation in the distributed architecture is inherently probabilistic. To ensure synchrony in stabilizer measurements across the code lattice, we introduce a cut-off time, $t_\text{cut}$, which limits the maximum duration allowed for GHZ state generation before performing stabilizer measurements. A small $t_\text{cut}$ reduces the available syndrome information, impairing error detection, while a large $t_\text{cut}$ increases idle time, leading to greater decoherence. We therefore optimize $t_\text{cut}$ to maximize the error-correction threshold of the surface code.

\begin{figure}[hbtp]
    \centering
    \includegraphics[width=0.5\textwidth]{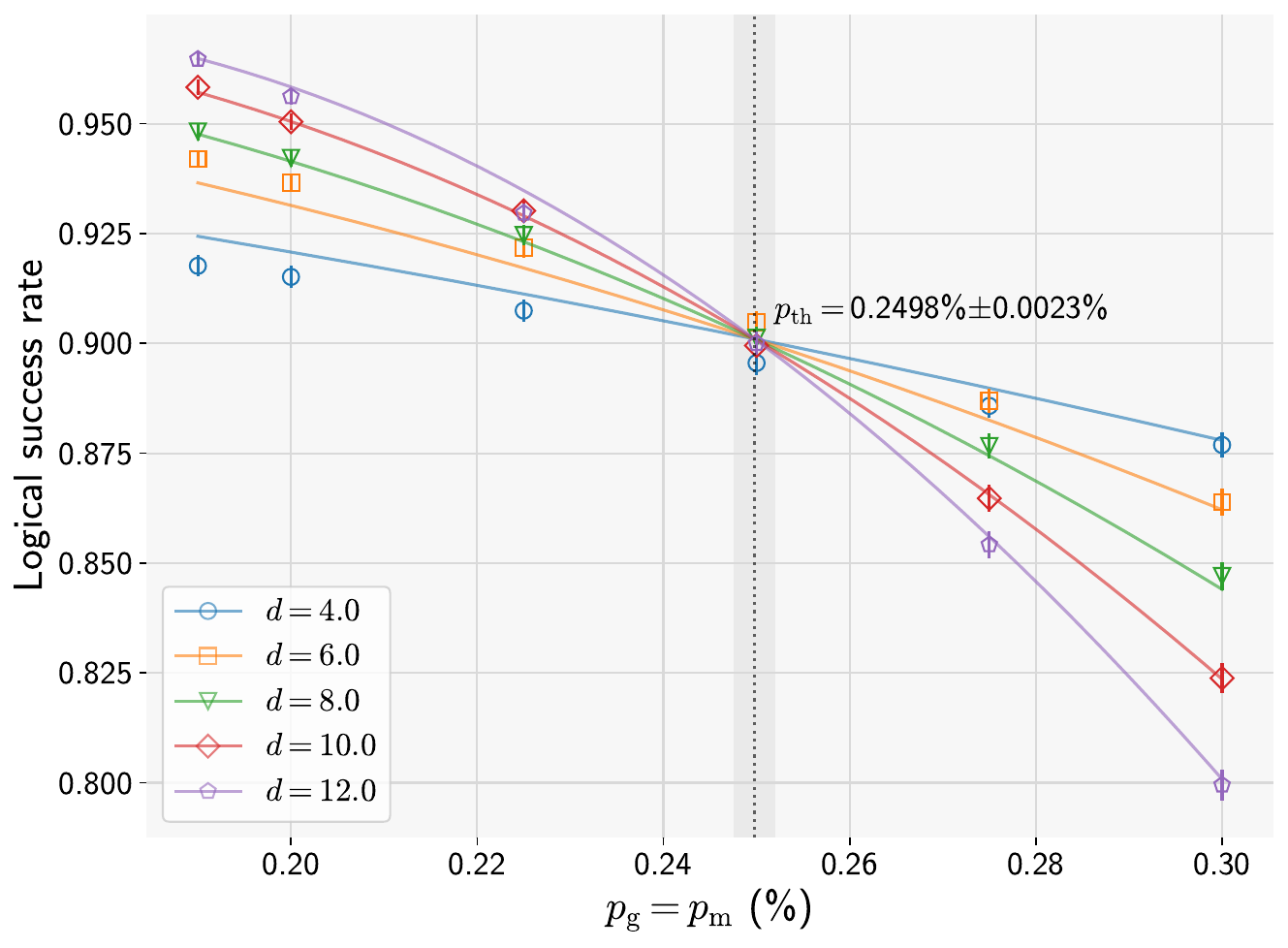}
    \caption{Threshold plot for OD-GHZ protocol with PNR detectors, for circuit-level noise. The threshold plot is evaluated for $T=10^6$, and hardware parameter set ES-2, leading to a toric surface code threshold of $p_\mathrm{th}=0.2498 \% \pm 0.0023 \%$.}
    \label{fig:threshold_DC_GHZ_PNR}
\end{figure}

We evaluate the threshold performance by scanning different hardware sets, fixing the coherence time to $T = 10^{6}$ and setting the bright-state parameter to $\alpha = 0.5$, which is optimal for the OD-GHZ protocol. We simulate logical success rates by sweeping the physical error rate $p$ (with $p = p_g = p_m$) for even code distances $d = 4, 6, 8, 10, 12$. The threshold extraction procedure is detailed in App.~\ref{app:thresholds}. For PNR detectors, we obtain a circuit-level noise threshold of $p_\mathrm{th} = 0.2498 \% \pm 0.0023 \%$ using hardware set ES-2. This exceeds the threshold of $0.13\%$ reported in Ref.~\cite{singh2024modulararchitecturesentanglementschemes} for a Bell-pair fusion protocol with non-PNR detectors and ES-1 hardware (labeled FP in Ref.~\cite{singh2024modulararchitecturesentanglementschemes}). This improvement is attributed to reduced double-excitation errors and enhanced detection resolution in the single-shot emission-based hardware. The threshold curve is shown in Fig.~\ref{fig:threshold_DC_GHZ_PNR}. At the threshold value, the GHZ state has a success probability of $P_\mathrm{succ} = 6.7\times 10^{-5}$ and a fidelity of $F_\mathrm{GHZ} = 0.9820$. The high threshold at ES-2 is enabled by the relatively high success rate with PNR detectors and the low impact of decoherence due to short GHZ generation times. This threshold competes with the threshold for direct scattering-based schemes studied in Ref.~\cite{singh2024modulararchitecturesentanglementschemes}, which were typically reported to be around $0.35\%$.

\begin{figure}[hbtp]
    \centering
    \includegraphics[width=0.5\textwidth]{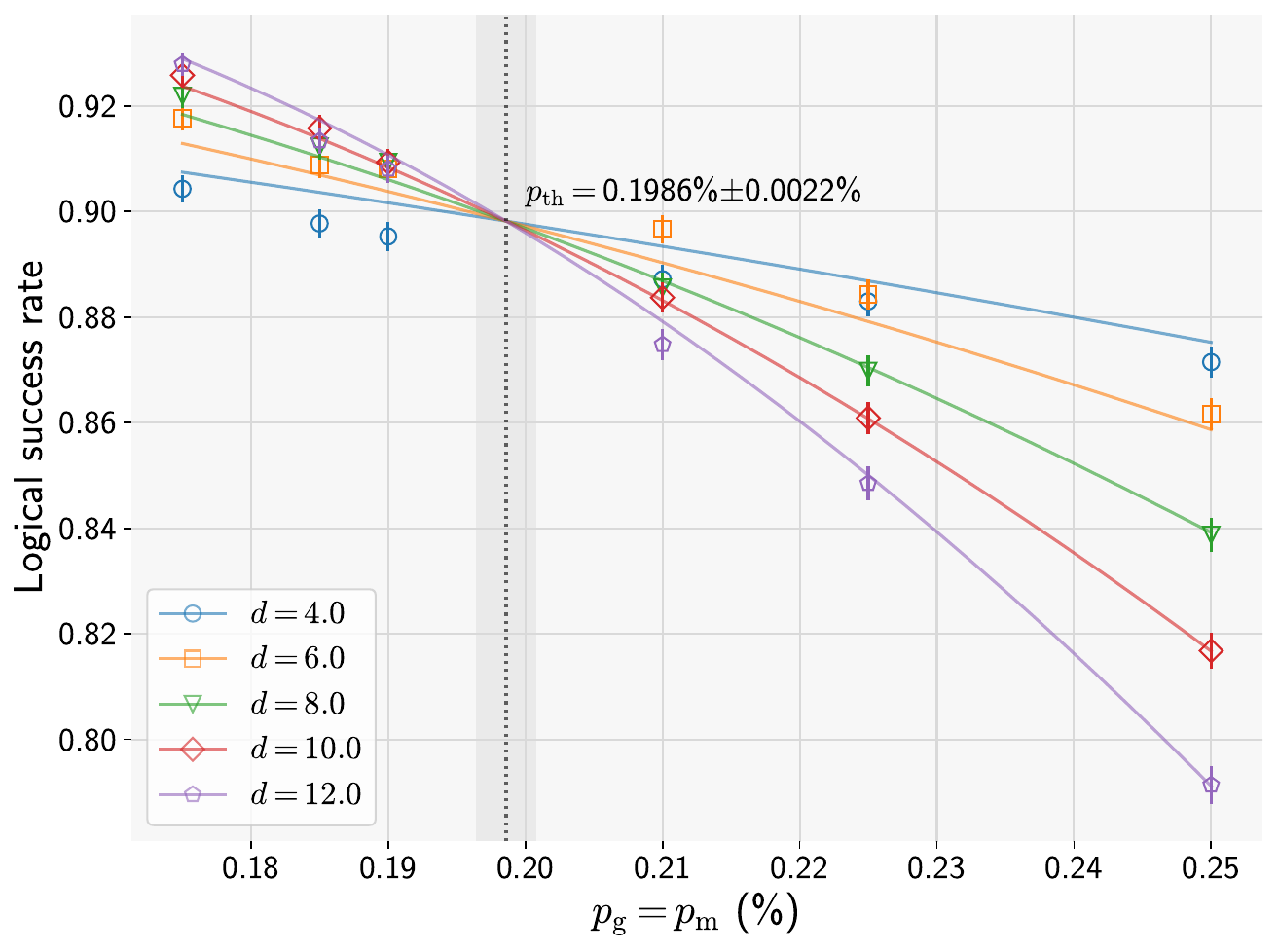}
    \caption{Threshold plot for OD-GHZ protocol with non-PNR detectors, for circuit-level noise. The threshold plot is evaluated for $T=10^6$, and hardware parameter set ES-5, leading to a toric surface code threshold of $p_\mathrm{th}=0.1986 \% \pm 0.0022 \%$.}
    \label{fig:threshold_DC_GHZ_non_PNR}
\end{figure}

For non-PNR detectors as well, we are able to find a threshold with ES-2. A threshold of  $p_\mathrm{th} = 0.1986\% \pm 0.0022\%$ is found. At this threshold, the OD-GHZ protocol with non-PNR detectors yields a GHZ success probability of $P_\mathrm{succ} = 3.38\times 10^{-5}$ and fidelity $F_\mathrm{GHZ} = 0.9824$. Due to the lower success rates in the non-PNR setting, additional hardware improvements are required to mitigate decoherence from longer GHZ generation times. Nevertheless, even with this modest hardware upgrade, the resulting threshold is sufficiently high and comparable to those achieved by fusion-based architectures reported in the literature~\cite{singh2024modulararchitecturesentanglementschemes,10.1116/5.0200190}. This is due to the high fidelity that we obtain.

To draw a direct comparison with the results of Ref.~\cite{singh2024modulararchitecturesentanglementschemes}, we can compare the stabilizer fidelity of hardware parameter set ES-1 (FP) of the previous best fusion-based scheme with the new OD-GHZ protocol with non-PNR detectors and PNR detectors, respectively. We do this since capturing the exact success rate of these complex scheduling fusion-based protocols is challenging, and stabilizer fidelity gives a good comparison of the input for threshold calculations. For the fusion-based protocol with $k=11$ Bell-pairs, the reported stabilizer fidelity was $0.9194$ at a reported threshold of $\approx 0.15\%$, and for the OD-GHZ protocol with PNR (non-PNR), we find a stabilizer fidelity of $0.9265$ at $p_\mathrm{th}=0.2498\%$ ($0.9255$ at $p_\mathrm{th}=0.1986\%$).

Thus, we affirmatively answer the questions posed at the outset: the memory-gate bottleneck of fusion-based schemes is eliminated by OD-GHZ protocols without sacrificing thresholds. Among near-term upgrades, PNR detectors provide the largest gain, while requiring only modest optical improvements.

\section{\label{sec:conclusion}Conclusion}

%In this work, we investigated fault-tolerant quantum error correction in a modular architecture using single-shot emission-based entanglement generation. 
In this work, we investigated the advantages of emission-based single-shot GHZ entanglement generation with respect to fusion-based schemes for modular architectures. Building on experimentally demonstrated hardware, we proposed and analyzed multiple entanglement generation protocols capable of creating high-fidelity GHZ states across four modules, for each stabilizer of the code. These protocols fall into two categories: memory-based distillation and fully optical distillation. We benchmarked their performance in terms of success probability and output GHZ state fidelity, under realistic noise and hardware parameters modeled after color centers in diamond.

Our results show that direct optical protocols, in particular the OD-GHZ protocol, can outperform memory-based protocols under modest hardware improvements, even in the absence of photon number-resolving detectors. 
One reason behind its performance is that it does not require two-qubit memory gates. Using realistic noise modeling and coherence constraints, we simulated distributed surface code operation and extracted fault-tolerance thresholds. Notably, we report a threshold of $0.24\%$ with PNR detectors at hardware set ES-2, and a threshold of $0.19\%$ for non-PNR detectors at ES-5. These values exceed or match previous results obtained with Bell-pair fusion-based approaches. We have shown that the thresholds with single-shot emission hardware can scale with improving hardware, unlike the previously evaluated fusion-based protocols that demonstrated maximum thresholds of $\approx 0.16\%$ restricted due to the slow two-qubit memory gates~\cite{singh2024modulararchitecturesentanglementschemes}.

Overall, this work establishes the feasibility of implementing fault-tolerant surface codes in a distributed architecture using only demonstrated emission-based hardware. Our proposed protocols enable early threshold crossings with modest hardware requirements and open pathways to scalable modular quantum computing architectures.

Several open directions follow from this work. An immediate extension is to study the performance of other topological or LDPC codes—such as hypergraph product codes or balanced product codes—within the same emission-based distributed architecture. These codes typically require higher-weight stabilizer measurements, which could benefit from generalizations of the single-shot GHZ generation protocols to larger module counts. Finally, incorporating realistic feedback and routing constraints into the protocol design at the logical level will be essential for bringing such modular architectures closer to practical fault-tolerance.

\section*{Acknowledgements}

We gratefully acknowledge support from the joint research program “Modular Quantum Computers” by Fujitsu Limited and Delft University of Technology, co-funded by the Netherlands Enterprise Agency under project number PPS2007, JST Moonshot R\&D, Grant Number JPMJMS2061 and JPMJMS226C, and Program for the Advancement of Next Generation Research Projects by Keio University. The authors thank SURF (www.surf.nl) for the support in using the Dutch National Supercomputer Snellius.

%\section*{Author Contributions}

%S.S. developed the theory of the noisy single-shot emission scheme, including all entanglement generation protocols, under the guidance of D.B. and D.E. He performed all the analytical and numerical simulations and wrote the manuscript. S.S. and D.B. conceptualized the idea of the DC GHZ/W protocol. R.K. designed the W-to-GHZ distillation protocol, K.T. evaluated the POVMs for the raw and W states under the supervision of W.R. and M.T., and wrote their appendix. D.B. developed the theory for the W-to-W distillation protocols. M.T. and D.E. together supervised the project. All authors reviewed and approved the final manuscript.

\section*{Source code and data availability}
The source code and simulation data for this research’s findings are openly available in 4TU.ResearchData repository~\cite{source_code_data}.

%%%%%%%%%%%%%%%%%%%%%%%%%%%%%%%
\section*{Appendix}
\appendix

%%%%%%%%%%%%%%%%%%%%%%%%%%%%%%%%%%%%%%%%%%%%%%
\section{\label{app:Noise_model}Noise model}
\subsection{\label{appsubsec:circuit_level_noise}Circuit-level noise}
Circuit-level noise~\cite{cao2025exactdecodingrepetitioncode,PhysRevX.13.031007} is applied to state preparation, single- and two-qubit gates, and measurements. Note that the actual circuits we run only use the native gateset (as shown in Tab.~\ref{tab:coherence_operation_times}), including the single or two-qubit gates for the communication and/or memory qubits, and measurements are only possible directly for communication qubits. State preparation, idling, and single-qubit gates are modeled as the ideal operation followed by a depolarizing channel with error probability $p_g$:
\begin{equation}
    \mathcal{N}_{\mathrm{1q}}(\rho) = (1-p_g)\rho + \frac{p_g}{3}\sum_{P_j\in\{X,Y,Z\}} P_j\rho P_j .
\end{equation}
Two-qubit gates are followed by a two-qubit depolarizing channel:
\begin{equation}
    \mathcal{N}_{\mathrm{2q}}(\rho) = (1-p_g)\rho + \frac{p_g}{15}\sum_{\substack{P_j,P_k\in\{I,X,Y,Z\} \\ \neq (I,I)}} (P_j\otimes P_k)\rho (P_j\otimes P_k) .
\end{equation}
The measurement error probability $p_m$ is modeled as the probability of flipping the measurement outcome. In simulations, we set $p_g = p_m = p$ and sweep $p$ to characterize the effect of circuit-level noise.

We model decoherence as a continuous-time process using characteristic times $T_1$ (energy relaxation) and $T_2$ (pure dephasing). The probability to undergo decoherence over time $t$ is $1 - e^{-t/T^{\mathrm{dec}}}$. During entanglement generation attempts, memory qubits have coherence time $T_\text{link}^{\mathrm{dec}}$, while in idle periods all qubits have $T_\text{idle}^{\mathrm{dec}}$, with $T_\text{idle}^{\mathrm{dec}}\geq T_\text{link}^{\mathrm{dec}}$ for the emission-based hardware considered~\cite{Reilly2015,Gold2021,Crawford2023}.

We implement decoherence channels in Kraus form:
\begin{equation}
    \mathcal{N}^\mathrm{dec}(\rho) = \sum_{j=1}^\kappa K_j \rho K_j^\dagger , \quad \sum_{j=1}^\kappa K_j^\dagger K_j = \mathbb{I} .
\end{equation}

The $T_1$ process is modeled by the generalized amplitude damping (GAD) channel:
\begin{equation}
\begin{aligned}
    K^\mathrm{GAD}_1 &= \frac{1}{\sqrt{2}}\begin{bmatrix} 1 & 0 \\ 0 & \sqrt{1-\gamma_1} \end{bmatrix}, \quad
    K^\mathrm{GAD}_2 = \frac{1}{\sqrt{2}}\begin{bmatrix} 0 & \sqrt{\gamma_1} \\ 0 & 0 \end{bmatrix}, \\
    K^\mathrm{GAD}_3 &= \frac{1}{\sqrt{2}}\begin{bmatrix} \sqrt{1-\gamma_1} & 0 \\ 0 & 1 \end{bmatrix}, \quad
    K^\mathrm{GAD}_4 = \frac{1}{\sqrt{2}}\begin{bmatrix} 0 & 0 \\ \sqrt{\gamma_1} & 0 \end{bmatrix},
\end{aligned}
\end{equation}
where $\gamma_1 = 1 - e^{-t/T_1}$.

The $T_2$ process is modeled by phase damping (PD):
\begin{equation}
    K^\mathrm{PD}_1 = \begin{bmatrix} 1 & 0 \\ 0 & \sqrt{1-\gamma_2} \end{bmatrix}, \quad
    K^\mathrm{PD}_2 = \begin{bmatrix} 0 & 0 \\ 0 & \sqrt{\gamma_2} \end{bmatrix},
\end{equation}
where $\gamma_2 = 1 - e^{-t/T_2}$.

When both channels act together, they yield an effective depolarizing process. All operations, including link generation, idling, and measurements, are assigned durations consistent with the main text, and decoherence is applied accordingly. In our simulations, this effective noise process is a time function and applied accordingly with the duration of the operations in action, in a fully time-scheduled modeling for all qubits.

\subsection{\label{appsubsec:hardware_noise}Emission-based hardware noise}
\subsubsection{\label{appsubsubsec:bright_state}Bright-state parameter and state preparation}
The bright-state parameter $\alpha$ characterizes the population of the communication qubit in the optically excited (``bright'') state $\ket{1}$ relative to the ground (``dark'') state $\ket{0}$ at the time of photon emission.  
The communication qubit, typically realized as the electronic spin of a nitrogen-vacancy (NV$^-$) center in diamond~\cite{dohertyNitrogenvacancyCentreDiamond2013}, is initialized by optical pumping into the $\ket{0}$ state ($m_s = 0$), which fluoresces upon resonant excitation, in contrast to $\ket{1}$ ($m_s = -1$), which exhibits a substantial probability of non-radiative decay to a metastable state~\cite{toganQuantumEntanglementBetween2010,bernienHeraldedEntanglementSolid2013}.  
The parameter $\alpha \in [0,1]$ is experimentally tunable by controlling the relative amplitude and duration of the optical or microwave pulses that prepare the initial superposition state
\begin{equation}
\sqrt{1-\alpha}\,\ket{0} + \sqrt{\alpha}\,\ket{1}.
\end{equation}

Upon application of a short, spin-selective resonant laser pulse (typically $\sim 2\,\mathrm{ns}$~\cite{robledoHighfidelityProjectiveReadout2011}) addressing the $\ket{1} \to \ket{1_e}$ transition, only the bright component $\ket{1}$ is excited to the optically excited state $\ket{1_e}$, which decays within $\tau_e \approx 20\,\mathrm{ns}$ to emit a single photon.  
This process generates the emitter–photon entangled state
\begin{equation}
\sqrt{1-\alpha}\,\ket{0}\ket{0_{\mathrm{ph}}} + \sqrt{\alpha}\,\ket{1}\ket{1_{\mathrm{ph}}},
\end{equation}
where $\ket{0_{\mathrm{ph}}}$ and $\ket{1_{\mathrm{ph}}}$ denote the zero- and one-photon Fock states, respectively.

Experimentally, the value of $\alpha$ plays a dual role:

\begin{enumerate}
    \item \textbf{Impact on entanglement fidelity.}  
    In the high-photon-loss regime ($\eta \ll 1$), the fidelity of the heralded two-qubit state with respect to the ideal maximally entangled state scales approximately linearly with $\alpha$ for low $\alpha$~\cite{Pompili_2021}.  
    At large $\alpha$, fidelity drops sharply due to the increased relative weight of noise counts (dark counts, stray light, leakage from excitation pulses), as noise events falsely herald entanglement. Preparation errors in the unitary rotation or off-resonant excitations can exacerbate this drop.
    
    \item \textbf{Impact on heralding probability.}  
    The per-attempt success probability (for a single-click Bell-pair) $p_{\mathrm{click}}$ scales linearly with $\alpha$ in the high-loss regime:
    \begin{equation}
        p_{\mathrm{succ}} \approx \eta \alpha.
    \end{equation}
    Imbalances in $\eta$ can be compensated by adjusting $\alpha$ to maximize fidelity for a given $p_{\mathrm{succ}}$~\cite{Pompili_2021}.
\end{enumerate}

In practice, $\alpha$ is optimized based on the specific trade-off between fidelity and rate for a given hardware configuration.

State-of-the-art NV center platforms report bright-state populations and preparation fidelities exceeding $99\%$~\cite{hensenLoopholefreeBellTest2015,kalbEntanglementDistillationSolid2017}, corresponding to SPAM (state preparation and measurement) errors below $10^{-3}$. Such performance is essential for meeting the stringent error thresholds required for fault-tolerant operation of distributed quantum error correction schemes.

The bright-state parameter is closely linked to the state-preparation fidelity $F_{\mathrm{prep}}$, which characterizes the accuracy of preparing the intended qubit state before photon emission. Imperfect preparation of the desired superposition state (e.g., due to off-resonant excitation or control pulse errors) leads to phase uncertainty in the emitter state, which in our simulations is modeled as a pure dephasing channel acting on the qubit before photon emission:
\begin{equation}
\mathcal{E}_{\mathrm{prep}}(\rho) = F_{\mathrm{prep}}\rho +(1-F_{\mathrm{prep}}) Z\rho Z,
\end{equation}
This channel captures the reduction in emitter coherence due to imperfect initialization, mapping directly to the experimentally measured $F_{\mathrm{prep}}$ values. The channel is applied to the emitter density matrix $\rho$ immediately after state preparation, prior to optical excitation.

Experimental reports of $F_{\mathrm{prep}}$ for NV-center platforms vary depending on the initialization protocol, optical pumping parameters, and environmental noise, with typical values in the range $0.95$–$0.99$ for optimized low-temperature setups~\cite{robledoHighfidelityProjectiveReadout2011, kalbEntanglementDistillationSolid2017, Pompili_2021}. In our numerical modeling, $F_{\mathrm{prep}}$ and $\alpha$ jointly determine the quality of the emitter–photon entangled state and, consequently, the performance of the overall entanglement-generation protocol.

\subsubsection{\label{appsubsubsec:path_length}Path-length difference}

In an ideal entanglement-generation protocol, the Bell state $\ket{\Psi^{\pm}} = \frac{\ket{01} \pm \ket{10}}{\sqrt{2}}$ is created with a well-defined relative phase between the two terms, depending on the detector click.  
In practice, optical path-length mismatches, residual frequency detuning between emitters, or imperfect phase stabilization introduce an additional phase $\phi$ in the state:
\begin{equation}
\ket{\Psi_\phi} = \frac{\ket{01} + e^{i\phi}\ket{10}}{\sqrt{2}}.
\end{equation}
If $\phi$ fluctuates between the attempts, the ensemble-averaged state acquires phase noise, which can be modeled as a dephasing process in the $\{\ket{01}, \ket{10}\}$ subspace. The amount of dephasing is determined by the variance $\sigma_\phi^2$ of the phase instability. Following Refs.~\cite{Humphreys2018,PhysRevA.99.052330,10.1116/5.0200190}, the corresponding dephasing parameter is
\begin{equation}
\lambda = \frac{1}{2}\left[1 + \frac{I_1(\sigma_\phi^{-2})}{I_0(\sigma_\phi^{-2})}\right],
\end{equation}
where $I_0$ and $I_1$ are modified Bessel functions of the first kind of order $0$ and $1$, respectively. This yields an effective single-qubit dephasing channel acting on one of the modes (e.g., emitter $A$'s photon) before detection:
\begin{equation}
\mathcal{E}_{\phi}(\rho) = \lambda \rho + (1-\lambda) Z \rho Z.
\end{equation}

Generalization to a four-module setting: For networks with $N>2$ modules, such as our four-module architecture, each optical link $(i,j)$ can have its own random phase offset $\phi_{ij}$ arising from independent path-length fluctuations and local oscillator phases.  
In the absence of active phase tracking, each $\phi_{ij}$ is drawn from a zero-mean distribution with variance $\sigma_{ij}^2$.  
The joint state $\rho$ of the network then experiences correlated dephasing across all photon-pair subspaces:
\begin{equation}
\mathcal{E}_{\phi,\text{multi}}(\rho) = \int \prod_{(i,j)} p_{ij}(\phi_{ij}) \; U_{\phi_{ij}}^{(i,j)} \rho \, {U_{\phi_{ij}}^{(i,j)}}^\dagger \; \mathrm{d}\phi_{ij},
\end{equation}
where $U_{\phi_{ij}}^{(i,j)}$ applies a phase rotation $e^{i\phi_{ij}}$ on the $\ket{10}$ component of the $(i,j)$ subspace.  
If all links share the same source laser and stabilization system, these phases can be partially correlated (however, we can also consider them to be fully independent, just synced timely); a simplified model is then to assign a single Gaussian-distributed global phase $\phi_g$ plus independent link noise $\delta\phi_{ij}$.

Neglecting this effect in our simulations: In our current proposal, the dominant source of phase fluctuations is slow relative to the entanglement-generation time.  
Measured values for NV-based optical links give $\sigma_\phi \approx 14.3^\circ$~\cite{Humphreys2018}, corresponding to $\lambda \approx 0.984$—a small reduction in off-diagonal coherence (for two module setup), about $\sim0.16$ dephasing factor. Given that this noise is much weaker than decoherence and hardware loss, we neglect this error source in our main simulations and assume phase tracking and corrections are available for small errors. We note that for long-term averaging or multi-round protocols without active phase tracking, this noise channel would need to be explicitly included, and we leave this for a possible future research direction.

\subsubsection{\label{appsubsubsec:double_excitation}Double-excitation error}
A critical non-ideal process in emission-based entanglement generation is double optical excitation.  
This occurs when, during a finite-duration optical excitation pulse, an emitter is re-excited after emitting a photon, leading to the sequential emission of two photons in a single attempt. In a heralded entanglement protocol, where one photon detection is used to project two distant module emitters into an entangled emitter state, the undetected photon from a double excitation event leaks information to the environment, reducing the coherence of the spin-photon state and thereby lowering the fidelity of the heralded emitter-emitter entanglement~\cite{Humphreys2018,Hermans2022,kalbEntanglementDistillationSolid2017}.

Consider an emitter initialized in its ground state $\ket{0}$ and driven by a resonant optical $\pi$-pulse. If the pulse duration is comparable to or longer than the excited-state lifetime $T_1$, there exists a finite probability that the emitter spontaneously emits a photon during the pulse and is then re-excited by the remaining optical field.  
This leads to a second emission event. In the high-loss regime typical of remote entanglement experiments, the probability of detecting both photons at the beam splitter is negligible; instead, one photon is typically lost, causing dephasing in the remaining spin-photon state.

Let $p_\mathrm{EE}$ denote the conditional probability that a given heralded detection is accompanied by an undetected extra photon emitted by the same emitter.  
Operationally, we define~\cite{Hermans_2023}:
\begin{equation}
p_\mathrm{EE} = \frac{P_2}{P_1 + P_2},
\end{equation}
where $P_1$ is the probability of a single-photon emission and $P_2$ is the probability of a two-photon emission within one attempt, conditioned on at least one photon being detected in the heralding window.  
Since the detection efficiencies $\eta_\mathrm{Z}$ (for zero-phonon line - ZPL photons) and $\eta_\mathrm{P}$ (for phonon-sideband photons - PSB) differ, the experimentally accessible quantities are:
\begin{align}
P_2' &= \eta_\mathrm{P} \eta_\mathrm{Z} P_2, \\
P_1' &= \eta_\mathrm{Z} P_1,
\end{align}
leading to the experimentally useful expression~\cite{Hermans_2023}:
\begin{equation}
p_\mathrm{EE} = \frac{N_\mathrm{coin}}{\eta_\mathrm{P} N_\mathrm{all} + (1-\eta_\mathrm{P}) N_\mathrm{coin}},
\end{equation}
where $N_\mathrm{coin}$ is the number of coincidences between ZPL and PSB detectors, and $N_\mathrm{all}$ is the total number of ZPL detection events in the heralding window over $n$ repetitions.

Error channel representation: From the perspective of quantum channel modeling, an undetected photon corresponds to tracing over a photonic degree of freedom entangled with the spin.  
This yields a pure dephasing channel on the emitter-photon state:
\begin{equation}
\mathcal{E}_\mathrm{EE}(\rho) = (1 - p_\mathrm{EE}) \rho + p_\mathrm{EE} \, Z \rho Z,
\end{equation}
where $Z$ acts on the qubit basis $\{\ket{0}, \ket{1}\}$.  
This noise channel is applied immediately after the emission step in our superoperator-based simulations.

Model assumptions and mitigation: Reducing $p_\mathrm{EE}$ requires shortening the excitation pulse relative to $T_1$, thus minimizing the chance of re-excitation within the pulse duration. 
However, excessively short pulses can increase spectral bandwidth, leading to unwanted excitation of off-resonant transitions.  
Optimizing pulse length and shape is therefore essential to balance the entanglement generation rate and the final fidelity. While the double-excitation process described above can, in principle, degrade the fidelity of heralded states, in our simulations, we neglect this effect, consistent with the assumptions stated in the main text. 
Experimental studies of NV center remote entanglement report $p_\mathrm{EE}$ values in the range $4\%-7\%$~\cite{Humphreys2018,Hermans2022}. More elaborate models, such as in~\cite{PhysRevA.104.052604}, combine $p_\mathrm{EE}$ with off-resonant excitation probabilities into a single ``excitation error'' parameter $p_\mathrm{EE}$, but for our purposes $p_\mathrm{EE}$ alone captures the dominant contribution in the parameter regime relevant for our single-shot protocols. Following Ref.~\cite{PRXQuantum.5.010202}, we assume that the emitter is embedded in an optical cavity, the excitation pulse is shorter than the cavity-enhanced excited-state lifetime, and the pulse is optimally shaped such that nearby unwanted transitions undergo a complete Rabi cycle and return to the ground state~\cite{PhysRevA.104.052604,mirambell2019fidelity}.  
Under these conditions, the probability of an undetected extra photon emission (\(p_\mathrm{EE}\)) becomes negligible.  
This is supported by experimental and theoretical studies showing that, for NV centers under such optimized driving, both double-excitation and off-resonant excitation probabilities are strongly suppressed~\cite{Nemoto_2014,Robledo_2011}.  
Therefore, in the parameter regime relevant to this work, \(p_\mathrm{EE}\) can be safely ignored without impacting the accuracy of our quantum error correction performance estimates.

\subsubsection{\label{appsubsubsec:photon_indistinguishability}Photon distinguishability}
A key requirement for high-fidelity entanglement generation via photon interference is that the photons emitted by different nodes are `indistinguishable' in all degrees of freedom except the quantum information they encode. To describe and test the indistinguishability of two photons one can make use of the so-called Hong--Ou--Mandel (HOM) effect~\cite{PhysRevLett.59.2044}.
If two perfectly indistinguishable single photons impinge simultaneously on the two input ports of a balanced beam splitter, the HOM effect dictates that they always exit through the same output port due to the two-photon quantum interference.
This interference arises because the probability amplitudes for the two indistinguishable paths---both photons transmitted or both reflected---cancel exactly, resulting in photon bunching with zero coincidence probability.
Any distinguishability---arising from differences in central frequency, spectral width, polarization, temporal shape, or arrival time---reduces the HOM visibility, which makes the HOM visibility a good measure for the pairwise photon distinguishability.
%If two single photons impinge simultaneously on the two input ports of a balanced beam splitter and are perfectly indistinguishable in all degrees of freedom, the Hong--Ou--Mandel (HOM) effect~\cite{PhysRevLett.59.2044} dictates that they always exit through the same output port due to the two-photon quantum interference.
%This interference arises because the probability amplitudes for the two indistinguishable paths---both photons transmitted or both reflected---cancel exactly, resulting in photon bunching with zero coincidence probability.
%When combined with coincidence detection and post-selection in a Bell state measurement, HOM interference enables the projection of remote qubits into entangled states, as employed in entanglement swapping and fusion-based protocols.
%Any distinguishability---arising from differences in central frequency, spectral width, polarization, temporal shape, or arrival time---reduces the HOM visibility and therefore the achievable entanglement fidelity~\cite{PhysRevLett.93.070503}.

In our setup, photons from each node are collected into polarization-maintaining (PM) fibers and interfered on a fiber-coupled beam splitter. We assume that we align the polarizations to within better than 20~dB extinction~\cite{Hermans_2023}, rendering polarization mismatch negligible. Spatial mode mismatch is also negligible due to the single-mode fiber coupling~\cite{Hermans_2023}. The dominant residual distinguishability is thus attributed to spectral and temporal variations only, in particular, fluctuating frequency offsets between photons from the two emitters.

Modeling indistinguishability: We model imperfect indistinguishability with a phenomenological `visibility' parameter $\mu_I \in [0,1]$, defined as the HOM two-photon interference visibility: $\mu_I = 1$ corresponds to perfect indistinguishability and $\mu_I = 0$ to completely distinguishable photons.  
When the photons have normalized spectral amplitudes $\psi(\omega)$ and $\phi(\omega)$, the indistinguishability can be expressed as the squared modulus of their spectral overlap~\cite{PhysRevA.69.032305,Dahlberg2019}:
\begin{equation}
\mu_I = \left| \int_{-\infty}^{\infty} \psi^*(\omega) \phi(\omega) \, d\omega \right|^2.
\label{eq:mu_definition}
\end{equation}
We adopt the term `intensity-definition' of indistinguishability (or visibility) to distinguish it from the amplitude-based definition introduced in App.~\ref{app:povm}. The two notions are directly related via $\mu_I = |\mu|^2$ and exhibit equivalent noise sensitivity. In the main text, we consistently employ the intensity-based definition, as it corresponds to experimentally accessible measurements. Accordingly, all numerical values reported in the main text and used in simulations are based on this convention. For consistency with the POVM formalism, these values are internally converted to their square roots before being inserted into the POVM matrices, ensuring results remain directly comparable with experimental interpretations. In the case where both spectra are Gaussian with identical widths $\sigma$ and a relative central frequency offset $\Delta\omega$, Eq.~\eqref{eq:mu_definition} yields
\begin{equation}
\mu_I = \exp\!\left[ -\frac{(\Delta\omega)^2}{4\sigma^2} \right],
\label{eq:mu_gaussian}
\end{equation}
illustrating the exponential sensitivity of HOM visibility to spectral detuning.

In our simulations, $\mu_I$ directly scales the off-diagonal terms of the heralded two-qubit density matrix via the POVMs described in App.~\ref{app:povm}, thereby capturing the reduction in coherence due to residual distinguishability. A similar approach has been used in prior NV-center entanglement experiments~\cite{bernienHeraldedEntanglementSolid2013,Hermans2022}. All photonic state projections that correspond to the desired detection patterns are calculated using the Kraus operators derived from these POVM elements.

Future improvements in $\mu_I$ could be achieved by stabilizing the emitter environment to suppress spectral diffusion, implementing active feedback to correct frequency drifts, or using cavity-enhanced emission to spectrally filter the photons~\cite{PhysRevX.7.031040}.  
Reaching $\mu_I \to 1$ is critical for scaling up remote entanglement protocols, as the entanglement fidelity scales linearly with $\mu_I$ in the absence of other dominant errors.

\subsubsection{\label{appsubsubsec:photon_detection_efficiency}Effective photon detection efficiency}
In any optical entanglement generation protocol, the probability that a resonantly emitted photon is actually detected is a critical figure of merit.  
We capture this in the `effective photon detection efficiency', denoted $\eta$, which represents the probability that an emitter prepared in the bright state $|1\rangle$ produces a zero-phonon-line (ZPL) photon that successfully propagates to the detector and triggers a click within the selected detection time window.  
Formally, $\eta$ can be written as the product
\begin{equation}
\eta = p_\text{ZPL} \, p_\text{col} \, p_\text{trans} \, p_\text{det} \, p_{t_w},
\label{eq:eta_ph_definition}
\end{equation}
where:
\begin{itemize}
    \item $p_\text{ZPL}$: probability of emission into the ZPL rather than the phonon sideband (PSB).  
    For nitrogen-vacancy (NV) centers, $p_\text{ZPL} \approx 0.03$~\cite{Aharonovich2011,PhysRevLett.110.243602}, with the remainder going into the PSB and thus unsuitable for high-visibility interference.
    \item $p_\text{col}$: collection efficiency of the optical interface (e.g., lens or waveguide), including Fresnel losses at the diamond-air interface.  
    \item $p_\text{trans}$: transmission probability through the optical path to the beam splitter, given for fiber coupling by  
    $p_\text{trans} = 10^{-\ell_\text{fiber} \, \gamma_\text{trans} / 10}$,  
    where $\ell_\text{fiber}$ is the fiber length and $\gamma_\text{trans}$ the attenuation coefficient in dB per unit length.
    \item $p_\text{det}$: detector quantum efficiency at the ZPL wavelength.
    \item $p_{t_w}$: probability that the photon is emitted within the chosen time window $t_w$,  
    \begin{equation}
        p_{t_w} = \int_{t_\text{start}}^{t_\text{end}} \frac{1}{\tau} e^{-t/\tau} \, dt
        = e^{-t_\text{start}/\tau} - e^{-t_\text{end}/\tau},
    \end{equation}
    where $\tau$ is the excited-state lifetime.
\end{itemize}
In our simulations for short fiber links, $p_\text{trans}$ is close to unity and can be neglected, but we retain it in the general expression \eqref{eq:eta_ph_definition} for completeness.

Channel-based modeling: Photon loss and finite detection efficiency are modeled as an effective amplitude-damping (AD) channel acting independently on the photonic qubits emitted from each module.  
The Kraus operators for the AD channel with damping probability $1 - \eta$ are
\begin{equation}
K^\text{AD}_1=
\begin{bmatrix}
1 & 0\\
0 & \sqrt{\eta}
\end{bmatrix},
\quad
K^\text{AD}_2=
\begin{bmatrix}
0 & \sqrt{1-\eta}\\
0 & 0
\end{bmatrix}.
\end{equation}
Applying these operators to the photon mode density matrices before measurement yields the state incorporating photon loss, which then feeds into the heralding and entanglement-generation calculations.  
This modeling is equivalent to assuming that any lost photon corresponds to an erasure of the corresponding detection event~\cite{Nielsen_Chuang_2010}.

Outlook and mitigation strategies: Photon loss directly lowers the entanglement success probability without affecting the fidelity of successful events (since they are post-selected).  
Improving $\eta$ is therefore a central target for experimental optimization.  
Several strategies are available:
\begin{itemize}
    \item Enhancing $p_\text{ZPL}$ via resonant cavities or strain engineering to suppress PSB emission~\cite{PhysRevX.7.031040,PhysRevLett.109.033604}.
    \item Increasing $p_\text{col}$ using high numerical aperture optics, solid-immersion lenses, or integrated photonic structures~\cite{Hausmann2012}.
    \item Reducing transmission losses $p_\text{trans}$ by minimizing fiber lengths, using low-loss fibers, or placing beam splitters closer to the emitters.
    \item Employing detectors with higher quantum efficiency $p_\text{det}$ and lower dark counts~\cite{Marsili2013}.
    \item Optimizing $p_{t_w}$ by matching the detection window to the emitter lifetime while rejecting excitation-laser leakage.
\end{itemize}
With state-of-the-art techniques, single-NV $\eta$ values exceeding a few percent are achievable, and cavity-enhanced platforms have demonstrated order-of-magnitude improvements~\cite{PhysRevX.7.031040}.

\subsubsection{\label{appsubsubsec:pnr_non_pnr_detectors}(Non) Photon number resolving (PNR) detectors}
A photon-number-resolving (PNR) detector can discriminate between different numbers of incident photons within a given detection window, outputting not just a binary click/no-click event but an integer photon count. In contrast, a non-PNR detector only registers whether at least one photon arrived, producing a click regardless of whether one or multiple photons were incident \cite{10.1063/1.3610677}.

In the context of entanglement-generation protocols using single-photon interference, the distinction between PNR and non-PNR detection directly affects both the success rates and the output fidelity. In non-PNR detection, multi-photon events can lead to false positives — cases where the heralding signal indicates successful entanglement, but the resulting state fidelity is degraded due to contributions from unwanted higher-photon-number components. With PNR capability, such events can be rejected by conditioning only on single-photon detection outcomes, thereby improving the fidelity at the expense of a reduced success rate (this is how we improved the success rates of the OD-GHZ protocol by making use of extra patterns using PNR configuration). The optimal choice depends on the photon statistics of the source, the target fidelity, and the desired entanglement rate.

POVM-based modeling: We model the detection process using a set of positive operator-valued measure (POVM) elements  
$\{ E_n \}$, where $E_n$ corresponds to the projection onto an $n$-photon detection event at a given detector. Here, $E_n$ can be interpreted as a detection pattern with a total number of photons $n$ desired in the pattern. See App.~\ref{app:povm} for details on this. These satisfy the completeness relation
\begin{equation}
\sum_{n=0}^\infty E_n = \mathbb{I}.
\end{equation}

In the non-PNR case, the effective ``click'' POVM is given by the sum of all nonzero photon-number outcomes:
\begin{equation}
E_\text{click}^\text{(non-PNR)} = \sum_{n \ge 1} E_n.
\end{equation}
This models the fact that the detector cannot distinguish between one and multiple photons — all such cases are treated identically in the analysis.

For an ideal PNR detector, the POVM element corresponding to the detection of exactly $n$ photons is simply $E_n$, corresponding to the exact target pattern. When the protocol requires conditioning on single-photon detection events, postselection is performed on $n=1$ outcomes only, i.e.,
\begin{equation}
E_\text{click}^\text{(PNR, $n=1$)} = E_1.
\end{equation}

In a realistic detector, optical loss and spurious counts must be taken into account.  
Losses are characterized by detection efficiency $\eta_\text{det}$, while dark counts occur with probability $p_\text{dc}$ per detection window. These imperfections transform the ideal POVM elements $\Pi_n$ into effective ones, $\tilde{\Pi}_n(\eta_\text{det}, p_\text{dc})$~\cite{PhysRevA.85.023820}. For example, the effective ``single-photon'' POVM becomes a weighted mixture of the ideal $\Pi_0$, $\Pi_1$, and higher-order elements, with weights determined by $\eta_\text{det}$ and $p_\text{dc}$. However, for the scope of this work, we consider all the detectors to be identical with no dark counts. The effects of $\eta_\text{det}$ are fully absorbed in $\eta$.

While most quantum network experiments to date have relied on non-PNR detectors — typically avalanche photodiodes (APDs) or superconducting nanowire single-photon detectors (SNSPDs) — significant progress has been made toward high-efficiency PNR detection. Several experimental devices exist in this research area, such as transition-edge sensors \cite{Lita:08}, multiplexed SNSPD with high timing resolution and lower dark counts \cite{Mattioli_2015}, and superconducting kinetic-inductance detectors enabling scalable PNR systems \cite{PhysRevA.84.060301}. For quantum repeater and distributed computing architectures, PNR detectors enhance heralding fidelity in multi-photon interference and reduce background effects.

%%%%%%%%%%%%%%%%%%%%%%%%%%%%%%%%%%%%%%%%%%%%%%
\section{\label{app:povm}POVM calculation for single-shot multi-partite entanglement generation}
In this appendix, we present the derivation of positive operator-valued measures (POVMs) relevant for the heralded generation of W, GHZ, and Bell states, following the framework of Ref.~\cite{Dahlberg2019}. The construction is carried out in the basis of input photon Fock states and explicitly incorporates temporal-mode mismatch as well as multi-photon contributions arising in threshold detection. These POVMs serve as an analytical tool for modeling detector responses to specific click patterns, which, in turn, underpin the elementary-state generation protocols for the emitters, as discussed in the main text.

\subsection{General framework of derivation}
As described in the main text, the measurement module consists of the 4-by-4 beamsplitter network and four photon detectors in the output ports. The projection operator for $n_k$ photons at the $k$-th output port is defined as
\begin{align}
    \label{eq:n_photon_projector}
    \ket{n_k}_k\bra{n_k}=\frac{1}{n_k!}{\hat{r}_k^{\dagger {n_k}}}\ket{0}_k\bra{0} \hat{r}_k^{n_k},
\end{align}
where $\hat{r}_k^\dagger$ and $\hat{r}_k$ are creation and annihilation operators at the $k$-th port. The $n$-photon projector at the $k$-th port is transformed into a projector for input photons using the beamsplitter transformation given in Eq.~(\ref{eqn:4_beamsplitter_transformation}):
\begin{equation}
\begin{pmatrix}
\hat p_0^\dagger \\
\hat p_1^\dagger \\
\hat p_2^\dagger \\
\hat p_3^\dagger 
\end{pmatrix}
=
\frac{1}{2}
\begin{pmatrix}
1 & 1 & 1 &1 \\
1 & -1 & 1 & -1 \\
1 & 1 & -1 & -1 \\
1 & -1 & -1 & 1
\end{pmatrix}
\begin{pmatrix}
\hat r_0^\dagger \\
\hat r_1^\dagger \\
\hat r_2^\dagger \\
\hat r_3^\dagger 
\end{pmatrix} ,
\label{eq:app_beamsplitter}
\end{equation}
where $\hat p_j^\dagger$ is the creation operator at the $j$-th input mode. Note that, for mathematical simplicity, the subscripts in this appendix start from 0, whereas they start from 1 in the main text.
For simplicity, we denote the beamsplitter transformation by $\hat V$.
Then, the projector for the input modes is
\begin{align}
    \label{eq:projector_for_cw_input}
    &\ket{n_k}_k\bra{n_k}\nonumber\\
    &\xrightarrow{\hat V}\frac{1}{n_k!}\left(\sum_j V_{jk}\hat{p}_j^\dagger\right)^{n_k} \ket{0}_{\rm in}\bra{0}\left(\sum_j V_{jk}\hat{p}_j\right)^{n_k},
\end{align}
where $V_{jk}$ denotes the $(j,k)$-element of $\hat V$, and $\ket{0}_{\rm in}\bra{0}$ represents the vacuum state of all input modes.

Since there are four output ports, the projection operator for all output modes is given by taking the tensor product of the single-port projection operator $\ket{n_k}_k\bra{n_k}$ in Eq.~(\ref{eq:n_photon_projector}), for $k=0,...,3$:
\begin{align}
    &\bigotimes_{k=0}^3 \ket{n_k}_k\bra{n_k}=\bigotimes_{k=0}^3\left[ \frac{1}{n_k!}{\hat{r}_k^{\dagger n_k}}\ket{0}_k\bra{0} \hat{r}_k^{n_k}\right].
\end{align}
Then, the projection operator for input states is given by
\begin{align}
    \bigotimes_{k=0}^3\frac{1}{n_k!}\left(\sum_j V_{jk}\hat{p}_j^\dagger\right)^{n_k} \ket{0}_{\rm in}\bra{0}\left(\sum_j V_{jk}\hat{p}_j\right)^{n_k}.
\end{align}

Hereafter, we incorporate temporal modes into the above discussion. When the photon detectors have a flat spectral response, the $n_k$ photon projector at the $k$-th detector is defined as
\begin{align}
    \hat{\mathcal{P}}_k^{(n_k)}=\frac{1}{n_k!}\int d^{n_k}\bm{\Omega}_k \bigotimes_{l=0}^{n_k-1} \left[\hat{r}_k^\dagger (\Omega_{kl})\ket{0}_k\bra{0}\hat{r}_k (\Omega_{kl})\right],
\end{align}
where $\bm{\Omega}_k$ is a vector of length $n_k$, and $\Omega_{kl}$ is a $l$-th element in the vector $\bm{\Omega}_k$.
The creation and annihilation operators for the frequency mode $\Omega_{kl}$ satisfy the following commutation relations:
\begin{equation}
    \label{eq:commutation_relation_spectral_modes}
    \begin{aligned}
    \left[\hat{r}(\Omega_{kl}),\hat{r}^\dagger(\Omega'_{kl})\right]&=\delta(\Omega_{kl}-\Omega'_{kl}),\\
    \left[\hat{r}(\Omega_{kl}),\hat{r}(\Omega'_{kl})\right]&=0.
    \end{aligned}
\end{equation}

The projection operator corresponding to the detection of $\bm{n}=[n_0,...,n_3]$ photons in the four detectors is defined as
\begin{align}
    \hat{\mathcal{P}}_{\bm{n}}=\bigotimes_{k=0}^3 \hat{\mathcal{P}}_k^{(n_k)}.
\end{align}

Next, we consider the temporal modes of the input photons. When all input photons have a single frequency, the input Hilbert space is spanned by the following basis:
\begin{align}
    \ket{0000}_{\rm in}, \ket{0001}_{\rm in},\ \ket{0010}_{\rm in},\ \ldots,\ \ket{1111}_{\rm in}.
\end{align}
For simplicity, we represent these input states using a bit string $\bm m=m_0m_1m_2m_3$ as 
\begin{align}
    \ket{\bm m}_{\rm in}&=\ket{m_0m_1m_2m_3}_{\rm in}\\
    &=\bigotimes_{j=0}^3 \left(\hat{p}_j^\dagger\right)^{m_j}\ket{0}_{\rm in},
\end{align}
where $m_i = 0,1$ for $i=0,\ldots,3$.
By incorporating the temporal mode function $\phi_j(\omega)$ and the phase shift $e^{i\omega \tau_j}$ into the input states, the creation operator at the $j$-th input mode is expressed as
\begin{align}
    \hat{p}_j^\dagger \rightarrow \int d\omega\ \phi_j^*(\omega) e^{i\omega \tau_j}\hat{p}_j^\dagger(\omega).
\end{align}
Furthermore, the input annihilation operator is transformed by the beamsplitter matrix in Eq.~(\ref{eq:app_beamsplitter}) as
\begin{align}
    \int d\omega\ \phi_j^*(\omega) e^{i\omega \tau_j}\hat{p}_j^\dagger(\omega)\xrightarrow{\hat V}\sum_{k=0}^3 V_{jk}\hat{R}_{jk}^\dagger,
\end{align}
where $\hat R_{jk}^\dagger$ is the creation operator at the $k$-th output port associated with the temporal mode of the $j$-th input port:
\begin{align}
    \hat{R}_{jk}^\dagger=\int d\omega\ \phi_j^*(\omega) e^{-i\omega \tau_j}\hat{r}_k^\dagger(\omega).
\end{align}

To construct the POVMs on the input space spanned by $\ket{\bm m}_{\rm in}$, we introduce the isometry $\hat{\mathcal{V}}$ defined as
\begin{align}
    \ket{\bm m}_{\rm in}&=\bigotimes_{j=0}^3 \left(\hat{p}_j^\dagger\right)^{m_j}\ket{0}_{\rm in}\nonumber\\
    &\xrightarrow{\hat{\mathcal{V}}}\bigotimes_{j=0}^3\left(\sum_{k=0}^3 V_{jk}\hat{R}_{jk}^\dagger\right)^{m_j}\ket{0}_{\rm out}\label{eq:def_vm}\\
    &\eqqcolon\ket{\mathcal{\hat{V}}\bm m}_{\rm out},
\end{align}
which is given by
\begin{equation}
    \hat{\mathcal{V}}=\sum_{\bm{m}} \bigotimes_{j=0}^3\left(\sum_{k=0}^3V_{jk}\hat{R}_{jk}^{\dagger}\right)^{m_j}\ket{0}_{\rm out}\bra{\bm{m}}_{\rm in}.
\end{equation}
Using this isometry, we obtain the POVMs acting on the input photons and incorporating temporal-mode mismatch effects:
\begin{align}
    \label{eq:POVM_n_photons}
    \hat{E}_{\bm{n}}&=\hat{\mathcal{V}}^\dagger \mathcal{\hat{P}}_{\bm{n}} \hat{\mathcal{V}}\nonumber\\
    &=\sum_{\bm{m},\bm{m}'}\ket{\bm{m}}_{\rm in}\bra{\bm{m}'}\nonumber\\
    &\quad\quad\int d^{N_{\rm in}}\bm{\Omega}\ M_{\bm{m}\bm{n}}(\bm{\Omega})M_{\bm{m}'\bm{n}}^*(\bm{\Omega}),
\end{align}
where we simply wrote the multiple integral over frequencies $\bm \Omega$ corresponding to $N_{\rm in}$ detected photons, and
\begin{align}
    M_{\bm{m}\bm{n}}(\bm \Omega)&={}_{\rm out}\bra{\mathcal{\hat{V}}\bm{m}}\bigotimes_{k'=0}^3\left[\bigotimes_{l=0}^{n_{k'}-1}\frac{1}{\sqrt{n_{k'}!}}\hat{r}_{k'}^\dagger(\Omega_{k'l})\right]\ket{0}_{\rm out}.
\end{align}
When the vacuum state is injected in the $j$-th input port, the relevant terms in Eq.~(\ref{eq:def_vm}) become identity. Then, we can simplify $E_{\bm n}$ by removing the terms corresponding to the vacuum inputs. We introduce the new indices $\bm j$ representing the input ports of injected single photons, defined as
\begin{align}
    \bm j=\left\{j\in \{0,1,2,3\}|m_j \neq 0\right\}.
\end{align}
Since the bit string $\bm m$ representing the input photon number is in one-to-one correspondence with the input port indices $\bm j$, the input state vector $\ket{\bm m}_{\rm in}$ can be replaced with $\ket{\bm j}$.
For example, $\ket{\bm{m}}_{\rm in}=\ket{0110}_{\rm in}$ leads to $\ket{\bm{j}}=\ket{1,2}$.
Then, the simplified POVMs are given by
\begin{align}
    \label{eq:simplified_POVM}
    \hat{E}_{\bm n}=\sum_{\bm j, \bm j'}\ket{\bm j}_{\rm in}\bra{\bm j'}\int d^{N_{\rm in}}\bm{\Omega} M_{\bm {j n}}(\bm \Omega)M_{\bm {j' n}}^*(\bm \Omega),
\end{align}
where
\begin{widetext}
    \begin{align}
        \label{eq:elements_of_POVM}
        M_{\bm{jn}}(\bm{\Omega})&={}_{\rm out}\bra{0}\bigotimes_{j\in\bm{j}}\left(\sum_{k=0}^3V_{jk}\hat{R}_{jk}\right)\bigotimes_{k'=0}^3\left[\bigotimes_{l=0}^{n_{k'}-1}\frac{1}{\sqrt{n_{k'}!}}\hat{r}_{k'}^\dagger(\Omega_{k'l})\right]\ket{0}_{\rm out}.
    \end{align}
\end{widetext}

\subsection{POVM for W states}
As described in the main text, the measurement module consists of the 4-by-4 beamsplitter network and four photon detectors, and it heralds W-state generation when a single photon is detected in one of the four detectors \cite{rogaEfficientDickestateDistribution2023}. In this appendix, we consider the case where the photon clicks in the first detector, which generates the following four-partite W state:
\begin{align}
    \ket{W_4}=\frac{1}{2}\left(\ket{1000}+\ket{0100}+\ket{0010}+\ket{0001}\right).
\end{align}
When the threshold detectors are used, it is impossible to distinguish between single-photon detection and multi-photon detection.
Since at most four photons are injected, the number of detected photons is also at most four. Accordingly, the following four projectors capture all output patterns associated with $\ket{W}$ generation:
\begin{align}
    \hat{\mathcal{P}}_{1000}&=\int d\Omega_0\ \hat{r}_0^\dagger (\Omega_0)\ket{0}_0\bra{0}\hat{r}_0(\Omega_0), \\
    \mathcal{P}_{2000}&=\frac{1}{2}\iint d\Omega_0d\Omega_1\nonumber\\ 
    &\qquad\quad \hat{r}_0^\dagger (\Omega_0)\hat{r}_0^\dagger (\Omega_1)\ket{0}_0\bra{0}\hat{r}_0(\Omega_0)\hat{r}_0(\Omega_1),\\
    \hat{\mathcal{P}}_{3000}&=\frac{1}{6}\int d\Omega_0d\Omega_1d\Omega_2\ \nonumber\\
    &\qquad\quad\bigotimes_{l=0}^2 \hat{r}_0^\dagger (\Omega_l)\ket{0}_0\bra{0}\bigotimes_{l=0}^2 \hat{r}_0(\Omega_l),\\
    \hat{\mathcal{P}}_{4000}&=\frac{1}{24}\int d\Omega_0d\Omega_1d\Omega_2\ d\Omega_3\ \nonumber\ \\
    &\qquad\quad\bigotimes_{l=0}^3 \hat{r}_0^\dagger (\Omega_l)\ket{0}\bra{0}\bigotimes_{l=0}^3 \hat{r}_0(\Omega_l).
\end{align}
We derive the POVM operators $\hat{E}_{1000},\ \hat{E}_{2000},\ \hat{E}_{3000},$ and $\hat{E}_{4000}$ in the following subsections.

\subsubsection{\texorpdfstring{$\hat{E}_{1000}$}{E1000}}
We derive the POVM $\hat{E}_{1000}$ corresponding to the projector $\hat{\mathcal{P}}_{1000}$, i.e., the case where $n_0=1$ and $n_1=n_2=n_3=0$ in Eq.~(\ref{eq:elements_of_POVM}). In this case, only one photon is injected into the central entangling module, represented by the following state vectors:
\begin{align}
    \label{eq:set_j_one_photon}
    \ket{\bm j}=\ket{j}\in\left\{\ket{0},\ \ket{1},\ \ket{2},\ \ket{3}\right\}.
\end{align}
Then, the elements $M_{\bm j1000}$ of the POVM $E_{1000}$ are given by
\begin{align}
    M_{\bm j1000}(\Omega)={}_{\rm out}\bra{0}V_{j0}\hat{R}_{j0}\hat{r}_0^\dagger(\Omega)\ket{0}_{\rm out}.
\end{align}
As shown in Eq.~(\ref{eq:app_beamsplitter}), all elements in the $0$-th column of the beamsplitter matrix $V_{j0}$ are equal to 1/2. Using the commutation relation given in Eq.~(\ref{eq:commutation_relation_spectral_modes}), 
\begin{align}
     &M_{\bm j1000}(\Omega)\nonumber\\
     =&\frac{1}{2}{}_{\rm out}\bra{0}\hat{R}_{j0}\hat{r}_0^\dagger(\Omega)\ket{0}_{\rm out}\nonumber\\
    =&\frac{1}{2}{}_{\rm out}\bra{0}\int d\omega\ \phi_j^*(\omega)e^{i\omega \tau_j}\hat{r}_0(\omega)\hat{r}_0^\dagger(\Omega)\ket{0}_{\rm out}\nonumber\\
    =&\frac{1}{2}\phi_j^*(\Omega)e^{i\Omega \tau_j}.
    \label{eq:M_j}
\end{align}
Substituting Eq.~(\ref{eq:M_j}) into Eq.~(\ref{eq:simplified_POVM}), we obtain
\begin{align}
    \hat{E}_{1000}&=\frac{1}{4}\sum_{\bm{j},\bm{j}'}\ket{\bm{j}}_{\rm in}\bra{\bm{j}'}\nonumber\\
    &\quad\qquad\int d\Omega\ \phi_j^*(\Omega)\phi_{j'}(\Omega)e^{-i\Omega (\tau_{j'}-\tau_j)},
\end{align}
which is simplified to
\begin{align}
    \hat{E}_{1000}&=\frac{1}{4}\sum_{\bm{j},\bm{j}'}\mu_{jj'}\ket{\bm{j}}_{\rm in}\bra{\bm{j}'},
\end{align}
where $\ket{\bm j}$ and $\ket{\bm{j}'}$ belong to the same set in Eq.~(\ref{eq:set_j_one_photon}), and $\mu_{jj'}$ denotes the visibility parameter between the photons from the $j$-th and $j'$-th input ports:
\begin{align}
    \mu_{jj'}=\int d\Omega\ \phi_j^*(\Omega)\phi_{j'}(\Omega)e^{-i\Omega (\tau_{j'}-\tau_j)}.
\end{align}
We refer to this as the `amplitude-definition' of indistinguishability or visibility. This should not be confused with the `squared' definition of indistinguishability as defined in App.~\ref{appsubsubsec:photon_indistinguishability}, where we define it based on intensity, and is the one that is considered for the numerical simulations in the main text.

\subsubsection{\texorpdfstring{$\hat{E}_{2000}$}{E2000}}
As in the previous subsection, when two photons are detected at the first detector, the photon-number indices are $n_0=2$ and $n_1=n_2=n_3=0$ in Eq.~(\ref{eq:elements_of_POVM}). In this case, the corresponding input states are given by
\begin{align}
    \label{eq:set_j_two_photon}
    \ket{\bm{j}}=\ket{j_0j_1}\in&\left\{\ket{01},\ket{02},\ket{03},\right.\nonumber\\
    &\left.\ket{12},\ket{13},\ket{23}\right\}.
\end{align}
The elements $M_{\bm{j}2000}$ of the POVM $E_{2000}$ are given by
\begin{align}
    &M_{\bm{j}2000}(\bm{\Omega})\nonumber\\=&\frac{1}{4\sqrt{2}}\ {}_{\rm out}\bra{0}\hat{R}_{j_0 0}\hat{R}_{j_1 0}\hat{r}_0^\dagger (\Omega_0)\hat{r}_0^\dagger (\Omega_1)\ket{0}_{\rm out}.
\end{align}
Using the commutation relation in Eq.~(\ref{eq:commutation_relation_spectral_modes}), we obtain
\begin{align}
    \label{eq:commutation_relation_two_terms}
    &{}_{\rm out}\bra{0}\hat{r}(\omega_0)\hat{r}(\omega_1)\hat{r}^\dagger(\Omega_0)\hat{r}^\dagger(\Omega_1)\ket{0}_{\rm out}\nonumber\\
    =&\delta(\omega_0-\Omega_0)\delta(\omega_1-\Omega_1)\nonumber\\
    +&\delta(\omega_0-\Omega_1)\delta(\omega_1-\Omega_0),
\end{align}
and we find
\begin{align}
    &M_{\bm{j}2000}(\bm{\Omega})\nonumber\\
    =&\frac{1}{4\sqrt{2}}\left\{\phi_{j_0}^*(\Omega_0)\phi_{j_1}^*(\Omega_1)\exp\left[i\left(\Omega_{0}\tau_{j_0}+\Omega_{1}\tau_{j_1}\right)\right]\right.\nonumber\\
    &\left.+\phi_{j_1}^*(\Omega_0)\phi_{j_0}^*(\Omega_1)\exp\left[i\left(\Omega_{1}\tau_{j_0}+\Omega_{0}\tau_{j_1}\right)\right]\right\}\nonumber\\
    =&\frac{1}{4\sqrt{2}}\sum_{\sigma\in S_2}\prod_{k=0}^{1}\phi^*_{j_{\sigma(k)}}(\Omega_k)\exp(i\Omega_k\tau_{j_{\sigma(k)}}).
\end{align}
Here, $\sigma$ is a permutation on two elements, and $S_2$ is the symmetric group on the set $\{0,1\}$.
Therefore, the POVM operator $\hat{E}_{2000}$ is given by
\begin{align}
    \hat{E}_{2000}=\frac{1}{32}\sum_{\bm{j},\bm{j}'}\left[\sum_{\sigma,\sigma'\in S_2}\prod_{k=0}^1 \mu_{j_{\sigma(k)}j'_{\sigma'(k)}}\right]\ket{\bm{j}}_{\rm in}\bra{\bm{j}'},
\end{align}
where $\ket{\bm j}$ and $\ket{\bm j'}$ belong to the same set in Eq.~(\ref{eq:set_j_two_photon}).

\subsubsection{\texorpdfstring{$\hat{E}_{3000}$}{E3000}}
When three photons are detected at the first detector, the photon-number indices are $n_0=3$ and $n_1=n_2=n_3=0$ in Eq.~(\ref{eq:elements_of_POVM}). In this case, photons are injected into the three input ports, represented by:
\begin{align}
    \label{eq:set_j_three_photon}
    \ket{\bm j}=\ket{j_0j_1j_2}\in\left\{\ket{012},\ket{013},\ket{023},\ket{123}\right\}.
\end{align}
The elements $M_{\bm{j}3000}$ of POVM $\hat{E}_{3000}$ are given by
\begin{align}
    M_{\bm{j}3000}(\bm{\Omega})&=\frac{1}{8\sqrt{6}}\ {}_{\rm out}\bra{0}\bigotimes_{k=0}^2\hat{R}_{j_k 0}\hat{r}_0^\dagger (\Omega_k)\ket{0}_{\rm out}.
\end{align}
Using the commutation relation in Eq.~(\ref{eq:commutation_relation_spectral_modes}), we obtain
\begin{align}
    \label{eq:commutation_relation_three_terms}
    &{}_{\rm out}\bra{0}\bigotimes_{k=0}^2\hat{r}_0(\omega_k)\hat{r}_0^\dagger(\Omega_k)\ket{0}_{\rm out}\nonumber\\
    &=\sum_{\sigma\in S_3}\prod_{k=0}^2\delta(\omega_k-\Omega_{\sigma(k)}),
\end{align}
and we find
\begin{align}
    M_{\bm{j}3000}(\bm{\Omega})=\frac{1}{8\sqrt{6}}\sum_{\sigma\in S_3}\prod_{k=0}^2\phi_{j_{\sigma(k)}}^*(\Omega_k)\exp(i\Omega_k\tau_{j_{\sigma(k)}}),
\end{align}
where $S_3$ denotes the symmetric group on three elements, namely $\{0,1,2\}$.
Therefore, the POVM operator $\hat{E}_{3000}$ is given by
\begin{align}
    \hat{E}_{3000}=\frac{1}{384}\sum_{\bm{j},\bm{j}'}\left[\sum_{\sigma,\sigma'\in S_3}\prod_{k=0}^2 \mu_{j_{\sigma(k)}j'_{\sigma'(k)}}\right]\ket{\bm{j}}_{\rm in}\bra{\bm{j}},
\end{align}
where $\ket{\bm j}$ and $\ket{\bm j'}$ belong to the same set in Eq.~(\ref{eq:set_j_three_photon})
\subsubsection{\texorpdfstring{$\hat{E}_{4000}$}{E4000}}
When four photons are detected at the first detector, the photon-number indices are $n_0=4$ and $n_1=n_2=n_3=0$ in Eq.~(\ref{eq:elements_of_POVM}). In this case, the photons are injected into all four input ports, represented by the state vector $\ket{\bm j}=\ket{0123}$. Then, the corresponding elements $M_{\bm{j}4000}$ of POVM $\hat{E}_{4000}$ are given by
\begin{align}
    M_{\bm{j}4000}(\bm{\Omega})&=\frac{1}{16\sqrt{4!}}\ {}_{\rm out}\bra{0}\bigotimes_{k=0}^3\hat{R}_{k 0}\hat{r}_0^\dagger (\Omega_k)\ket{0}_{\rm out}.
\end{align}
Using the commutation relation in Eq.~(\ref{eq:commutation_relation_spectral_modes}), we obtain
\begin{align}
    &{}_{\rm out}\bra{0}\bigotimes_{k=0}^3\hat{r}_0(\omega_k)\hat{r}_0^\dagger(\Omega_k)\ket{0}_{\rm out}\nonumber\\
    &=\sum_{\sigma\in S_4}\prod_{k=0}^3\delta(\omega_k-\Omega_{\sigma(k)}),
\end{align}
and we find
\begin{align}
    M_{\bm{j}}(\bm{\Omega})=\frac{1}{16\sqrt{4!}}\sum_{\sigma\in S_4}\prod_{k=0}^3\phi_{\sigma(k)}^*(\Omega_k)\exp(i\Omega_k\tau_{\sigma(k)}),
\end{align}
where $S_4$ is the symmetric group on the four letters, $0,\ldots,3$.
Therefore, the POVM operator $\hat{E}_{4000}$ is given by
\begin{align}
    \hat{E}_{4000}=\frac{1}{6144}\sum_{\sigma,\sigma'\in S_4}\prod_{k=0}^3 \mu_{\sigma(k)\sigma'(k)}\ket{0123}_{\rm in}\bra{0123}.
\end{align}

\subsection{POVM for GHZ states}

\begin{table}
  \caption{Generated GHZ states depending on detection pattern}
  \label{tab:GHZ_detection_pattern}
  \centering
  \begin{tabular}{cc}
    \hline
    Detection pattern & Generated state  \\
    \hline \hline
    0,1 & $\frac{1}{\sqrt{2}} (\ket{1010}-\ket{0101})$ \\
    \hline
    0,2 & $\frac{1}{\sqrt{2}} (\ket{1100}-\ket{0011})$ \\
    \hline
    0,3 & $\frac{1}{\sqrt{2}} (\ket{1001}-\ket{0110})$ \\
    \hline
    1,2 & $\frac{1}{\sqrt{2}} (-\ket{1001}+\ket{0110})$ \\
    \hline
    1,3 & $\frac{1}{\sqrt{2}} (-\ket{1100}+\ket{0011})$ \\
    \hline
    2,3 & $\frac{1}{\sqrt{2}} (-\ket{1010}+\ket{0101})$ \\
    \hline
  \end{tabular}
\end{table}
The four-partite GHZ state is generated by the same interferometer used for the W-state generation, but two-photon detection is required \cite{Shimizu2025}. Table~\ref{tab:GHZ_detection_pattern} lists the detector indices corresponding to the measurement outcomes and generated GHZ states. In this appendix, we consider the first row in Tab.~\ref{tab:GHZ_detection_pattern} as an example. The corresponding projectors at the output side, including multi-photon detection, are as follows:
\begin{align}
    \hat{\mathcal{P}}_{1100}&=\hat{\mathcal{P}}_0^{(1)}\otimes\hat{\mathcal{P}}_1^{(1)},\\
    \hat{\mathcal{P}}_{1200}&=\hat{\mathcal{P}}_0^{(1)}\otimes\hat{\mathcal{P}}_1^{(2)},\\
    \hat{\mathcal{P}}_{2100}&=\hat{\mathcal{P}}_0^{(2)}\otimes\hat{\mathcal{P}}_1^{(1)},\\
    \hat{\mathcal{P}}_{2200}&=\hat{\mathcal{P}}_0^{(2)}\otimes\hat{\mathcal{P}}_1^{(2)},\\
    \hat{\mathcal{P}}_{1300}&=\hat{\mathcal{P}}_0^{(1)}\otimes\hat{\mathcal{P}}_1^{(3)},\\
    \hat{\mathcal{P}}_{3100}&=\hat{\mathcal{P}}_0^{(3)}\otimes\hat{\mathcal{P}}_1^{(1)},
\end{align}
where
\begin{align}
    \hat{\mathcal{P}}_k^{(1)}&=\int d\Omega_k\  \hat{r}_k^\dagger (\Omega_k)\ket{0}_k\bra{0}\hat{r}_k (\Omega_k),\\
    \hat{\mathcal{P}}_k^{(2)}&=\frac{1}{2}\iint d\Omega_{k0}d\Omega_{k1}\  \nonumber\\
    &\qquad\hat{r}_k^\dagger (\Omega_{k0})\hat{r}_k^\dagger (\Omega_{k1})\ket{0}_k\bra{0}\hat{r}_k (\Omega_{k0})\hat{r}_k (\Omega_{k1}),\\
    \hat{\mathcal{P}}_k^{(3)}&=\frac{1}{6}\int d^3\bm{\Omega}_k \bigotimes_{l=0}^2 \hat{r}_k^\dagger (\Omega_{kl})\ket{0}_k\bra{0}\hat{r}_k (\Omega_{kl}).
\end{align}
We derive the POVM operators $E_{1100},\ E_{1200},\ E_{2100},\ E_{2200},\ E_{1300},\ $ and $E_{3100}$ in the rest of this subsection.

\subsubsection{\texorpdfstring{$\hat{E}_{1100}$}{E1100}}
We derive the POVM $\hat{E}_{1100}$ corresponding to the projector $\hat{\mathcal{P}}_{1100}$. In Eq.~(\ref{eq:elements_of_POVM}), the corresponding measurement outcomes are $n_0=n_1=1$ and $n_2=n_3=0$, while the input states are given by Eq.~(\ref{eq:set_j_two_photon}). The elements $M_{\bm j1100}$ of the POVM $\hat{E}_{1100}$ are given by
\begin{align}
    &M_{\bm j 1100}(\Omega_0,\Omega_1)\nonumber\\
    &={}_{\rm out}\bra{0}\left(\sum_{k=0}^1 V_{j_0 k}\hat{R}_{j_0 k}\right)\left(\sum_{k=0}^1 V_{j_1 k}\hat{R}_{j_1 k}\right)\nonumber\\
    &\qquad\qquad \hat{r}_0^\dagger(\Omega_0)\hat{r}_1^\dagger(\Omega_1)\ket{0}_{\rm out}.
\end{align}
The elements of the beamsplitter matrix $\hat{V}$~(\ref{eq:app_beamsplitter}) satisfy the following relations:
\begin{equation}
    \label{eq:V_ij_two_columns}
    \begin{aligned}
        V_{j0}&=\frac{1}{2},\\
        V_{j1}&=\frac{(-1)^j}{2},
    \end{aligned}
\end{equation}
which leads to
\begin{align}
    &M_{\bm j 1100}(\Omega_0,\Omega_1)\nonumber\\
    &=\frac{1}{4}{}_{\rm out}\bra{0}\left[(-1)^{j_1}\hat{R}_{j_0 0}\hat{R}_{j_1 1}+(-1)^{j_0}\hat{R}_{j_0 1}\hat{R}_{j_1 0}\right]\nonumber\\
    &\qquad\qquad \hat{r}_0^\dagger(\Omega_0)\hat{r}_1^\dagger(\Omega_1)\ket{0}_{\rm out}\nonumber\\
    &=\frac{1}{4}{}_{\rm out}\bra{0}\sum_{\sigma\in S_2}(-1)^{j_{\sigma(1)}}\hat{R}_{j_{\sigma(0)} 0}\hat{R}_{j_{\sigma(1)} 1}\nonumber\\
    &\qquad\qquad \hat{r}_0^\dagger(\Omega_0)\hat{r}_1^\dagger(\Omega_1)\ket{0}_{\rm out}.
\end{align}
Using the commutation relation in Eq.~(\ref{eq:commutation_relation_spectral_modes}),
\begin{align}
    &{}_{\rm out}\bra{0}\hat{R}_{j_{\sigma(0)}0}\hat{r}_0^\dagger(\Omega_0)\ket{0}_{\rm out}\nonumber\\
    =&\phi_{j_{\sigma(0)}}^*(\Omega_0)\exp\left[i\Omega_0\tau_{j_{\sigma(0)}}\right],
\end{align}
we obtain
\begin{align}
    \label{eq:GHZ_M_j0j1}
    &M_{\bm j 1100}(\Omega_0,\Omega_1)\nonumber\\
    =&\frac{1}{4}\sum_{\sigma\in S_2}(-1)^{j_{\sigma(1)}}\prod_{k=0}^1 \phi_{j_{\sigma(k)}}^*(\Omega_k)\exp\left[i\Omega_k\tau_{j_{\sigma(k)}}\right].
\end{align}
Substituting Eq.~(\ref{eq:GHZ_M_j0j1}) into (\ref{eq:POVM_n_photons}), we find
\begin{align}
    \hat{E}_{1100}&=\frac{1}{16}\sum_{\bm{j},\bm{j}'}\ket{\bm{j}}_{\rm in}\bra{\bm{j'}}\nonumber\\
    &\left[\sum_{\sigma,\sigma'\in S_2}(-1)^{j_{\sigma(1)}+j'_{\sigma'(1)}}\prod_{k=0}^1\mu_{j_{\sigma(k)}j'_{\sigma'(k)}}\right],
\end{align}
where $\ket{\bm j}$ and $\ket{\bm j'}$ belong to the same set given in Eq.~(\ref{eq:set_j_two_photon}). 

\subsubsection{\texorpdfstring{$\hat{E}_{1200}$}{12000} and \texorpdfstring{$\hat{E}_{2100}$}{21000}}
The POVM $\hat{E}_{1200}$ corresponds to the measurement outcomes $n_{0}=1$, $n_{1}=2$, and $n_{2}=n_{3}=0$. In this case, the input states are given by Eq.~(\ref{eq:set_j_three_photon}). The elements $M_{\bm{j}1200}$ of POVM $\hat{E}_{1200}$ are given by
\begin{align}
    M_{\bm j1200}(\bm \Omega)&=\frac{1}{\sqrt{2}}{}_{\rm out}\bra{0}\bigotimes_{i=0}^2\left(\sum_{k=0}^1 V_{j_i k}\hat{R}_{j_i k}\right)\nonumber\\
    &\qquad\qquad\hat{r}_0^\dagger(\Omega_{00})\hat{r}_1^\dagger(\Omega_{10})\hat{r}_1^\dagger(\Omega_{11})\ket{0}_{\rm out}.
\end{align}
Using Eq.~(\ref{eq:V_ij_two_columns}), this reduces to
\begin{align}
    &M_{\bm j1200}(\bm \Omega)\nonumber\\
    =&\frac{1}{8\sqrt{2}}{}_{\rm out}\bra{0}\left[(-1)^{j_1+j_2}\hat{R}_{j_0 0}\hat{R}_{j_1 1}\hat{R}_{j_2 1}\right.\nonumber\\
    &\qquad\qquad+(-1)^{j_0+j_1}\hat{R}_{j_0 1}\hat{R}_{j_1 1}\hat{R}_{j_2 0}\nonumber\\
    &\quad\qquad\left.+(-1)^{j_0+j_2}\hat{R}_{j_0 1}\hat{R}_{j_1 0}\hat{R}_{j_2 1}\right]\nonumber\\
    &\qquad\qquad\qquad\hat{r}_0^\dagger(\Omega_{00})\hat{r}_1^\dagger(\Omega_{10})\hat{r}_1^\dagger(\Omega_{11})\ket{0}_{\rm out}.
\end{align}
Applying the commutation relations in Eqs.~(\ref{eq:commutation_relation_spectral_modes}) and (\ref{eq:commutation_relation_two_terms}), we obtain
\begin{align}
    M_{\bm j1200}(\bm \Omega)&=\frac{1}{8\sqrt{2}}\sum_{\sigma\in S_3}(-1)^{j_{\sigma{(1)}}+j_{\sigma{(2)}}}\nonumber\\
    &\qquad\qquad\prod_{k=0}^2\phi_{j_{\sigma{(k)}}}^*(\Omega_k)\exp{i\Omega_k\tau_{j_{\sigma(k)}}}.
\end{align}
Therefore, the POVM $\hat{E}_{1200}$ is given by
\begin{align}
    \label{eq:E_1200}
    \hat{E}_{1200}&=\frac{1}{128}\sum_{\bm{j},\bm{j}'}\ket{\bm j}_{\rm in}\bra{\bm j'}\nonumber\\
    &\left[\sum_{\sigma,\sigma'\in S_3}(-1)^{j_{\sigma(1)}+j_{\sigma(2)}+j'_{\sigma'(1)}+j'_{\sigma'(2)}}\prod_{k=0}^2\mu_{j_{\sigma(k)}j'_{\sigma'(k)}}\right].
\end{align}
where $\ket{\bm j}$ and $\ket{\bm j'}$ belong to the same set in Eq.~(\ref{eq:set_j_three_photon}).

For the POVM $\hat{E}_{2100}$, indices of output ports are different from those in $\hat{E}_{1200}$, while the number of detected photons is the same. Thus, the structure of $\hat{E}_{2100}$ is identical to that of $\hat{E}_{1200}$ except for the phase factor. Since only one photon is detected in the first port, two of the four $(-1)$-exponent terms in Eq.~(\ref{eq:E_1200}) are removed, resulting
\begin{align}
    \label{eq:E_2100}
    \hat{E}_{2100}&=\frac{1}{128}\sum_{\bm{j},\bm{j}'}\ket{\bm j}_{\rm in}\bra{\bm j'}\nonumber\\
    &\left[\sum_{\sigma,\sigma'\in S_3}(-1)^{j_{\sigma(2)}+j'_{\sigma'(2)}}\prod_{k=0}^2\mu_{j_{\sigma(k)}j'_{\sigma'(k)}}\right].
\end{align}

\subsubsection{\texorpdfstring{$\hat{E}_{2200}$}{22000}}
The POVM $\hat{E}_{2200}$ corresponds to the measurement outcomes $n_{0}=n_{1}=2$ and $n_{2}=n_{3}=0$. Since photons are injected into all input ports, the input state is given by $\ket{\bm j}=\ket{1234}$. The elements $M_{\bm{j}2200}$ of POVM $\hat{E}_{2200}$ are given by
\begin{align}
    M_{\bm j2200}(\bm \Omega)&=\frac{1}{2}{}_{\rm out}\bra{0}\bigotimes_{i=0}^3\left(\sum_{k=0}^1 V_{j_i k}\hat{R}_{j_i k}\right)\nonumber\\
    &\quad\hat{r}_0^\dagger(\Omega_{00})\hat{r}_0^\dagger(\Omega_{01})\hat{r}_1^\dagger(\Omega_{10})\hat{r}_1^\dagger(\Omega_{11})\ket{0}_{\rm out}.
\end{align}
Using Eq.~(\ref{eq:V_ij_two_columns}) and the commutation relation in Eq.~(\ref{eq:commutation_relation_two_terms}), this reduces to
\begin{align}
    M_{\bm j2200}(\bm \Omega)&=\frac{1}{2^5}\sum_{\sigma\in S_4}(-1)^{\sigma{(2)}+\sigma{(3)}}\nonumber\\
    &\qquad\quad\prod_{k=0}^3\phi_{\sigma{(k)}}^*(\Omega_k)\exp{i\Omega_k\tau_{\sigma(k)}},
\end{align}
where the subscripts are replaced with $0,...,3$ for simplicity.
Therefore, the POVM $\hat{E}_{2200}$ is given by
\begin{align}
    \hat{E}_{2200}&=\frac{1}{2^{10}}\ket{1234}_{\rm in}\bra{1234}\nonumber\\
    &\sum_{\sigma,\sigma'\in S_4}(-1)^{\sigma(2)+\sigma(3)+\sigma'(2)+\sigma'(3)}\prod_{k=0}^3 \mu_{\sigma(k),\sigma'(k)}.
\end{align}

\subsubsection{\texorpdfstring{$\hat{E}_{1300}$}{1300} and \texorpdfstring{$\hat{E}_{3100}$}{3100}}
The POVM $\hat{E}_{1300}$ corresponds to the measurement outcomes $n_0=1$, $n_1=3$, and $n_2=n_3=0$. 
Since photons are injected into all input ports, the input state is given by $\ket{\bm j}=\ket{1234}$. The elements $M_{\bm{j}1300}$ of POVM $\hat{E}_{1300}$ are expressed as
\begin{align}
    &M_{\bm j1300}(\bm \Omega)\nonumber\\
    =&\frac{1}{\sqrt{6}}{}_{\rm out}\bra{0}\bigotimes_{i=0}^3\left(\sum_{k=0}^1 V_{j_i k}\hat{R}_{j_i k}\right)\nonumber\\
    &\hat{r}_0^\dagger(\Omega_{00})\hat{r}_1^\dagger(\Omega_{10})\hat{r}_1^\dagger(\Omega_{11})\hat{r}_1^\dagger(\Omega_{12})\ket{0}_{\rm out}.
\end{align}
Using Eq.~(\ref{eq:V_ij_two_columns}) and the commutation relation in Eq.~(\ref{eq:commutation_relation_three_terms}), we obtain
\begin{align}
    M_{\bm j1300}(\bm \Omega)&=\frac{1}{16\sqrt{6}}\sum_{\sigma\in S_4}(-1)^{\sigma{(1)}+\sigma{(2)}+\sigma{(3)}}\nonumber\\
    &\prod_{k=0}^3\phi_{\sigma(k)}^*(\Omega_k)\exp{i\Omega_k\tau_{\sigma(k)}}.
\end{align}
where the subscripts are replaced with $0,...,3$ for simplicity.
The POVM $\hat{E}_{1300}$ becomes
\begin{align}
    \label{eq:E_1300}
    \hat{E}_{1300}&=\frac{1}{2^8\cdot 6}\ket{1234}_{\rm in}\bra{1234}\nonumber\\
    &\sum_{\sigma,\sigma'\in S_4}(-1)^{\sum_{i=1}^3\sigma{(i)}+\sigma'{(i)}}
    \prod_{k=0}^3\mu_{\sigma(k)\sigma'(k)}.
\end{align}
As in the derivation of $\hat{E}_{2100}$, the POVM $\hat{E}_{3100}$ has the same structure as $\hat{E}_{1300}$ except for the phase factors. Since only one photon is measured in the first output port for $\hat{E}_{3100}$, four of the six $(-1)$-exponent terms in Eq.~(\ref{eq:E_1300}) are removed, leading to
\begin{align}
    \hat{E}_{3100}&=\frac{1}{2^8\cdot 6}\ket{1234}_{\rm in}\bra{1234}\nonumber\\
    &\sum_{\sigma,\sigma'\in S_4}(-1)^{\sigma{(3)}+\sigma'{(3)}}
    \prod_{k=0}^3\mu_{\sigma(k)\sigma'(k)}.
\end{align}

\subsection{POVM for Bell states}
In order to create a Bell state on two emitters, either A-B or C-D, we adopt a set of positive operator-valued measures (POVMs) $\{\hat{E}_{ij}\}$ that have been previously derived in the literature and are directly applicable to our construction~\cite{Dahlberg2019}. Explicitly, the elements of the POVM are given as follows (assuming pairwise indistinguishability as $\mu$):

\begin{align}
\hat{E}_{00} &= 
\begin{pmatrix}
1 & 0 & 0 & 0 \\
0 & 0 & 0 & 0 \\
0 & 0 & 0 & 0 \\
0 & 0 & 0 & 0
\end{pmatrix}, \\[6pt]
\hat{E}_{10} &= \tfrac{1}{2}
\begin{pmatrix}
0 & 0 & 0 & 0 \\
0 & 1 & \mu & 0 \\
0 & \mu^* & 1 & 0 \\
0 & 0 & 0 & 0
\end{pmatrix}, \\[6pt]
\hat{E}_{01} &= \tfrac{1}{2}
\begin{pmatrix}
0 & 0 & 0 & 0 \\
0 & 1 & -\mu & 0 \\
0 & -\mu^* & 1 & 0 \\
0 & 0 & 0 & 0
\end{pmatrix}, \\[6pt]
\hat{E}_{11} &= \tfrac{1}{2}
\begin{pmatrix}
0 & 0 & 0 & 0 \\
0 & 0 & 0 & 0 \\
0 & 0 & 0 & 0 \\
0 & 0 & 0 & 1 - |\mu|^2
\end{pmatrix}, \\[6pt]
\hat{E}_{20} &= \tfrac{1}{4}
\begin{pmatrix}
0 & 0 & 0 & 0 \\
0 & 0 & 0 & 0 \\
0 & 0 & 0 & 0 \\
0 & 0 & 0 & 1 + |\mu|^2
\end{pmatrix}, \\[6pt]
\hat{E}_{02} &= \tfrac{1}{4}
\begin{pmatrix}
0 & 0 & 0 & 0 \\
0 & 0 & 0 & 0 \\
0 & 0 & 0 & 0 \\
0 & 0 & 0 & 1 + |\mu|^2
\end{pmatrix}.
\end{align}
The above operators satisfy the completeness relation
\[
\sum_{i,j} E_{ij} = \mathbb{I},
\]
ensuring a valid POVM. When we create an elementary Bell state on the emitters, these POVMs are employed for the detectors, pair-wise, either for the PNR or the non-PNR case, as described in App.~\ref{appsubsubsec:pnr_non_pnr_detectors}.

%%%%%%%%%%%%%%%%%%%%%%%%%%%%%%%%%%%%%%%%%%%%%%
\section{\label{app:superoperator}Mapping hardware noise to QEC performance}
\subsection{\label{appsubsec:superoperator}Superoperator  and Pauli twirling}
Let $\mathcal{S}$ denote the noisy quantum map that implements a stabilizer measurement cycle on the data qubits' subsystem $\mathcal{H}_D$ (including the interaction with communication qubits and their projective readout), producing a (measurement outcome‑conditioned) reduced state on the data qubits. The key idea is that $\mathcal{S}$ has all the noise due to the hardware, and then we want to map its action to a distribution of pure Pauli errors on the data qubits that can be applied to a surface code again. The Choi-Jamiolkowski isomorphism accomplishes this action as it sets up a relation between the state and the noisy quantum channel. The Choi matrix~\cite{Homa_2024,Frembs_2024} of $\mathcal{S}$ is
\begin{equation}
J_{\mathcal{S}} \;=\; (\mathcal{S}\otimes\mathbb{I})\big(\ket{\Phi}\!\bra{\Phi}\big),
\qquad
\ket{\Phi}=\frac{1}{\sqrt{2^{n_D}}}\sum_{i=0}^{2^{n_D}-1}\ket{i}\otimes\ket{i},
\end{equation}
where $n_D$ is the number of data qubits in the stabilizer. Here, $|i\rangle |i\rangle$ denote joint qubit computational basis states over the stabilizer unit-cell data qubits and the auxiliary system, which together form Bell-pairs as described above and can be expanded and also be written in the binary form. For instance, if $n_D=4$, then $\rho=\frac{1}{4}|\Phi_+\rangle^{\otimes 4}=\frac{1}{4}(|0000\rangle|0000\rangle+\hdots +|1111\rangle|1111\rangle)$. The map action can be recovered from $J_{\mathcal{S}}$ via
\begin{equation}
\mathcal{S}(\rho)=\mathrm{Tr}_{A}\big[(\mathbb{I}\otimes\rho^{T})\,J_{\mathcal{S}}\big],
\end{equation}
or equivalently in vectorized form $\operatorname{vec}[\mathcal{S}(\rho)]=\mathcal{S}_{\text{super}}\operatorname{vec}[\rho]$ with $\mathcal{S}_{\text{super}}$ obtained from $J_{\mathcal{S}}$. Here $\mathrm{Tr}_A[\cdot]$ is the partial-trace over sub-system A.

A convenient operator‑sum (Kraus) representation is obtained by diagonalizing $J_{\mathcal{S}}$,
\begin{equation}
J_{\mathcal{S}}=\sum_j \lambda_j \ket{v_j}\!\bra{v_j},\qquad
K_j=\sqrt{\lambda_j}\,\operatorname{mat}(\ket{v_j}),
\end{equation}
where $\operatorname{mat}(\cdot)$ reshapes the Choi eigenvector into an operator on $\mathcal{H}_D$. The channel is then $\mathcal{S}(\rho)=\sum_j K_j \rho K_j^\dagger$. Which means that we can see the action of $\mathcal{S}$ in terms of its decomposition into Kraus operators. We can either find this decomposition or select a valid choice that we pre-compute.

Since $\mathcal{S}$ includes distinct classical outcomes (e.g. measurement result $m=\pm1$, GHZ success/failure flag $g$), we build separate conditioned Choi matrices $J_{\mathcal{S}}^{(g,m)}$ and repeat the procedure for each $(g,m)$ branch; the final tabulation is organized per branch in separate columns of various possible combinations with their weights denoting the probabilities.

Pauli twirling maps a general CPTP (completely positive and trace preserving) map into a Pauli channel by averaging conjugation over the $n$‑qubit Pauli group $\mathcal{P}_n$~\cite{katabarwa2017dynamicalinterpretationpaulitwirling,Geller_2013}. Concretely, the twirled channel is
\begin{equation}
\mathcal{S}_{\text{PT}}(\rho)
=\frac{1}{4^{n_D}}\sum_{P\in\mathcal{P}_{n_D}} P\,\mathcal{S}(P\,\rho\,P)\,P .
\end{equation}
The twirl eliminates off‑diagonal operator terms in a Pauli operator expansion and yields a Pauli channel of the form
\begin{equation}
\mathcal{S}_{\text{PT}}(\rho)=\sum_{e} q_e \,E_e \,\rho\, E_e^\dagger,
\qquad E_e\in\mathcal{P}_{n_D},
\end{equation}
with nonnegative probabilities $q_e$ satisfying $\sum_e q_e = 1$ (for trace‑preserving conditioned maps, the normalization may be subunit and interpreted as conditional probability for configurations in $g,m$ variables).

A numerically robust route to compute $\{q_e\}$ is the following (used in our code for noisy circuit-simulation to superoperator decomposition):

\begin{enumerate}
  \item Diagonalize the Choi matrix and obtain Kraus operators $\{K_j\}$ as above (work with each conditioned Choi $J_{\mathcal{S}}^{(g,m)}$ separately).
  \item Expand each Kraus operator in the Pauli basis $\{E_e\}$, using the Hilbert–Schmidt inner product. For $w$ data qubits the Pauli strings obey $\Tr(E_e^\dagger E_{e'})=2^{w}\delta_{ee'}$. The expansion coefficients are
  \[
    c_{j,e}=\frac{1}{2^{w}}\Tr\!\big(E_e^\dagger K_j\big),
    \qquad K_j=\sum_e c_{j,e} E_e.
  \]
  \item The Pauli‑twirled channel retains only diagonal terms in this basis, and the Pauli error weights are
  \begin{equation}
    q_e \;=\; \sum_j |c_{j,e}|^2 .
  \label{eq:qe_def}
  \end{equation}
  Equivalently, expanding the channel $\mathcal{S}(\rho)=\sum_{e,e'} \alpha_{e,e'} E_e\rho E_{e'}^\dagger$, the twirl sets $\alpha_{e,e'}\mapsto 0$ for $e\neq e'$ and $q_e=\alpha_{e,e}$.
  \item Enforce numerical trace preservation / normalization if needed (due to numerical errors and precision): for conditioned maps (e.g. conditioned on an observed measurement outcome) the $q_e$ sum to the conditional trace $\Tr[\mathcal{S}(\openone/2^{w})]$; we divide by this trace if a normalized conditional probability vector is required.
\end{enumerate}

The procedure above is stable and avoids explicit analytic evaluation. This procedure is repeated for many iterations for both X- and Z-type distributed stabilizer circuits. It produces the operationally meaningful probabilities (or error configuration weights) $\{q_e\}$ that approximate the original noisy map (directly mapped to hardware parameters) by the Pauli channel that is closest under the Pauli‑twirl ansatz.

The twirled decomposition is directly used to build the compact table consumed by the surface code QEC simulator in the next step. For each conditioned branch (GHZ success/failure $g$, measurement outcome $m$), we list all Pauli strings $E_e$ acting on the $w$ data qubits together with their probability weight $q_e^{(g,m)}$. The effective (noisy) stabilizer projector then reads
\begin{equation}
\begin{aligned}
\Pi_{m}^{\mathrm{eff}}(\rho_D)
=&\sum_e q_e^{ (\text{succ},m)}\,E_e\,\Pi_{m}^{\mathrm{ideal}}(\rho_D)\,E_e^\dagger
\\ &
+\sum_e q_e^{ (\text{fail},m)}\,E_e\,\Pi_{m}^{\mathrm{ideal}}(\rho_D)\,E_e^\dagger
\\ & \;
+\sum_e q_e^{ (\text{succ},\bar m)}\,E_e\,\Pi_{\bar m}^{\mathrm{ideal}}(\rho_D)\,E_e^\dagger \\ & \;
+\sum_e q_e^{ (\text{fail},\bar m)}\,E_e\,\Pi_{\bar m}^{\mathrm{ideal}}(\rho_D)\,E_e^\dagger,
\end{aligned}
\end{equation}
where $\bar m$ denotes the complementary measurement outcome when the stabilizer measurement circuit fails. The first (second) two summations correspond to the branch producing outcome with GHZ success and failure, respectively, for $m$ (and outcome with measurement error $\bar m $ for error-configuration $E_e$). All these numerical computations are achieved by our source code \href{https://github.com/siddhantphy/emission_direct_scheme}{\textsc{CircuitSimulator}} tailored for the single-shot emission scheme. The output is the table format superoperators that can be used as explained in the next subsection.

\subsection{\label{appsubsec:monte_carlo_qec}Monte-Carlo simulations over superoperator for QEC performance}

This section describes the numerical procedure used to estimate logical success/error rates and fault‑tolerance thresholds by Monte Carlo sampling of the precomputed tabular superoperator and subsequent decoding. The recipe below is the exact implementation used in our \href{https://github.com/siddhantphy/qsurface}{\textsc{Qsurface}} workflow.

The tabular superoperator supplies a discrete probability distribution over triples \((E_e,g,m)\) where \(E_e\in\mathcal{P}_w\) is a Pauli string on the \(w\) data qubits of a stabilizer unit cell, \(g\in\{\text{succ},\text{fail}\}\) is the GHZ‑flag (or other branch flag), and \(m\in\{\pm1\}\) is the reported measurement outcome (possibly flipped by measurement error). The table provides probabilities \(q_e^{(g,m)}\) for each triple. These tables are provided in the source code and accompanying data as CSV files. For a visual representation of these superoperator tables, one can also refer to the Ref.~\cite{singh2024modulararchitecturesentanglementschemes}. A single Monte Carlo shot consists of sampling one row from the table for every stabilizer measurement event occurring in a QEC sub-round (of either X or Z type), applying the sampled errors to the code data qubits, assembling the syndrome record, and decoding.

Spatial layout and scheduling: Given a code of physical distance \(d\) we construct the lattice of data qubits and stabilizer unit cells for one QEC round. For distributed measurement circuits, we expand each stabilizer into the sequence of sub‑rounds (sub‑layers) required by the hardware scheduling (see main text). Each stabilizer measurement attempt corresponds to one draw from the table.

Sampling procedure: For each stabilizer measurement event (plaquette-Z or vertex-X, at a particular time / sub‑round), we:

\begin{enumerate}
  \item Select the relevant conditioned table: choose the rows corresponding to the stabilizer geometry (which fixes \(w\) qubits' ordering) and the measurement instrument branches (GHZ success/fail).
  \item Draw a random sample from the discrete distribution \(\{q_e^{(g,m)}\}\). Practically, we accumulate the cumulative distribution and perform a uniform random number generation \(u\in[0,1)\) to select the row where \(u\) falls.
  \item Interpret the sampled triple \((E_e,g,m)\): apply the Pauli string \(E_e\) to the specified data qubits (composition is multiplicative with prior errors), record the measurement outcome \(m\) as the stabilizer result for this sub‑round, and record the GHZ success flag \(g\).
\end{enumerate}

GHZ failure handling and cut‑off policy: If \(g=\text{fail}\) for a stabilizer sub‑round, the code employs the decision rule for aborting that stabilizer measurement: the stabilizer measurement is aborted and the previous valid outcome is reused (or a default outcome is substituted) depending on the chosen protocol. This logic is implemented exactly during sampling by maintaining a per‑stabilizer memory of the last accepted outcome and updating it only on successful GHZ events or according to the protocol's replacement rule.

Order of application: When multiple sampled Pauli strings act on overlapping data qubits (from different stabilizer attempts within a QEC cycle), we apply them in the same chronological order as the physical protocol: for each sub‑round we (i) sample and apply data‑qubit Pauli from that stabilizer, (ii) update the stored stabilizer outcome, then advance to the next stabilizer. This preserves time‑ordering and correctly models idling noise via the decoherence scheduling described in App.~\ref{app:Noise_model}.

Syndrome assembly: After all sub‑rounds of a QEC cycle are sampled and applied, the simulation produces a set of stabilizer outcomes per round (including reused or substituted outcomes where GHZ failed). These outcomes form the space–time syndrome record (a 3D syndrome graph over sub-rounds in time layers).

Decoder input: We pass the syndrome record to the chosen decoder (Union‑Find~\cite{Delfosse_2021,wu2022interpretationunionfinddecoderweighted,Chan_2023} or MWPM-Minimum Weight Perfect Matching~\cite{Edmonds_1965,Higgott_2025}). The decoder inputs are simply binary stabilizer outcomes per stabilizer location and time; it does not require knowledge of the underlying physical error mechanisms because the superoperator sampling already embeds those mechanisms into the syndrome. Note that we use a weighted UF decoder version for the thresholds reported in this work.

Logical outcome and success metric: The decoder returns a correction operator \(C\). We apply \(C\) (classically) to the simulated data‑qubit error configuration and check whether the total error combined with \(C\) implements a logical operator on the code (i.e. whether the net logical Pauli is nontrivial). A Monte Carlo shot is declared a logical failure if a logical error occurs. Otherwise, it is a logical success. Formally:
\[
\text{logical failure} \;\Longleftrightarrow\; L_\text{net}\neq I,
\]
where \(L_\text{net}\) is the equivalence class of the net data‑qubit error after applying the decoder correction.

Estimating logical error/success rates (Repetition and statistics): For each tuple of hardware parameters and physical error rate \(p\), run \(N_\mathrm{shots}\) independent Monte Carlo trials (typical \(N_\mathrm{shots}=5\times 10^4\)). The empirical logical error rate is
\[
p_\text{L} \;=\; \frac{1}{N_\mathrm{shots}}\sum_{i=1}^{N_\mathrm{shots}} \mathbf{1}\{\text{failure}_i\},
\]
and the success rate is \(1-p_\text{L}\). The standard error is \(\sqrt{p_\text{L}(1-p_\text{L})/N_\mathrm{shots}}\); choose \(N_\mathrm{shots}\) to ensure desired confidence per data point.

Confidence and rare events: When logical error probabilities are very small, increase \(N_\mathrm{shots}\) or use importance sampling/splitting methods. For our reported thresholds, we used \(N_\mathrm{shots}=10^4\), which yields sub‑percent statistical uncertainty in the threshold estimates reported.

Threshold extraction recipe: To obtain a threshold \(p_{\mathrm{th}}\) for a given protocol and hardware set:

\begin{enumerate}
  \item Fix other parameters (coherence times, \(\alpha\), or hardware parameter set ES-$x$).
  \item For a set of physical error rates \(p\) spanning a region around the expected threshold, simulate logical error rates \(p_\text{L}(p,d)\) for several code distances \(d\) (we use \(d\in\{4,6,8,10,12\}\) as it makes the distributed toric surface code simulations easier, and the planar surface code performance is expected to be similar).
  \item For each \(d\) plot \(p_\text{L}(p,d)\) vs.\ \(p\). The threshold is the critical \(p\) where logical error curves cross / become approximately distance‑independent. Numerically find \(p_{\mathrm{th}}\) by locating the intersection of \(p_\text{L}(p,d)\) curves (or by fitting finite‑size scaling forms to the data and extracting the critical point).
  \item Estimate uncertainty by bootstrap resampling of Monte Carlo trials or by repeating the entire sweep with different random number generator seeds and reporting the standard deviation of the extracted \(p_{\mathrm{th}}\).
\end{enumerate}
The method for optimizing for cut-off times and fitting procedure for threshold estimation for the logical success rate curve is explained in App.~\ref{app:thresholds}. The recipe in this appendix allows us to link the hardware parameters directly to QEC performance.

%%%%%%%%%%%%%%%%%%%%%%%%%%%%%%%%%%%%%%%%%%%%%%
\section{\label{app:trade_off_alpha}Sweeping bright-state parameter for base and distilling state}
\begin{figure*}
    \centering
    \includegraphics[width=0.8\textwidth]{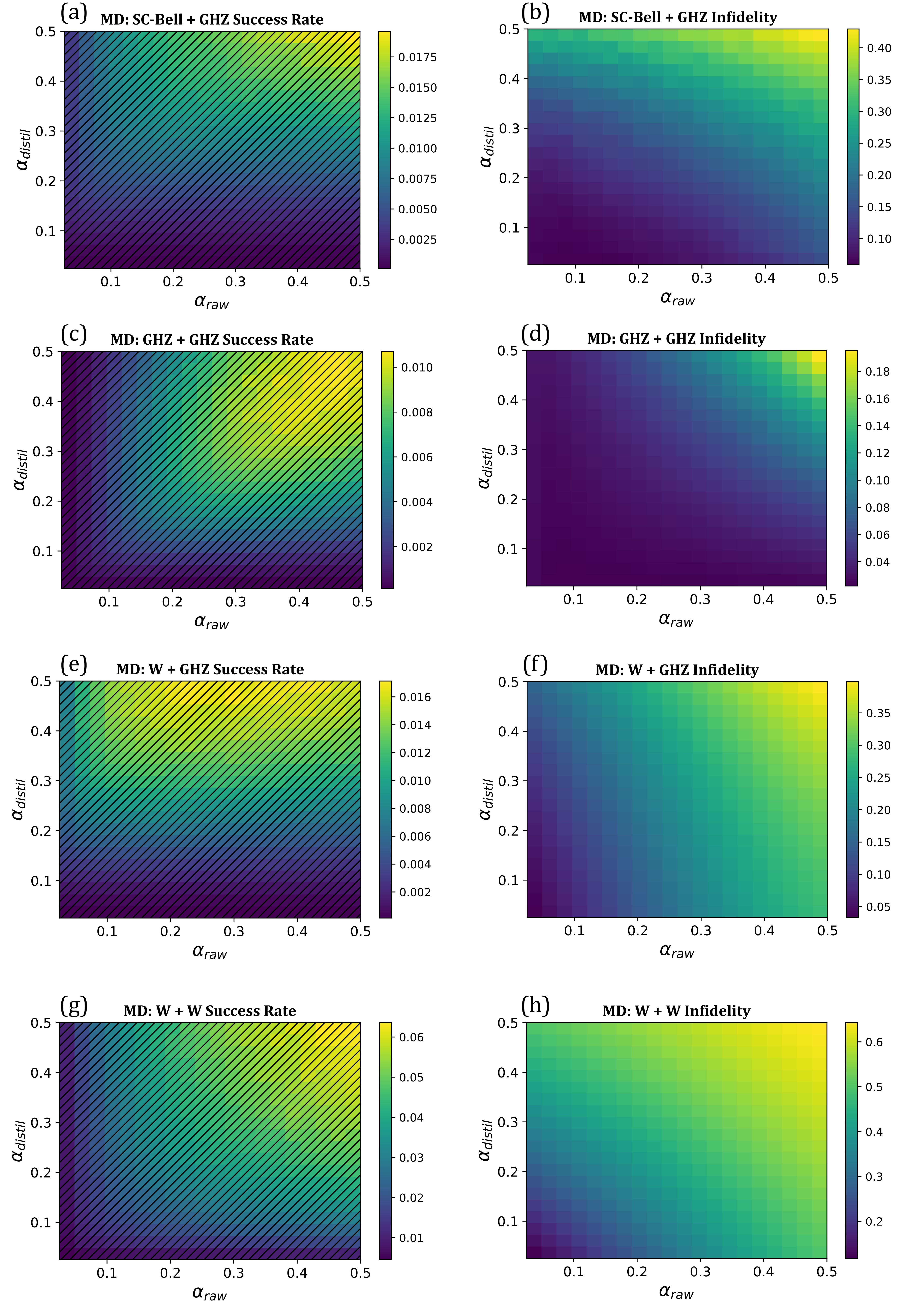}
    \caption{Bright-state parameter tradeoff for protocols with distinct $\alpha$ for base state and distilling state, with PNR detectors. Evaluated with coherence times of $T=10^6$, hardware parameter set ES-2, and physical error-rates of $p=10^{-3}$. Fault-tolerance regions are shaded with black lines.}
    \label{fig:trade_off_alpha}
\end{figure*}
For the memory distillation protocols discussed in the main text, we initially assumed that the bright-state parameter $\alpha$ was identical for both the base state and the distilling state. In general, this assumption is not required, and it is worthwhile to investigate potential trade-offs when $\alpha$ is varied independently for the two states. Intuitively, one might expect that choosing lower values of $\alpha$ could maximize fidelity. However, the impact of the distillation quantum circuit on both the final success probability and fidelity may reveal nontrivial trade-offs. 

Fig.~\ref{fig:trade_off_alpha} presents the results for the four memory distillation protocols. Here, we fix the coherence time to $T=10^6$, employ the ES-2 hardware parameter set, consider a gate error of $p=10^{-3}$, and assume photon-number-resolving (PNR) detectors. The fault-tolerance regions are indicated by the black line-shaded areas. While the output infidelity for this set of input parameters does not reach fault-tolerance—consistent with Fig.~\ref{fig:hardware_infidelity} for the ES-2 platform—the plots illustrate the variation of the output infidelity with respect to $\alpha_\mathrm{base}$ and $\alpha_\mathrm{distil}$. 

If the desired infidelity threshold is attainable, there exists flexibility in the choice of $\alpha_\mathrm{base}$ and $\alpha_\mathrm{distil}$. For symmetric protocols, where the base and distilling states are of the same type (e.g., distilling GHZ with GHZ or W with W), it is possible to select a relatively high value of $\alpha$ for one of the states. This flexibility can be leveraged to relax experimental requirements, enabling optimization of success probabilities while still meeting infidelity constraints.

%%%%%%%%%%%%%%%%%%%%%%%%%%%%%%%%%%%%%%%%%%%%%%
\section{\label{app:heatmap_coherence_hardware}Heatmaps for hardware and coherence parameters}

\begin{figure*}
    \centering
    \includegraphics[width=0.8\textwidth]{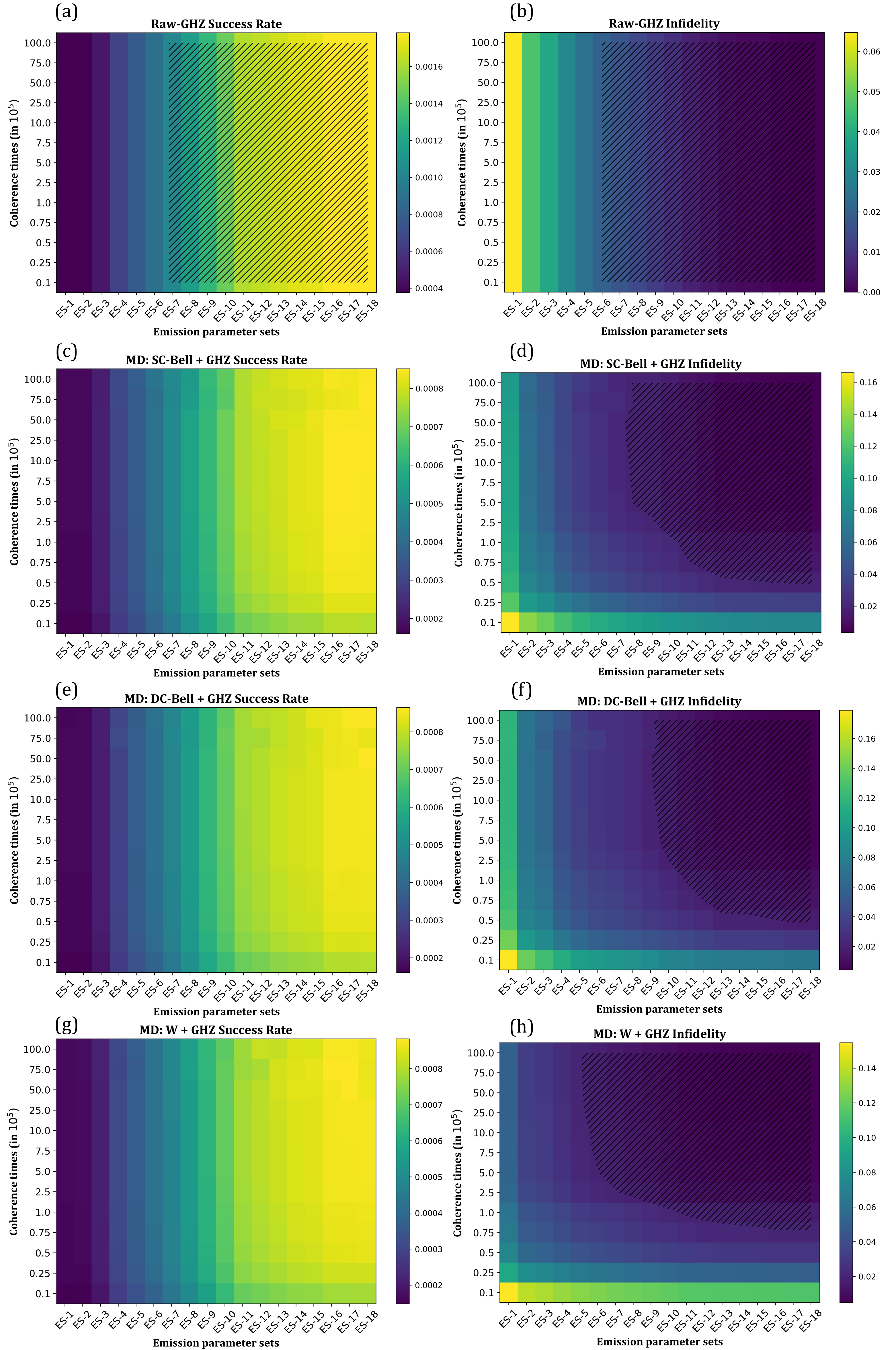}
    \caption{(1 of 2) Heatmaps for varying coherence times and hardware parameters with PNR detectors, physical error rate of $p=10^{-3}$, and $\alpha=0.025$. OD-GHZ is evaluated at $\alpha=0.5$. Fault-tolerance regions are shaded with black lines.}
    \label{fig:heatmaps_coherence_hardware_1}
\end{figure*}

\begin{figure*}
    \centering
    \includegraphics[width=0.8\textwidth]{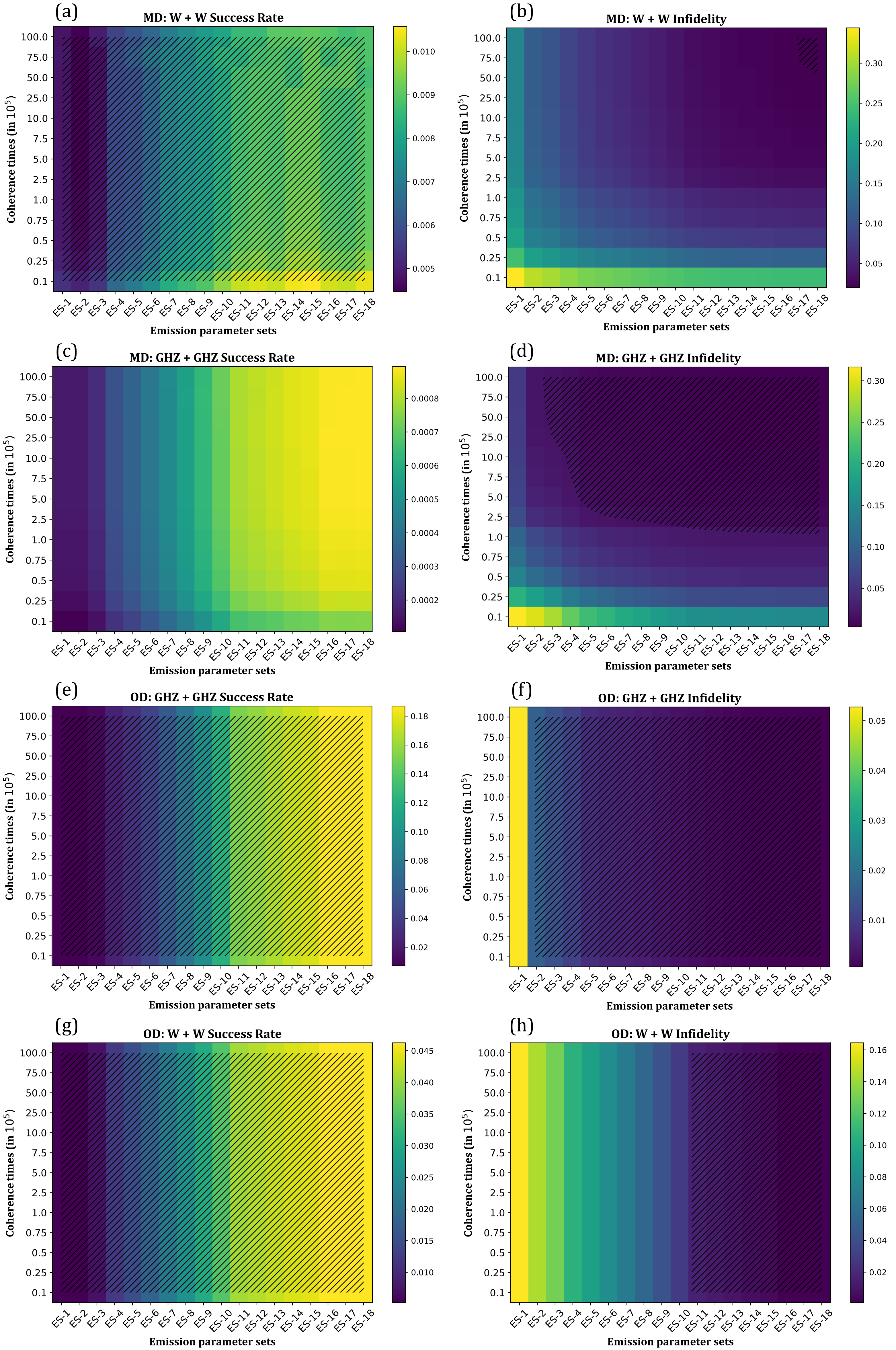}
    \caption{(2 of 2) Heatmaps for varying coherence times and hardware parameters with PNR detectors, physical error rate of $p=10^{-3}$, and $\alpha=0.025$. OD-GHZ is evaluated at $\alpha=0.5$. Fault-tolerance regions are shaded with black lines.}
    \label{fig:heatmaps_coherence_hardware_2}
\end{figure*}

In a similar spirit to App.~\ref{app:trade_off_alpha}, we perform a heatmap analysis over coherence times and hardware parameters to gain deeper insights into the precise fault-tolerance requirements of all protocols. In the main text, for numerical tractability, we pre-selected the most relevant parameters for the two-dimensional plots. Here, we extend this analysis by presenting three-dimensional heatmaps where both coherence times are varied for all hardware parameter sets. The results are shown in Fig.~\ref{fig:heatmaps_coherence_hardware_1} and Fig.~\ref{fig:heatmaps_coherence_hardware_2} in two parts, covering all raw, optical, and memory distillation protocols.

Throughout this analysis, we employ photon-number-resolving (PNR) detectors, a physical error rate of $p=10^{-3}$, and set $\alpha=0.025$, except for the OD-GHZ protocol, where we use $\alpha=0.5$ to achieve optimal performance as noted in the main text. The coherence times are swept from modest present-day values through near-term improvements to ambitious, futuristic targets. This sweep also covers hardware sets~1, 2, and 3 considered in Ref.~\cite{singh2024modulararchitecturesentanglementschemes}, and therefore, the results can be directly compared.

The resulting heatmaps (in Fig.~\ref{fig:heatmaps_coherence_hardware_1} and Fig.~\ref{fig:heatmaps_coherence_hardware_2}) provide additional insight and a roadmap for attaining fault tolerance with single-shot emission-based hardware. Our findings indicate that the primary challenge lies in meeting the stringent infidelity requirements, while the target generation success rates are largely achievable with realistic hardware parameters.

%%%%%%%%%%%%%%%%%%%%%%%%%%%%%%%%%%%%%%%%%%%%%%
\section{\label{app:thresholds} Threshold simulations}
\subsection{\label{appsubsec:cutoff_times}Cut-off time optimization}
As outlined in the main text, the cut-off time \(t_\text{cut}\) sets the maximum duration allowed for GHZ-state generation during a stabilizer measurement attempt. Any entanglement generation RUS attempts exceeding this time are aborted, and the corresponding stabilizer measurement is handled according to the protocol’s failure policy (e.g., outcome substitution or reuse of the previous measurement). 

We define \(t_\text{cut}^{100\%}\) as the average time required to achieve a successful GHZ state across all trials (i.e., corresponding to the full success probability over the allowed attempts). More generally, we introduce \(t_\text{cut}^{x\%}\), the cut-off time that captures the average duration required for success within the fastest \(x\%\) of attempts. The value \(t_\text{cut}^{100\%}\) is precomputed in a prior simulation stage (when sweeping and scanning the range of physical error rates $p$); the shorter \(t_\text{cut}^{x\%}\) values are obtained by truncating the distribution of generation times at the \(x\)-th percentile.

The search for an optimal cut-off proceeds as follows:
\begin{enumerate}
    \item Choose an initial percentile \(x\) --- typically \(x \in [95\%, 99\%]\).
    \item Determine \(t_\text{cut}^{x\%}\) from the simulated distribution of GHZ-generation times. The \textsc{CircuitSimulator} outputs this based on the $x$ input.
    \item Run the noisy stabilizer-measurement simulator using this fixed \(t_\text{cut}^{x\%}\) to estimate the logical error rate \(p_\text{L}\) over a sweep of physical error rates \(p\).
    \item If \(p_\text{L} \lesssim 0.2\) for some \(p\), perform a full threshold search (varying code distance) to estimate \(p_\mathrm{th}\). If \(p_\text{L} > 0.80\) for all tested \(p\), discard this \(x\) value as the probability of finding a threshold is negligible (verified numerically) with such high logical error rate.
\end{enumerate}

In general, \(t_\text{cut}^{x\%}\) grows approximately exponentially with \(x\). At low \(x\), increasing \(x\) (and thus \(t_\text{cut}\)) improves the threshold \(p_\mathrm{th}\), as more stabilizer outcomes are successfully collected per QEC cycle. Beyond a certain point, however, \(p_\mathrm{th}\) saturates: additional collection time yields negligible gains in syndrome information but introduces more decoherence noise due to longer waiting times. For very high \(x\), the increased decoherence dominates and \(p_\mathrm{th}\) drops sharply.

We therefore perform a binary search over \(x\) to locate the optimal percentile that maximizes \(p_\mathrm{th}\) within statistical error bars. The reported code threshold for each architecture is the highest \(p_\mathrm{th}\) obtained in this search, using the corresponding entanglement-generation scheme. In the next subsection, we highlight the exact fitting procedure that we use to estimate thresholds from the logical success rates for varying code distances.

\begin{figure}[hbtp]
    \centering
    \includegraphics[width=1.0\linewidth]{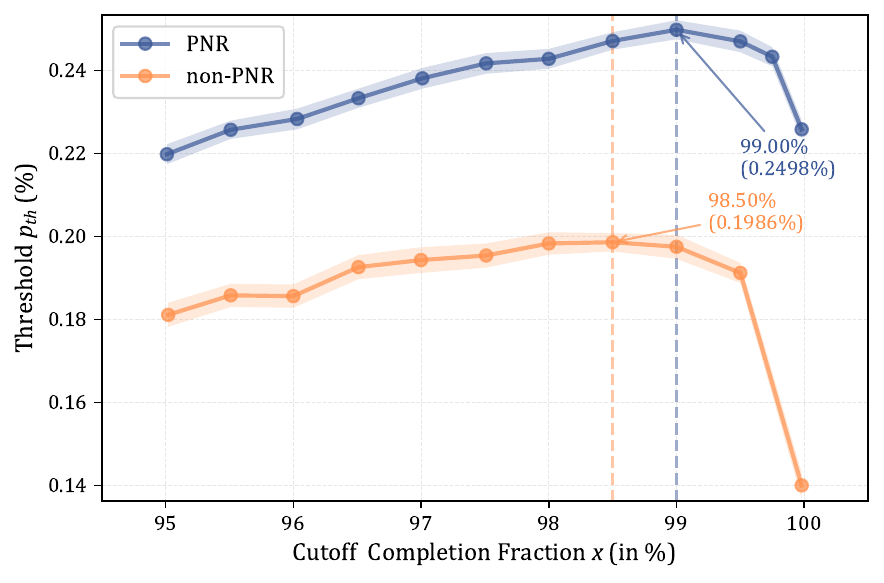}
    \caption{Code threshold versus cutoff completion fraction for the OD-GHZ protocol under PNR (blue) and non-PNR (orange) detection. Optimal thresholds $p_\mathrm{th}^{\text{PNR}} = 0.2498\%$ and $p_\mathrm{th}^{\text{non-PNR}} = 0.1986\%$ are attained at $x = 99.00\%$ and $x=98.50 \%$ (marked by dashed lines), respectively. The decline beyond the optimal completion fraction reflects the dominance of decoherence accumulation on idle qubits over gains in syndrome information, establishing the necessity for precise cutoff optimization in distributed surface-code.}
    \label{fig:cutoff_threshold}
\end{figure}

Fig.~\ref{fig:cutoff_threshold} demonstrates the fundamental trade-off between syndrome information fidelity and memory decoherence in optimizing the GHZ-generation cutoff completion fraction. For both photon-number-resolving (PNR, blue) and non-PNR (orange) detectors, the code threshold $p_\mathrm{th}$ initially increases with the completion fraction $x$ due to improved stabilizer measurement success, reaching optimal values of $p_\mathrm{th}^{\text{PNR}} = 0.2498\%$ and $p_\mathrm{th}^{\text{non-PNR}} = 0.1986\%$ at $x = 99.00\%$ and $x=98.50 \%$, respectively (marked by dashed lines). Beyond this point, the threshold degrades sharply as the prolonged waiting times for rare GHZ-generation events accumulate decoherence on memory qubits faster than the marginal improvement in syndrome outcomes can compensate, reflecting the exponential cost of extending $t_\text{cut}^{x\%}$ at high percentiles. The $\sim 26\%$ superiority of PNR over non-PNR detection underscores the critical importance of photon-number resolution for fault-tolerant performance in distributed emission-based architectures.

\subsection{\label{appsubsec:Threshold_fitting}Threshold curve fitting and estimation}

In this sub-section, we outline the procedure used to determine the threshold values for our quantum error-correcting code simulations, along with their associated uncertainties. The fitting methodology combines Monte Carlo sampling of logical success rates with a weighted non-linear regression model, followed by statistical error estimation. This numerical recipe is identical to the one followed in Ref.~\cite{10.1116/5.0200190}.

\subsubsection{Data acquisition and notation}
For a fixed set of physical error parameters, denoted by $p$, we compute an effective superoperator via the procedure in App.~\ref{app:superoperator}. Using this superoperator, we run Monte Carlo simulations for various surface code distances $d$ to obtain the logical success probability $1-p_L(p,d)=r_L(p,d)$.  

Let each simulation setting be indexed by $i \in \{1, \dots, n_T\}$, where $n_T$ is the total number of $(p,d)$ combinations simulated. The observed logical success rate is
\begin{equation}
    r_i = \frac{M_i}{N_i},
\end{equation}
where $N_i$ is the number of Monte Carlo trials and $M_i$ is the number of trials with no logical failure.  

Statistical uncertainty is assumed to follow a binomial distribution, giving the standard deviation
\begin{equation}
    \sigma_i = \sqrt{ \frac{ r_i(1-r_i) }{ N_i } }.
\end{equation}
Uncertainty decreases with increasing $N_i$, so higher-statistics points naturally carry more weight in the fit.

\subsubsection{Regression model}
Following standard finite-size scaling theory for code thresholds, we fit the logical success rates to~\cite{WANG200331}
\begin{equation}
    \hat{r}(p,d) =
        a + b\,(p - p_\mathrm{th}) d^{1/\kappa}
        + c\,(p - p_\mathrm{th})^2 d^{2/\kappa}
        + e\,d^{-1/\zeta},
    \label{eq:fitting_model}
\end{equation}
where $a,b,c,e,p_\mathrm{th},\kappa,\zeta$ are free parameters determined by the fit. Here, $p_\mathrm{th}$ is the threshold of interest, $\kappa$ is the finite-size scaling exponent, and $\zeta$ governs subleading corrections.

For each data point $(p_i, d_i)$, the residual is
\begin{equation}
    \epsilon_i = r_i - \hat{r}(p_i, d_i).
\end{equation}
The parameters are found by minimizing the weighted sum of squared residuals
\begin{equation}
    Q = \sum_{i=1}^{n_T} \left( \frac{ \epsilon_i }{ \sigma_i } \right)^2,
    \label{eq:q_statistic}
\end{equation}
which ensures that data with higher statistical precision (smaller $\sigma_i$) contribute more strongly to the optimization.

\subsubsection{Weighted non-linear least squares}
We solve the minimization of Eq.~\eqref{eq:q_statistic} using the Gauss–Newton algorithm for non-linear regression~\cite{ruckstuhlIntroductionNonlinearRegression2010, constalesChapterExperimentalData2017}. Let the parameter vector at iteration $t$ be
\begin{equation}
    \boldsymbol{\beta}^{(t)} = 
    \begin{bmatrix}
        a^{(t)} & b^{(t)} & c^{(t)} & d^{(t)} &
        p_\mathrm{th}^{(t)} & \kappa^{(t)} & \zeta^{(t)}
    \end{bmatrix}^T,
\end{equation}
and define the parameter update $\Delta\boldsymbol{\beta}^{(t+1)} = \boldsymbol{\beta}^{(t+1)} - \boldsymbol{\beta}^{(t)}$.  

Linearizing Eq.~\eqref{eq:fitting_model} around $\boldsymbol{\beta}^{(t)}$ yields the iterative update rule
\begin{equation}
    \Delta\boldsymbol{\beta}^{(t+1)} =
        \left[ (\mathbf{J}^{(t)})^T \boldsymbol{\Sigma}^{-1} \mathbf{J}^{(t)} \right]^{-1}
        (\mathbf{J}^{(t)})^T \boldsymbol{\Sigma}^{-1} \boldsymbol{\epsilon}^{(t)},
    \label{eq:gn_update}
\end{equation}
where $\mathbf{J}^{(t)}$ is the Jacobian matrix of partial derivatives $\partial\hat{r}/\partial\beta_j$ evaluated at step $t$,  
$\boldsymbol{\epsilon}^{(t)}$ is the residual vector at step $t$, and $\boldsymbol{\Sigma} = \mathrm{diag}(\sigma_1^2, \dots, \sigma_{n_{\mathrm{C}}}^2)$.

The iteration proceeds until $\boldsymbol{\beta}^{(t)}$ converges within a chosen tolerance.

\subsubsection{Estimating parameter uncertainties}
Once convergence is reached, the covariance matrix of the fitted parameters is estimated as
\begin{equation}
    \mathrm{Var}(\hat{\boldsymbol{\beta}}) \approx
    \left[ \mathbf{J}^T \boldsymbol{\Sigma}^{-1} \mathbf{J} \right]^{-1},
\end{equation}
where $\mathbf{J}$ is the Jacobian matrix at the converged parameter values.

The quality of the fit is assessed using the reduced chi-squared statistic
\begin{equation}
    \chi^2_\nu = \frac{Q}{\nu}, \qquad \nu = n_T - n_P,
\end{equation}
where $n_P=7$ is the number of fitting parameters.  
A value $\chi^2_\nu \approx 1$ indicates a statistically consistent fit; $\chi^2_\nu > 1$ may suggest model inadequacy or underestimated $\sigma_i$.

If $\chi^2_\nu > 1$, we rescale $\boldsymbol{\Sigma}$ by $\chi^2_\nu$ to obtain more conservative uncertainty estimates:
\begin{equation}
    \mathrm{Var}(\hat{\boldsymbol{\beta}}) =
        \chi^2_\nu \left[ \mathbf{J}^T \boldsymbol{\Sigma}^{-1} \mathbf{J} \right]^{-1}.
\end{equation}

\subsection{Confidence intervals for \texorpdfstring{$p_\mathrm{th}$}{pth}}
The standard deviation of $\hat{p}_\mathrm{th}$ is given by the square root of the corresponding diagonal entry of $\mathrm{Var}(\hat{\boldsymbol{\beta}})$.  
Confidence intervals are obtained by multiplying this standard deviation by the Student’s $t$-factor $t_{\mathrm{ci}}$ corresponding to the desired confidence level and $\nu$ degrees of freedom.

For the 95\% confidence intervals shown in our threshold plots, $t_{\mathrm{ci}} \approx 1.96$ for large $\nu$, with slightly larger values for small sample sizes.

%%%%%%%%%%%%%%%%%%%%%%%%%%%%%%%

%%%%%%%%%%%%%%%%%%%%%%%%%%%%%%%
% \bibliographystyle{apsrev4-2}
\bibliography{references}
%%%%%%%%%%%%%%%%%%%%%%%%%%%%%%%
\end{document}